\documentclass[3p,times]{elsarticle}

\usepackage{amssymb}
\usepackage{amsmath,graphicx,bm}
\usepackage{booktabs, multirow,pgfplots,makecell,float,threeparttable,tikz,color}
\usepackage[ruled]{algorithm2e} 
\usepackage[pagewise, displaymath, mathlines]{lineno}

\UseRawInputEncoding
\usetikzlibrary{arrows.meta}
\journal{CMAME}
\newcommand{\mc}{\textbf{\emph{c}}}
\newcommand{\mpp}{\textbf{\emph{p}}}
\newcommand{\mq}{\textbf{\emph{q}}}
\newcommand{\mx}{\textbf{\emph{x}}}

\newcommand{\mt}{\textbf{\emph{t}}}
\newcommand{\mv}{\textbf{\emph{v}}}
\newcommand{\mss}{\textbf{\emph{S}}}

\newcommand{\ma}{\textbf{\emph{a}}}
\newcommand{\muu}{\textbf{\emph{u}}}
\newcommand{\meps}{\bm{\varepsilon}}
\newcommand{\mkap}{\bm{\kappa}}
\newcommand{\mn}{\textbf{\emph{n}}}
\newcommand{\mm}{\textbf{\emph{m}}}

\definecolor{mygreen}{rgb}{0.47, 0.67, 0.19}
\definecolor{mypurple}{rgb}{0.49 0.18 0.56}
\newtheorem{remark}{Remark}

\usepackage[labelformat=simple]{subcaption}

\usetikzlibrary{shapes.geometric,shapes.symbols}
\begin{document}
\begin{frontmatter}



\title{Kirchhoff-Love shell representation and analysis using triangle configuration B-splines}

\author[author1,author2]{Zhihao Wang}
\author[author1,author2]{Juan Cao\corref{mycorrespondingauthor}}
\cortext[mycorrespondingauthor]{Corresponding author}
\ead{juancao@xmu.edu.cn}
\author[author5]{Xiaodong Wei}
\author[author6]{Zhonggui Chen}
\author[author7]{Hugo Casquero}
\author[author8]{Yongjie Jessica Zhang}
\address[author1]{School of Mathematical Sciences, Xiamen University, Xiamen, Fujian, 361005, China}
\address[author2]{Fujian Provincial Key Laboratory of Mathematical Modeling and High-Performance Scientific Computation,\\ Xiamen University, Xiamen, Fujian, 361005, China}
\address[author5]{University of Michigan-Shanghai Jiao Tong University Joint Institute, Shanghai Jiao Tong University, Shanghai, China}
\address[author6]{School of Informatics, Xiamen University, Xiamen, Fujian, 361005, China}
\address[author7]{Mechanical Engineering Department, University of Michigan - Dearborn, 4901 Evergreen Road, Dearborn, MI 48128-1491, USA}
\address[author8]{Department of Mechanical Engineering, Carnegie Mellon University, Pittsburgh, PA 15213, USA}


\begin{abstract}
 This paper presents the application of triangle configuration B-splines (TCB-splines) for representing and analyzing the Kirchhoff-Love shell in the context of isogeometric analysis (IGA). The Kirchhoff-Love shell formulation requires global $C^1$-continuous basis functions. The nonuniform rational B-spline (NURBS)-based IGA has been extensively used for developing Kirchhoff-Love shell elements. However, shells with complex geometries inevitably need multiple patches and trimming techniques, where stitching patches with high continuity is a challenge. On the other hand, due to their unstructured nature, TCB-splines can accommodate general polygonal domains, have local refinement, and are flexible to model complex geometries with $C^1$ continuity, which naturally fit into the Kirchhoff-Love shell formulation with complex geometries. Therefore, we propose to use TCB-splines as basis functions for geometric representation and solution approximation. We apply our method to both linear and nonlinear benchmark shell problems, where the accuracy and robustness are validated. The applicability of the proposed approach to shell analysis is further exemplified by performing geometrically nonlinear Kirchhoff-Love shell simulations of a pipe junction and a front bumper represented by a single patch of TCB-splines.

\end{abstract}




\end{frontmatter}


\section{Introduction}
Shell structures have been ubiquitous in applications such as civil, automotive, and aerospace engineering, all attributed to their optimal load-carrying behavior~\cite{Bischoff:2004:ECM}. Therefore, the analysis of shell structures has been the subject of intense interest over the past couple of decades~\cite{Kiendl:2010:CMAME}. Kirchhoff\--{}Love (KL)~\cite{Chapelle:1998:CS,Kiendl:2009:CMAME} and Reissner\--{}Mindlin (RM)~\cite{Simo:1989:CMAME,Simo:1990:CMAME,Benson:2010:CMAME} theories are the most widely used formulations for thin- and thick-shell analysis, respectively. The KL formulation requires the use of shape functions with global $C^1$ continuity, which limits the use of traditional Lagrangian shape functions. Achieving $C^1$ continuity for arbitrarily shaped cells is challenging even with higher-order polynomials. 

In recent years, IGA has become a popular method to analyze shell structures. Many engineering analysis objects can be idealized as smooth surfaces,  which can be accurately represented by NURBS. A KL shell theory can also be discretized using the same NURBS basis functions, resulting in a geometrically exact discrete formulation. However, NURBS-based IGA has some limitations when it comes to the analysis of shells with complex geometries. These difficulties mainly stem from the tensor product nature inherent in NURBS. For example, shells with complex geometries often lead to trimming or multi-patch descriptions. The trimmed patches usually have no independent representation to pass to the analyst. Shells with multi-patches cannot be directly analyzed either because higher continuity across the patch connections is not automatically preserved, and different patches have different parameterizations. Especially in the case of trimmed surfaces, the multi-patches are often jointed with gaps and overlaps~\cite{Sederberg:2008:TOG}, which fails to meet the  $C^0$ continuity requirements for shell analysis. From an analysis perspective, the tensor product structure of NURBS proves to be inefficient, caused by the global refinement operation. Local refinement is desirable for high-precision and efficient simulations to reduce approximation errors with fewer degrees of freedom.

To address these limitations, various approaches have been proposed to enhance the analysis to be adaptable enough to cope with models with geometric flaws, including gaps, overlaps, trimmed boundaries, and non-conforming multi-patches~\cite{Hiemstra:2020:CMAME,Shepherd:2022:CMAME}. On the other hand, many contributions have been performed to circumvent these issues at their origin. Different surface descriptions, such as T-splines~\cite{Bazilevs:2010:CMAME,Casquero:2017:CAD}, subdivision surface (SubD)~\cite{Bandara:2018:CAD}, and rational triangular B\'{e}zier splines (rTBS)~\cite{Zareh:2019:CMAME} are employed to remodel shells with complex structures and avoid geometric flaws. However, their success heavily relies on the approximation properties of the basis functions used for describing the shapes. For example, both T-splines and subdivision surfaces encounter the problem of interior extraordinary points when representing complex geometries. For rTBS, stitching multiple patches together to form complex topologies is non-trivial if specific continuity requirements are to be maintained.

Recently, the so-called triangle configuration B-splines (TCB-splines) have been introduced to the IGA community to increase the robustness of IGA and allow for a broader application of analysis~\cite{Cao:2019:CMAME,Wang:2022:CMAME,Zhu:2022:TOG}. TCB-spline surfaces can be defined over general polygonal domains. A degree $k$ TCB-spline surface is naturally $C^{k-1}$-continuous (almost) everywhere, except for a small region near the concave vertices, which is $C^1$-continuous. Compared with the traditional quad-based splines, TCB-splines significantly increase the complexity of the topology and enable the use of complex geometry in IGA. It also allows for flexible local adaptive refinement without introducing undesirable propagation. The attractive theoretical properties of TCB-splines and their success in complex geometry modeling and optimization-free parameterization in IGA~\cite{Wang:2022:CMAME} inspire us further to explore its application in complex shell modeling and analysis. 

In this paper, we use the TCB-splines for KL shell modeling and develop an isogeometric framework for complex shell analysis. A given multi-patch and trimmed CAD shell mid-surface is first discretized into a 3D triangular mesh and then projected onto the 2D plane to form a planar triangular mesh. Then, a smooth and watertight TCB spline surface is constructed on a 2D polygonal region derived from the planar triangular mesh and fitted to the 3D triangular mesh. As a result, our method successfully approximates a smooth surface that is topologically equivalent to an open disk with a finite number of holes using a single TCB-spline piece, without the need for extraordinary points within the parametric domain~\cite{Wang:2022:CMAME}. The obtained TCB-spline surface naturally meets the $C^1$-continuous requirement of KL theory. Finally, we analyze the obtained shells based on the isogeometric concept. The numerical analysis results verify the accuracy and effectiveness of our TCB-based shell analysis framework. 

The remainder of this paper is organized as follows. In Section~\ref{realated_work}, a brief review of KL shell analysis based on the isogeometric concept is given. The definition of TCB-splines, including simplex splines and triangle configuration, is described in Section~\ref{TCB-splines}. Geometrically linear and nonlinear Kirchhoff-Love shell theory in curvilinear coordinates is presented  in Section~\ref{KL_theory}. The TCB-spline-based shell fitting method is presented in Section~\ref{surface_fitting}. We demonstrate the capabilities of our framework in solving a series of linear and nonlinear problems in Section~\ref{numerical_examples}. The section also considers two shells with complex geometry, including a pipe-junction and front bumper models. Conclusions and future work are drawn in Section~\ref{conclusions}.

\section{Related work \label{realated_work}}

Since the first shell theory was proposed~\cite{Kirchhoff:1850:CJ}, various works have been developed to analyze shell structures. In this section, we limit ourselves to the state-of-the-art complex shell analysis methods based on the isogeometric concept. These methods can be roughly classified into three groups: the multi-patch coupling method, the immersed boundary method, and the reparameterization method.  

\textbf{Multi-patch coupling method.} The underlying geometries for complex engineering structures usually consist of multiple patches, typically non-conforming along their interfaces. Various techniques for coupling multiple patches have been proposed to analyze the non-conforming multi-patch shell structures using IGA. The main categories are the penalty methods~\cite{Kiendl:2010:CMAME,Liu:2021:CMAME}, the Lagrange multiplier methods~\cite{Brivadis:2015:CMAME,Kamensky:2017:CMAME, Miao:2021:CMAME} and the Nitsche's methods~\cite{Ruess:2014:CMAME,Guo:2021:CMAME}. The penalty methods are attractive because they are simple and efficient and do not introduce additional degrees of freedom. The main challenge of penalty methods is to find suitable penalty factors to ensure accurate solutions and avoid numerical problems. The Lagrange multiplier methods do not introduce the penalty factors, but they often lead to a saddle point problem, which requires the Lagrange multiplier space to satisfy inf-sup stability. While Nitsche's method neither requires an empirical parameter nor additional degrees of freedom, they depend on the respective variational formulation that is not easy to perform~\cite{Apostolatos:2014:IJNME,Coradello:2021:CMAME}. For all multi-patch coupling methods, the evaluation of the interfacial integrals considerably increases the computational complexity~\cite{Hu:2021:AMS}.

\textbf{Immersed boundary method.} Trimming is one of the most fundamental operations in Computer Aided Geometric Design (CAGD), enabling the construction of complex geometries. However, it is also one of the most severe impediments to interoperability between design and analysis since the trimming operation cannot be implemented exactly in the CAGD field~\cite{Marussig:2018:ACME}. Various approaches have addressed the problems of incorporating trimmed geometries into an analysis; see the survey paper~\cite{Marussig:2018:ACME}. Most publications on isogeometric analysis of trimmed geometries employ the immersed boundary methods or similar methods with other names, such as the finite cell methods (FCM)~\cite{Schillinger:2012:CMAME,Rank:2012:CMAME} and the embedded domain method~\cite{Coradello:2020:CM}. These methods embed a trimmed model into a simple background mesh while the trimming curves determine the domain of interest. Immersed methods have been successfully applied to various frameworks of trimmed shell analysis and demonstrated their strong application capabilities. { The weak enforcement of boundary conditions within the immersed boundary method was studied in~\cite{Ruess:2013:IJNME,Buffa:2020:SJNA} and then was applied to geometrically nonlinear coupling KL shells~\cite{Guo:2018:CMAME}, shell buckling analysis~\cite{Guo:2019:IJNME} and membrane locking in the context of nonlinear multi-patch analyses~\cite{Guo:2021:CMAME}. When the cut cells obtained from the intersection of the background mesh and the physical domain contain a small volume fraction, the immersed boundary method is prone to severe ill-conditioning and stability problems. These problems hinder solving the resulting system of equations, and additional efforts are needed to address the robustness issue from a computational point of view~\cite{De:2020:CM}. }

\textbf{Reparameterization method.} The reparameterization method aims to resolve models' deficiencies by reconstructing the trimmed or multi-patch geometries suffering from flaw issues into a new representation suitable for analysis utilizing other surface representations. For example, the trimmed shell models are remodeled into a set of NURBS patches that are watertight, feature-aware, and boundary-conforming in~\cite{Hiemstra:2020:CMAME,Shepherd:2022:CMAME}. To support local refinements, PHT-splines~\cite{Nguyen:2011:CMAME}, T-splines~\cite{Bazilevs:2010:CMAME} and THB-splines~\cite{Kiss:2014:GM,Atri:2018:SCS,Wei:2017:CMAME} were employed for representing shells with a simple structure. Furthermore, considering the ability for both design and analysis, the analysis-suitable unstructured T-splines (AST-splines) were developed by allowing multiple extraordinary points within the same face and applied to model shells with arbitrary topology~\cite{Liu:2015:CMAME,Casquero:2020:CMAME,Wei:2022:CMAME}. However, to define the basis functions and achieve the desired smoothness around extraordinary points, some special treatments are required~\cite{Lai:2017:CMA,Toshniwal:2017:CMAME,Wen:2023:CMAME}.
Subdivision models also possess great flexibility in complex shape modeling and have been applied to remodel shells not restricted to a regular grid structure. The remodeled subdivision surface is gap-free but still contains extraordinary points. In~\cite{Zareh:2019:CMAME},  rTBS were used to analyze KL shells with complex topology. While the rTBS have efficient local mesh refinement, the linear continuity constraints have to impose on the adjacent elements.

The so-called TCB-splines are new modeling tools for representing models with complex geometry and topology~\cite{Cao:2019:CMAME,Wang:2022:CMAME}. They possess appealing properties such as accommodating the general polygonal domains, automatic smoothness and allowing local refinement without propagation, which make them a suitable basis for shell analysis. In this paper, KL shells with complex geometries were remodeled and analyzed using TCB-splines to show their potential in industrial applications.

\section{TCB-splines \label{TCB-splines}}
In our shell analysis framework, the complex shell geometries are modeled by TCB-splines defined over a general polygonal domain. In this section, we first briefly recall the two ingredients for defining TCB-splines, i.e., simplex splines and triangle configurations. Then we describe the construction of TCB-splines. Finally, we refer the reader to~\cite{Liu:2007:CG, Zhang:2017:cagd, Cao:2019:CMAME, Schmitt:2021:JCSS, Wang:2022:CMAME} for a detailed introduction to TCB-splines. 
\subsection{Simplex splines}

\begin{figure}
  \centering
   \graphicspath{{figures/}}
  \begin{subfigure}{0.22\textwidth}{\includegraphics[width=\textwidth]{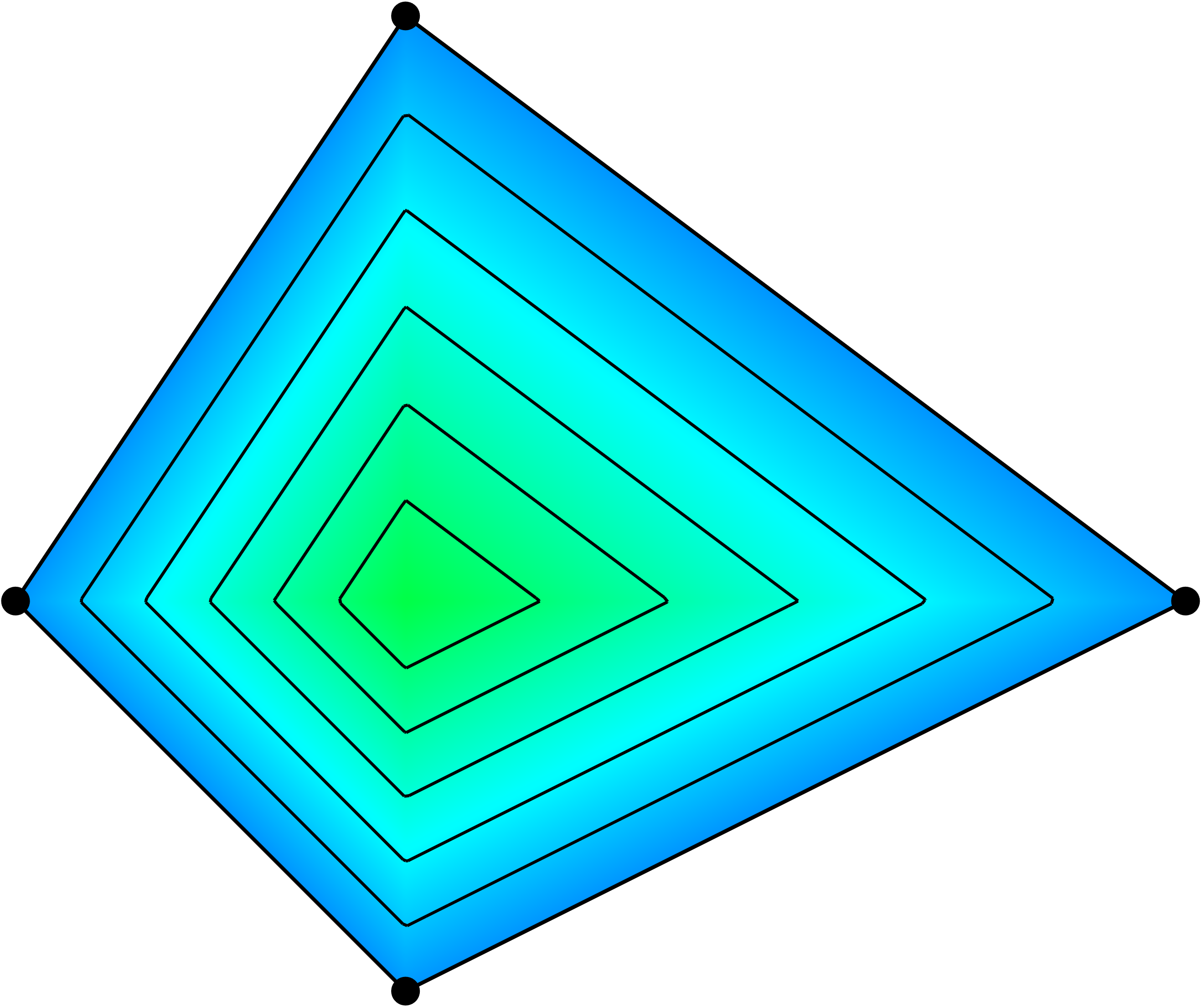}}   \caption{} \end{subfigure}\quad
  \begin{subfigure}{0.22\textwidth}{\includegraphics[width=\textwidth]{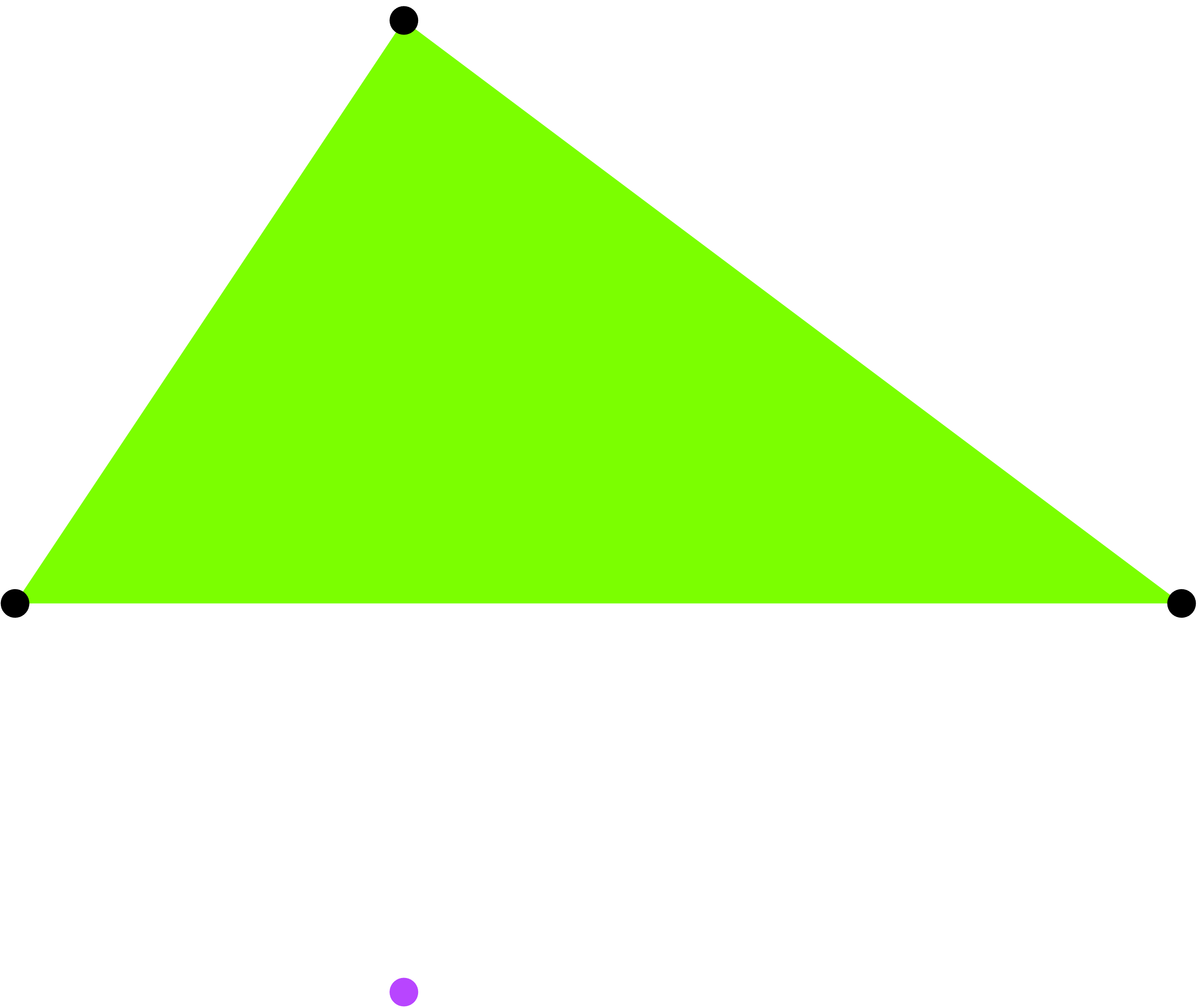}}   \caption{} \end{subfigure}\quad
  \begin{subfigure}{0.22\textwidth}{\includegraphics[width=\textwidth]{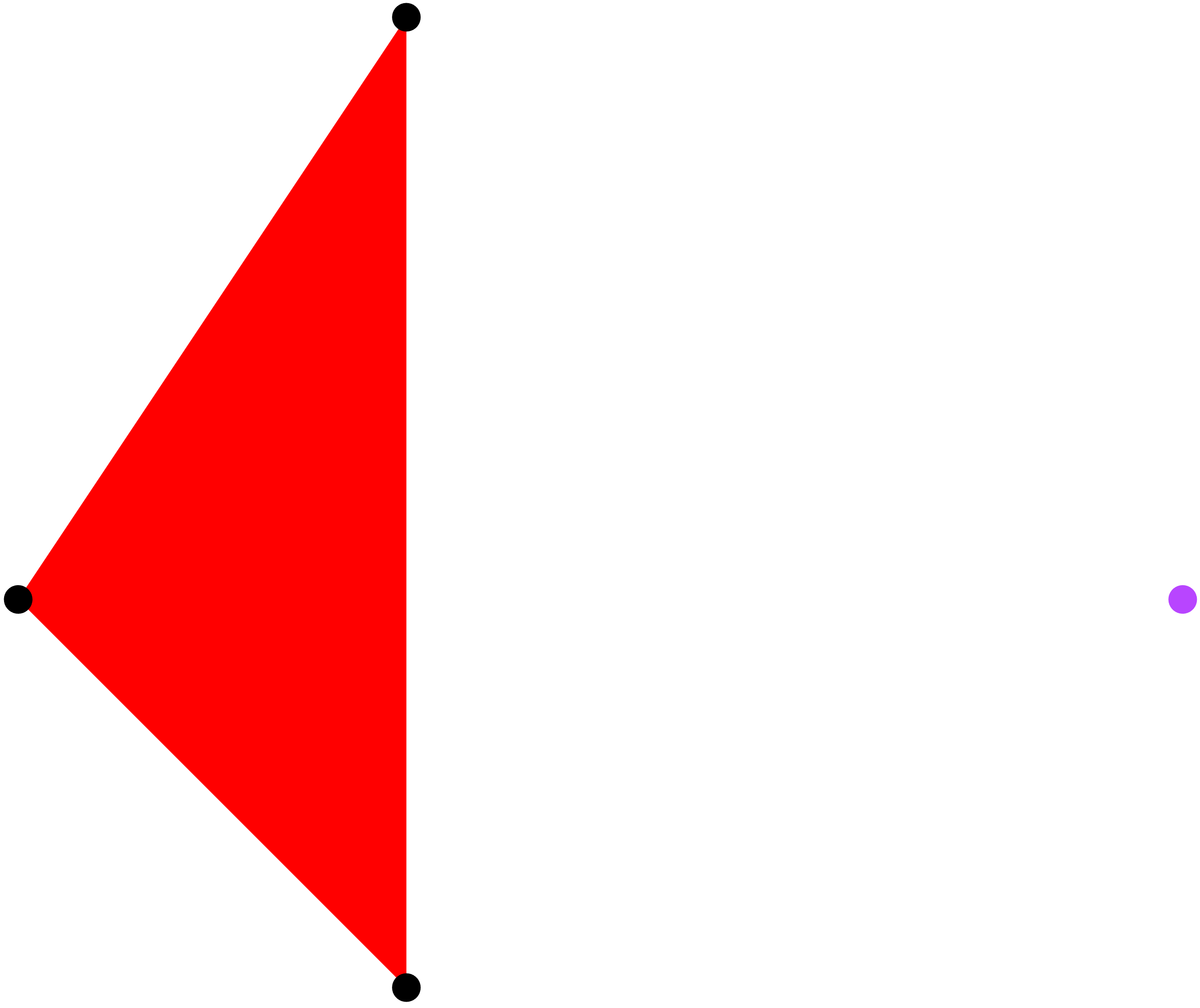}}   \caption{} \end{subfigure}\quad
  \begin{subfigure}{0.22\textwidth}{\includegraphics[width=\textwidth]{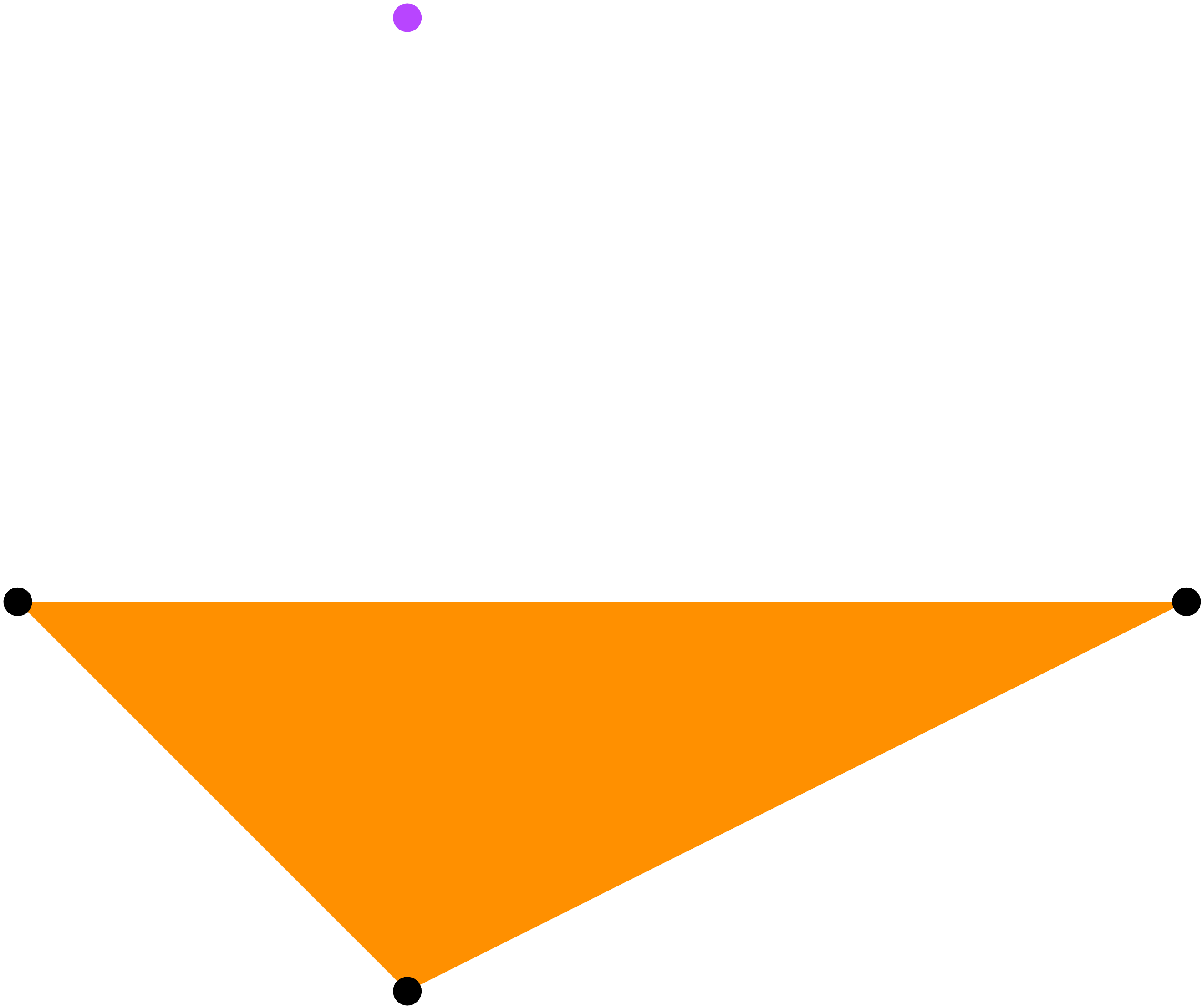}}   \caption{} \end{subfigure}\quad
  \begin{subfigure}{0.028\textwidth}{\includegraphics[width=\textwidth]{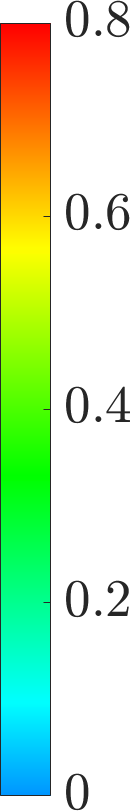}}  \end{subfigure}\quad

  \caption{Recursive evaluation of a degree-one simplex spline.  The degree-one simplex spline in (a) is computed by three degree-zero simplex splines in (b-d), where the solid black points mark the knots used to define the simplex splines.  The values of the simplex splines are color-coded, where red and blue colors at the top and bottom of the color bar represent the maximum and minimum values, respectively.}\label{fig:simplex}
\end{figure}

A degree-$k$ simplex spline is defined by a set of points $V=\{\mt_1, \mt_2, \cdots, \mt_{k+3}\} \subset  \mathbb{R}^2$, where each point $\mt_i\in V $ is referred to as a knot in the context of spline definition. We denote by $\textrm{Area}(\{\mt_{i,1}, \mt_{i,2}, \mt_{i,3}\})$ the oriented area of the oriented triangle formed by these three points $\{\mt_{i,1}, \mt_{i,2}, \mt_{i,3}\}$. Assume that the set of knots $V$ is non-degenerated, i.e., there exist at least three knots $\{\mt_{i,1}, \mt_{i,2}, \mt_{i,3}\} \subset V$ with non-zero oriented area, then the degree-$k$ simplex spline associated with a set of $k+3$ non-degenerated knots $V$ is a piece-wise polynomial that can be recursively defined as the linear combination of three degree-$k-1$ simplex splines as follows~\cite{Micchelli:1980}:
\begin{equation}
  M(\mt|V)=\sum_{j=1}^3 \zeta_j(\mt|\hat{V})M(\mt|V\setminus \{\mt_{i,j}\}), \quad \mt \in \mathbb{R}^2,
  \label{equation:simplex_splines}
\end{equation}
where $\hat{V}=\{\mt_{i,1}, \mt_{i,2}, \mt_{i,3}\}\subseteq V$ is any {\it non-degenerated} set, and $\{\zeta_j(\mt|\hat{V})\}$ are the barycentric coordinates of $\mt$ with respect to the triangle $\hat{V}$ as
\begin{equation*}
(\zeta_1(\mt|\hat{V}),\zeta_2(\mt|\hat{V}),\zeta_3(\mt|\hat{V}))  = \left(\frac{\textrm{Area}(\{\mt, \mt_{i,2}, \mt_{i,3}\})}{\textrm{Area}(\{\mt_{i,1}, \mt_{i,2}, \mt_{i,3}\})},\frac{\textrm{Area}(\{\mt_{i,1}, \mt, \mt_{i,3}\})}{\textrm{Area}(\{\mt_{i,1}, \mt_{i,2}, \mt_{i,3}\})},\frac{\textrm{Area}(\{\mt_{i,1}, \mt_{i,2}, \mt\})}{\textrm{Area}(\{\mt_{i,1}, \mt_{i,2}, \mt_{i,3}\})}\right).
\end{equation*}
For $k=0$, $$ M(\mt|V)=
\begin{cases}
0& \text{$\mt \notin [V)$},\\
\frac{1}{|\textrm{Area}(V)|}& \text{$\mt \in [V)$},
\end{cases}$$
where $[V)$ is the half-open convex hull of $V$~\cite{Franssen:1995}. Fig.~\ref{fig:simplex} shows an example of the evaluation of a degree-one simplex spline. Simplex splines have several properties that are important for the application in the context of this paper~\cite{Prautzsch:2002:Springer, Micchelli:1979:Springer}: 
\begin{itemize}
  \item[$\bullet$] Non-negativity: $M(\mt|V)\geq0$ for any $\mt \in \mathbb{R}^2$.
  \item[$\bullet$] Local support: If $\mt$ does not belong to the convex hull of $V$, then $M(\mt|V)=0$.
  \item[$\bullet$] Continuity: For a degree-$k$ simplex spline $M(\mt|V)$, if there are no three knots collinear in $V$, then $M(\mt|V)$ is $C^{k-1}$-continuous. Otherwise, if $n~(3\leq n\leq k+3)$ knot in $V$ are collinear, then $M(\mt|V)$ is $C^{k+1-n}$-continuous across the line formed by the collinear knots.
\end{itemize}


\subsection{Triangle configurations \label{sec:LTP}}
\begin{figure}
  \centering
 \graphicspath{{figures/}}
 \begin{subfigure}[b]{0.182\textwidth} \centering \includegraphics[width=\textwidth]{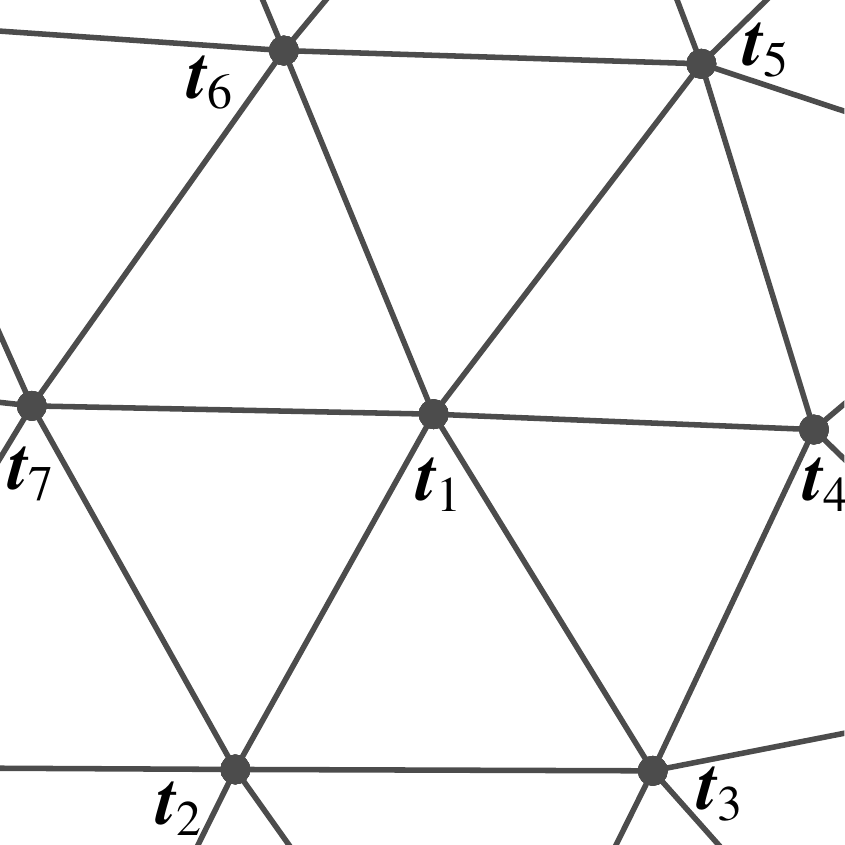}  \caption{} \end{subfigure} \quad
 \begin{subfigure}[b]{0.182\textwidth} \centering \includegraphics[width=\textwidth]{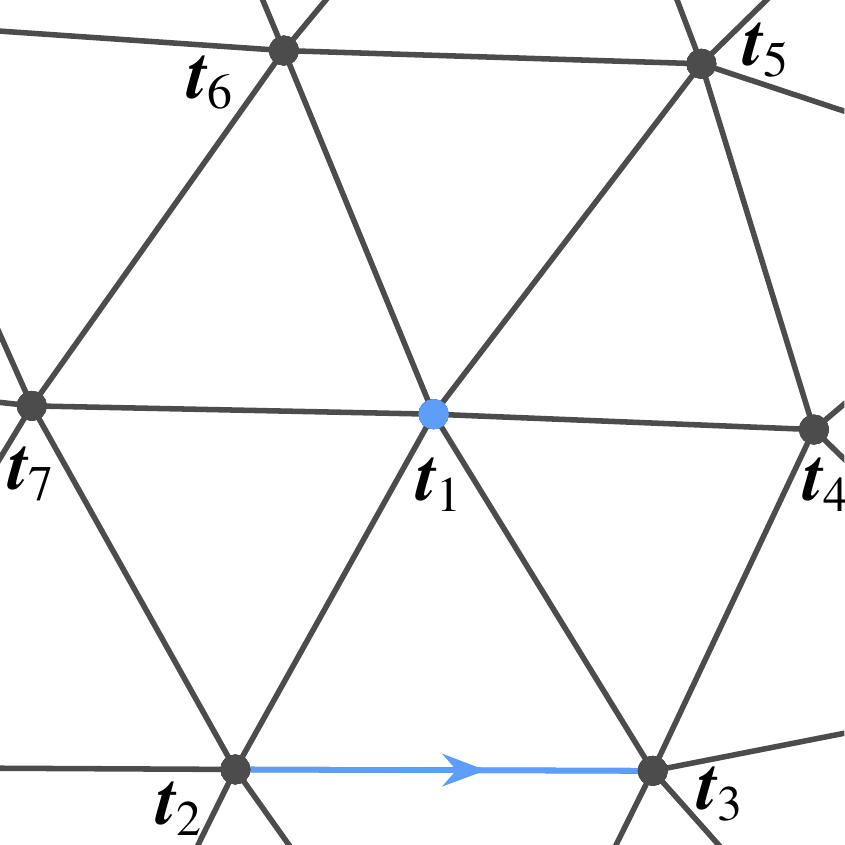}  \caption{}  \end{subfigure} ~
 \begin{subfigure}[b]{0.182\textwidth} \centering \includegraphics[width=\textwidth]{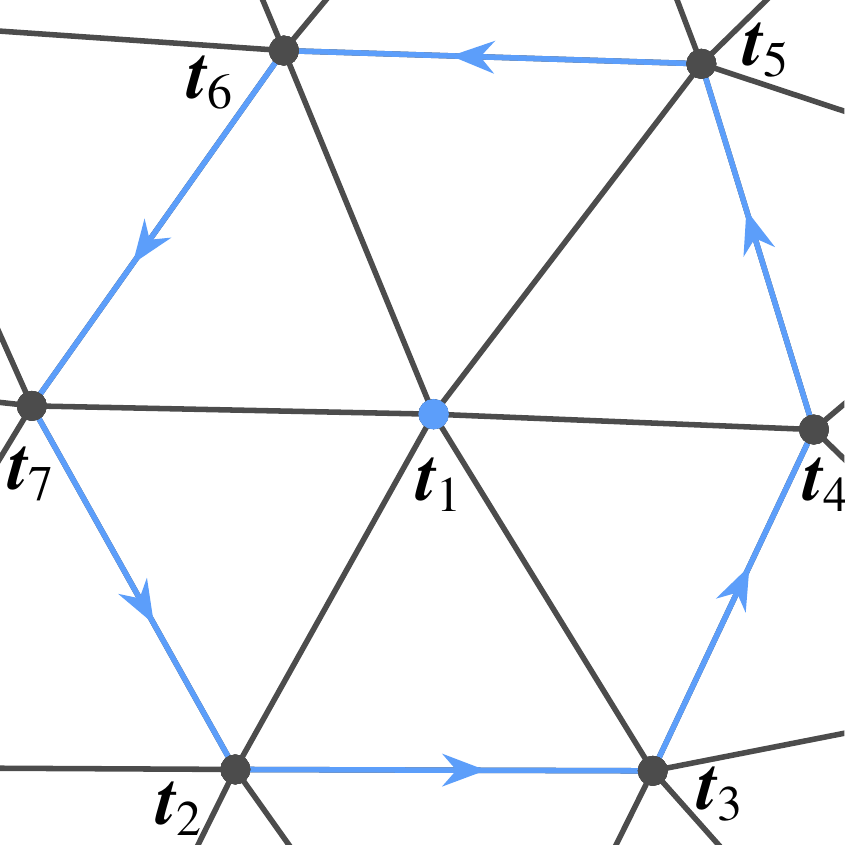}  \caption{}  \end{subfigure} ~
 \begin{subfigure}[b]{0.182\textwidth} \centering \includegraphics[width=\textwidth]{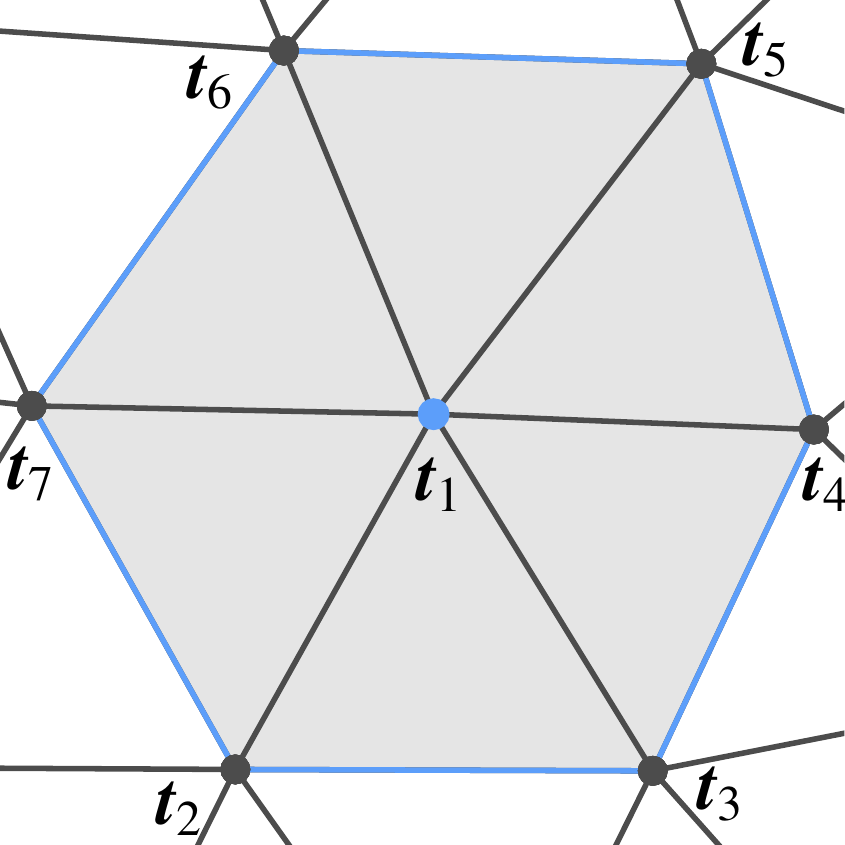}  \caption{} \end{subfigure} ~
 \begin{subfigure}[b]{0.182\textwidth} \centering \includegraphics[width=\textwidth]{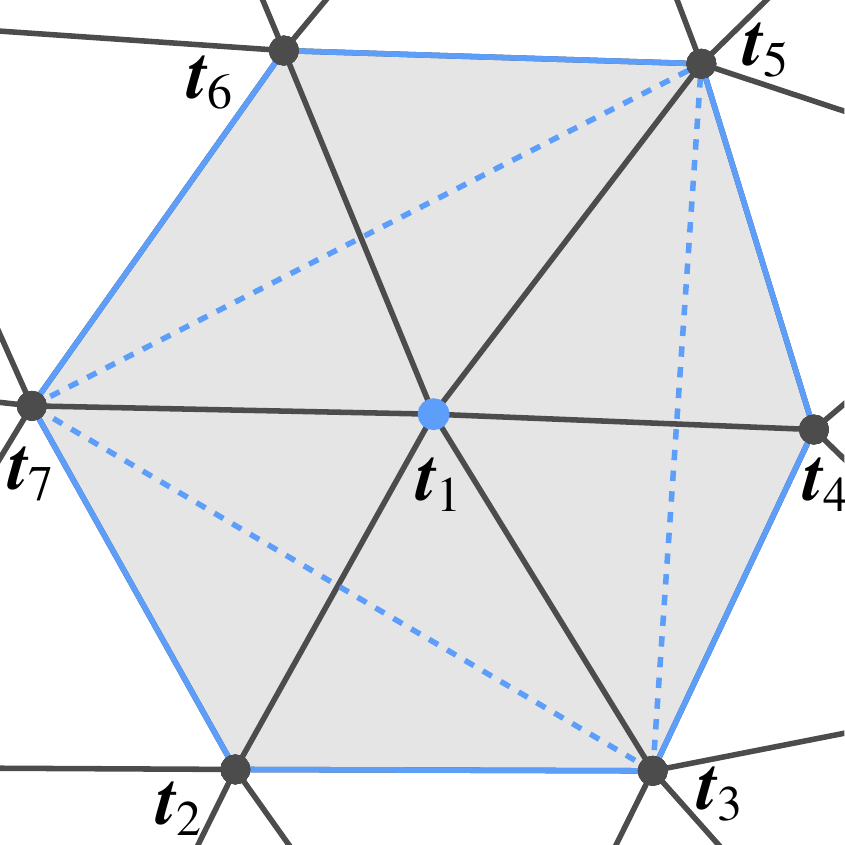}  \caption{} \end{subfigure} ~
\caption{Basic notations in t-configs. (a) Given a knot set and an initial triangulation, corresponding to a family of degree-zero t-configs $\{(\{\mt_1,\mt_2,\mt_3\},\emptyset), (\{\mt_1,\mt_3,\mt_4\},\emptyset), \cdots\}$;  (b) a vertex $\{\mt_1\} = \{\mt_1\}\cup \emptyset$ (the solid blue point) and the edge $\protect\overrightarrow{\mt_2\mt_3}$ (the blue arrow) obtained by degree-zero t-config $(\{\mt_1,\mt_2,\mt_3\},\emptyset)$
; (c) the edges of different degree-zero t-configs that correspond to the same vertex $\{\mt_1\}$; (d) the closed simple polygon $P_{\{\mt_1\}} = \{\mt_2, \mt_3, \mt_4, \mt_5, \mt_6, \mt_7\}$ shaded in gray; (e) a constrained triangulation of polygon $P_{\{\mt_1\}}$, marked by blue dashed lines, resulting in four degree-one t-configs $\{(\{\mt_2,\mt_3,\mt_7\},\{\mt_1\}), (\{\mt_3,\mt_4,\mt_5\},\{\mt_1\}), (\{\mt_3,\mt_5,\mt_7\},\{\mt_1\}),(\{\mt_5,\mt_6,\mt_7\},\{\mt_1\})\}$.}\label{fig:ltp}
\end{figure}
A degree-$k$ configuration of a knot set $A$ is a subset of $A$ with a size of $k+3$. A degree-$k$ simplex spline can be associated with a degree-$k$ configuration. To define a degree-$k$ simplex spline space over a finite knot set $A$, we first need to get a family of degree-$k$ configurations of $A$. Many methods for defining different configuration families have been proposed in the past~\cite{Neamtu:2001:VU,Liu:2007:CG,Schmitt:2021:JCSS}. In this paper, our spline space is defined based on the so-called triangle configurations (t-configs for short)~\cite{Schmitt:2021:JCSS}. A degree-$k$ t-config is composed of a pair of subset $(T, I)$ with $\#T=3,~\#I=k,~T,I\subset A$, where ``$\#$'' refers to the size of a set. To ease the discussion, we also denote by $T$ the counterclockwise oriented triangle with the three elements of a subset $T$ as vertices. Denote by
$\it\Gamma_k$ the degree-$k$ t-config family. We 
use the so-called link triangulation procedure (LTP) to construct valid t-config families~\cite{Liu:2007:CG,Schmitt:2021:JCSS}.

 Note that any triangulation $\mathcal{T}$ of the given knot set $A$ corresponds to a family of degree-zero t-configs ${\it\Gamma}_0$; see Fig.~\ref{fig:ltp}(a). We refer to  $\mathcal{T}$ as the knot mesh of the given knot set. LTP then starts with an arbitrary triangulation and recursively computes t-config families of higher degrees. 
We introduce concepts and necessary notations for computing higher degree t-configs here and refer the reader to ~\cite{Liu:2007:CG,Schmitt:2021:JCSS} for more details. For a degree-$k$ t-config $(T, I)$ and any $\mv\in T$,
\begin{itemize}
  \item [$\bullet$] the union of $\{\mv\}$ and $I$, i.e., $\{\mv\}\cup I$, is referred to as a vertex of the t-config; see Fig.~\ref{fig:ltp}(b).
  \item [$\bullet$] the oriented edge of $T$ opposite to $\mv$, denoted by $_{\mv}T$, is referred to as an edge of the t-config; see Fig.~\ref{fig:ltp}(b).
\end{itemize}
Note that the vertex and edge of a t-config come in pairs, and different t-configs may have a vertex in common. Let $V(\it\Gamma_k)$ denote the set of vertices of all degree-$k$ t-configs in $\it\Gamma_k$. For each vertex $\mathcal{V}\in V(\it\Gamma_k)$, the corresponding edges in the different degree-$k$ t-configs form either a closed or an open simple polygon of line segments that do not cross each other~\cite{Schmitt:2021:JCSS}; see Fig.~\ref{fig:ltp}(c). The open polygons can be further closed using the method in~\cite{Wang:2022:CMAME}. We denote $P_{\mathcal{V}}$ as the simple closed polygon derived from a vertex $\mathcal{V}\in V(\it\Gamma_k)$; see Fig.~\ref{fig:ltp}(d). If $P_{\mathcal{V}}$ is not degenerated, we compute a constrained triangulation of it; see Fig.~\ref{fig:ltp}(e). Then, each triangle $T^*$ of the triangulation leads to a degree-$k+1$ t-config $(T^*,\mathcal{V})$. The pseudo-code of LTP is shown in Algorithm~\ref{alg:LTP}.  Fig.~\ref{fig:ltp} illustrates the process of generating degree-one t-configs. Using different triangulation in each step throughout the LTP process results in different t-config families. In this paper, we use the Delaunay triangulation method to obtain the initial triangulation and partition of the polygon in LTP by default. 

\begin{algorithm}[t]
\SetAlgoNoLine
\KwIn{A family of degree-$k$ t-configs $\it\Gamma_{k}$.}
\KwOut{A family of degree-$k+1$ t-configs $\it\Gamma_{k+\textrm{1}}$.}
$\it\Gamma_{k+\textrm{1}} \leftarrow \emptyset$\;

  \For{each vertex  $\mathcal{V} \in V(\Gamma_k)$}
    {
      calculate the simple closed polygon {$P_{\mathcal{V}}$ derived from $\mathcal{V}$\;
      \If{$P_{\mathcal{V}}$ is non-degenerated }{
        compute a constrained triangulation of $P_{\mathcal{V}}$\;
         \For{each triangle $T^*$ of the triangulation}
    {
    $ \it\Gamma_{k+\textrm{1}} \leftarrow \it\Gamma_{k+\textrm{1}} \bigcup (T^*, \mathcal{V})$\;
    }
      }
        }
        }
\caption{Link triangulation procedure}
\label{alg:LTP}
\end{algorithm}

\subsection{TCB-spline basis functions and surfaces \label{TCB-surface}}

\begin{figure}
  \centering
 \graphicspath{{figures/}}
 \begin{subfigure}[b]{0.35\textwidth} \centering \includegraphics[width=\textwidth]{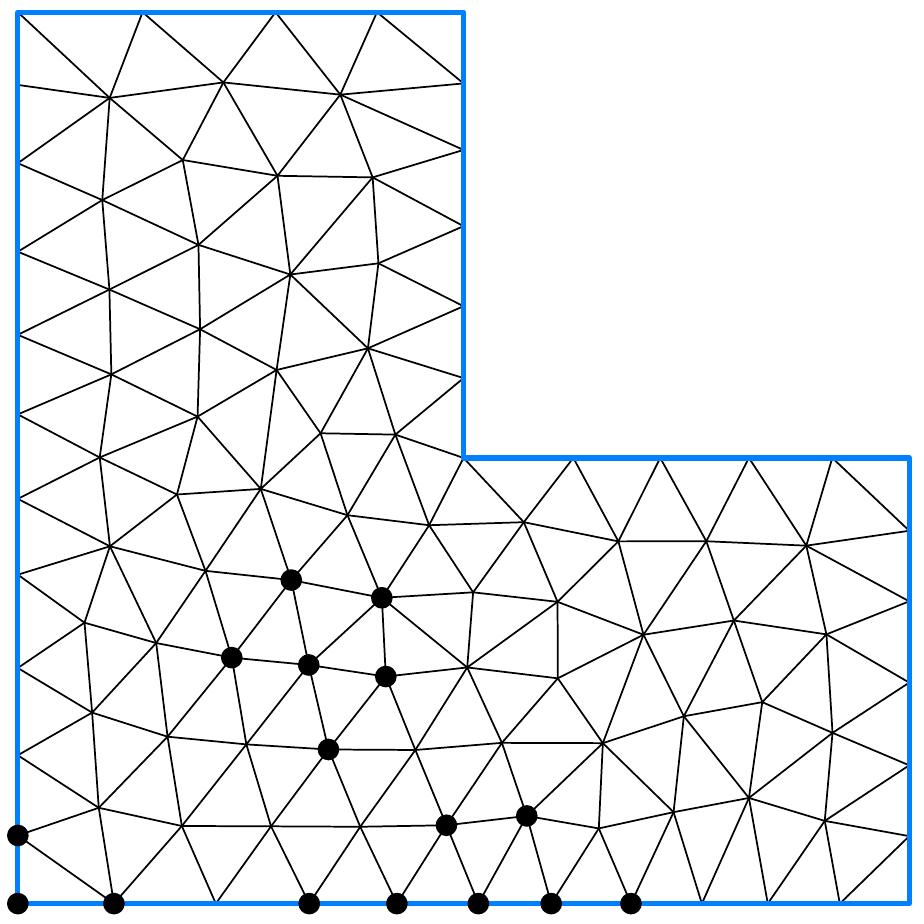} \caption{} \end{subfigure} ~
 \begin{subfigure}[b]{0.5\textwidth} \centering \includegraphics[width=\textwidth]{ 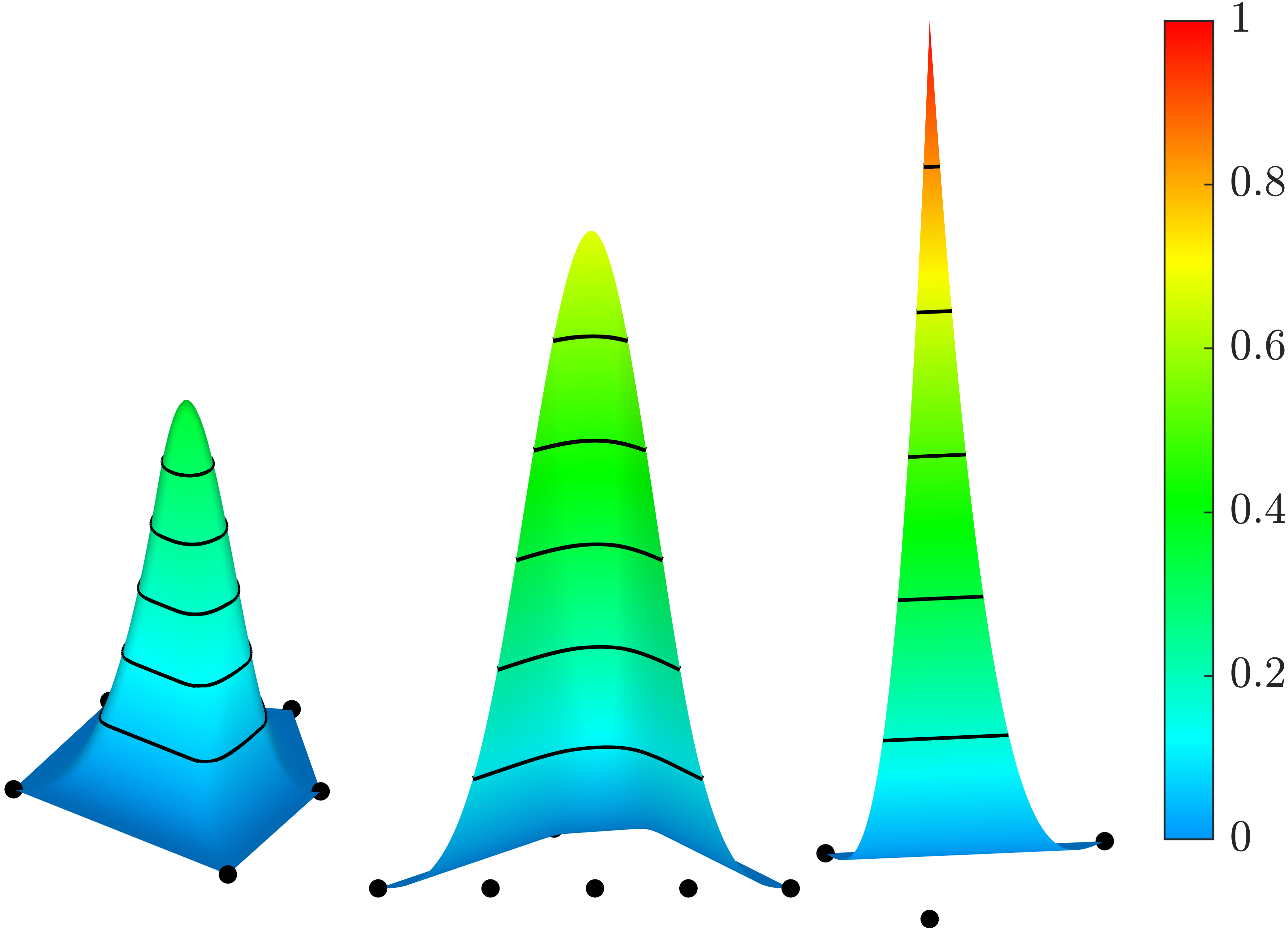} \caption{} \end{subfigure}~

 \begin{subfigure}[b]{0.33\textwidth} \centering \includegraphics[width=\textwidth]{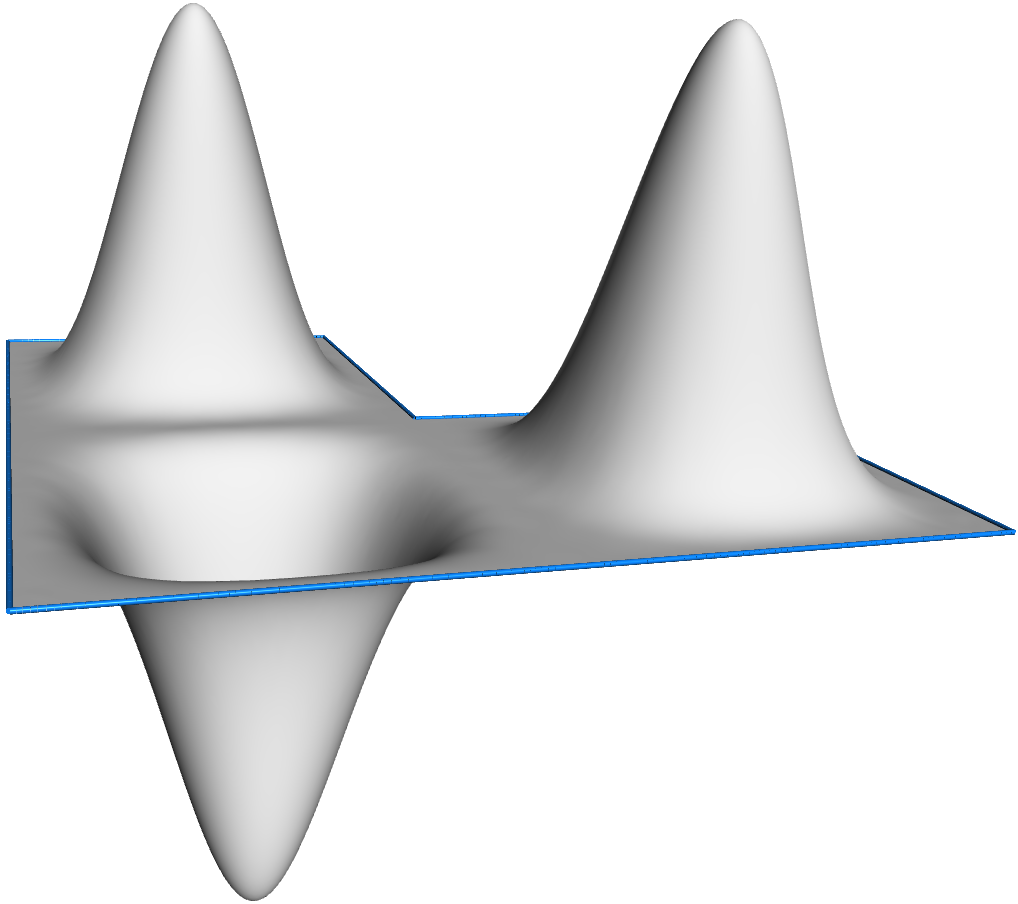} \caption{} \end{subfigure}
 \begin{subfigure}[b]{0.33\textwidth} \centering \includegraphics[width=\textwidth]{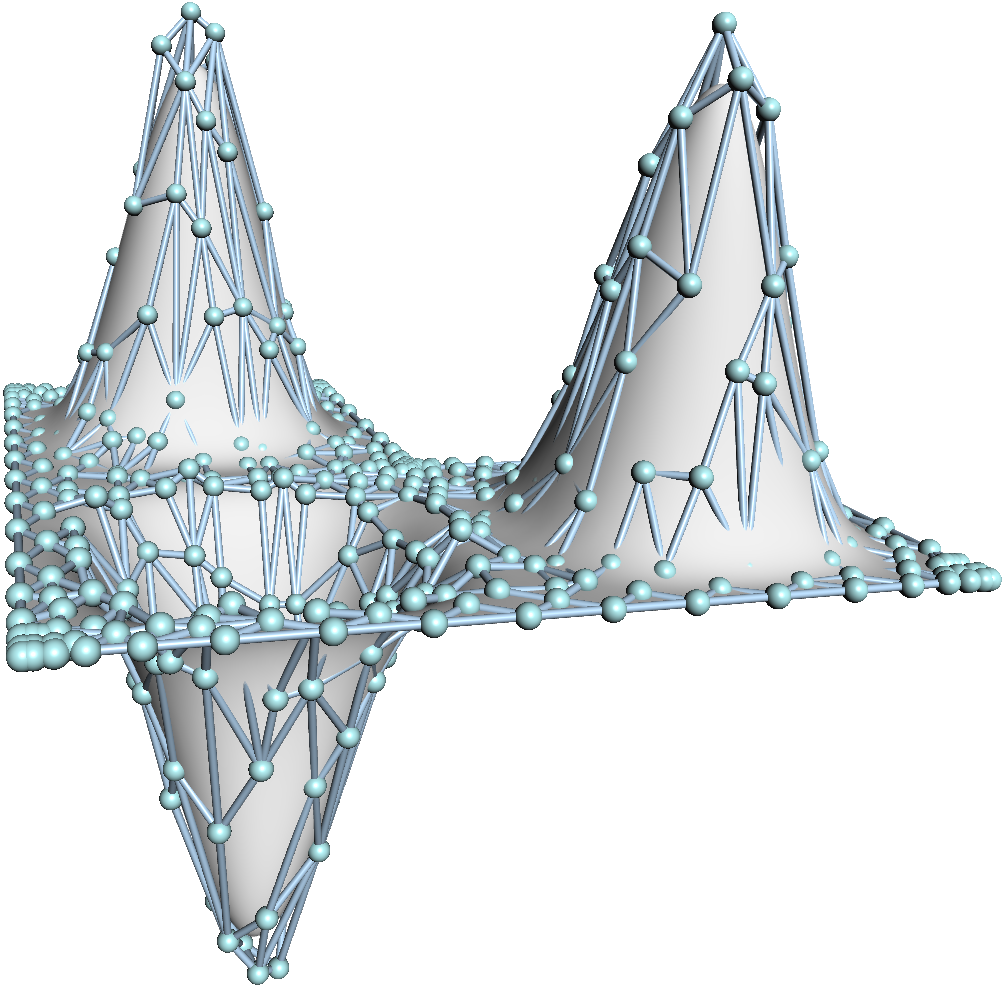} \caption{} \end{subfigure}
 \begin{subfigure}[b]{0.33\textwidth} \centering \includegraphics[width=\textwidth]{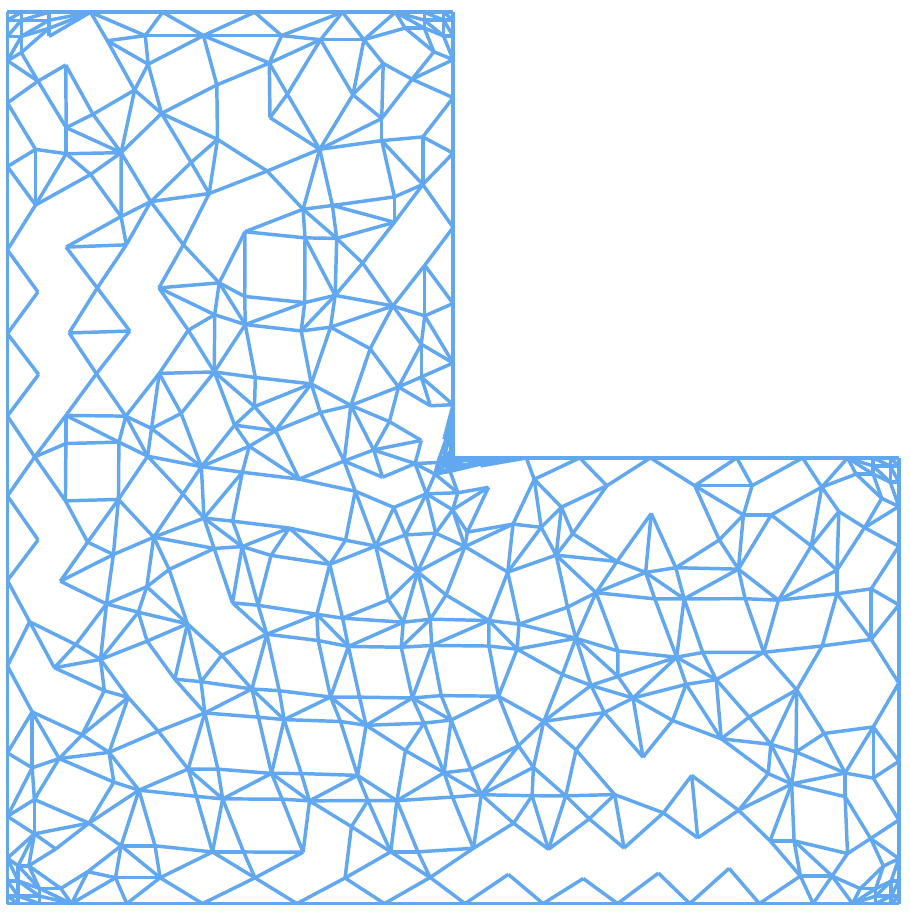} \caption{} \end{subfigure}

\caption{A cubic TCB-spline surface. (a) The initial triangulation of a knot set where the polygonal boundary is marked by blue lines; (b) from left to right: the cubic TCB-spline basis functions defined at the interior, boundary and a vertex; the cubic TCB-spline surface and its control net shown in (c) and (d), respectively; (e) the mesh generated by the Greville sites. }\label{fig:TCB_spline_surface}
\end{figure}

Let $\Omega_0$ be a polygonal domain in $\mathbb{R}^2$; see  Fig.~\ref{fig:TCB_spline_surface}(a). To define TCB-spline basis functions and surfaces over $\Omega_0$, we should place a set of knots over this domain. As an initialization, we first generate an evenly-distributed knot placement over $\Omega_0$ using the centroidal Voronoi tessellations (CVT) method~\cite{Du:1999:SIAM}, as did in~\cite{Wang:2022:CMAME}.  We can compute a CVT by Lloyd's relaxation, iteratively moving each seed point to the centroid of the corresponding cell~\cite{Du:1999:SIAM}. Moreover, the knots at the corner of $\Omega_0$ inserted $k+1$ times to maintain the partition of unity of the TCB-spline basis functions. This knot set may be further locally refined according to the surface fitting error and displacement magnitude. We deter the discussion of knot refinement till Section~\ref{sec:knot_refinement}. We can compute the degree-$k$ triangle configuration family $\it\Gamma_k= \{(T,I)\}$ of the knot set using the LTP procedure described above. Denote by $\mathcal{I}_k$ the collection of all the different second-knot subset $I$ in $\it\Gamma_k$ and $X_I \subset \it\Gamma_k$ the collection of t-configs with the same knot set $I$, i.e., $\mathcal{I}_k = \{I|(T,I)\in \it\Gamma_k\}, ~X_I =\{(T, \hat{I})|\hat{I} = I, ~(T, \hat{I}) \in \it\Gamma_k\} $. 
Then a degree-$k$ TCB-spline associated with a knot subset $I$ is defined as a linear combination of simplex splines as follows
\begin{equation*}B_I(\mt)=\sum_{(T, I)\in X_I} \textrm{Area}(T)M(\mt|T\cup I) \quad  \rm{with} \quad \mt\in\mathbb{R}^2.
\end{equation*}
Fig.~\ref{fig:TCB_spline_surface}(b) shows examples of TCB-spline basis functions at the interior, boundary, and vertex of $\Omega_0$, respectively. TCB-spline basis functions share many appealing properties with traditional B-splines. In particular, TCB-spline basis functions possess properties important for shell representation and analysis, such as non-negativity, partition of unity, local support, polynomial reproduction, and automatic smoothness (i.e., a degree-$k$ TCB-spline defined over a set of non-degenerated knots is $C^{k-1}$-continuous).

A TCB-spline surface is defined as
\begin{equation}\label{eq:tcbsurface}\mss(\mt)=\sum_{I\in\mathcal{I}_k}B_I(\mt)\mc_I, \mt\in\mathbb{R}^2,
\end{equation}
where $\mc_I$ is referred to as the control point associated with the TCB-spline basis function $B_I(\mt)$. The average of the knot subset $I$, denoted by $\bar{I}$, is referred to as the Greville site of function $B_I$~\cite{Liu:2007:phdthesis}. The TCB-spline basis function, or its Greville site, is a one-to-one correspondence with the control points. Unlike traditional tensor product splines, which typically have a control net consisting of a quad mesh, the TCB-spline has a control net with a general topology that is a hybrid mesh~\cite{Liu:2007:phdthesis,Liu:2007:CG,Zhang:2017:cagd}. Fig.~\ref{fig:TCB_spline_surface}(c\&d) shows a toy example of a cubic TCB-spline surface and its control net, respectively. The connectivity relationship of the control net is applied to the corresponding Greville sites on the 2D domain to illustrate the control net's topology better; see Fig.~\ref{fig:TCB_spline_surface}(e). Denote by $\mathcal{I}_k^b$ a subset of $\mathcal{I}_k$ with corresponding Greville site $\bar{I}$ in the domain boundary, i.e., $\mathcal{I}_k^b = \{I|I\in\mathcal{I}_k, \bar{I}\in\partial\Omega_0\}$. Then, the restriction of a degree-$k$ TCB-spline surface in Eq.~(\ref{eq:tcbsurface}) to the parametric domain boundary is the classical degree-$k$ clamped B-spline curves passing through the two end control points~\cite{Wang:2022:CMAME}:
$$ \mss(\mt)|_{\partial\Omega_0} = \sum_{I\in\mathcal{I}_k^b}B_I(\mt)\mc_I, \mt\in\mathbb{R}^2.$$

A degree-$k$ TCB-spline surface is almost $C^{k-1}$-continuous everywhere except around the concave corners, which may only have $C^0$ continuity~\cite{Cao:2019:CMAME}. To ensure global $C^1$ continuity, some special treatment is required. Following the method~\cite{Cao:2019:CMAME},  we control the topology of triangulations in LTP near concave corners of the parametric domain, such that the $C^0$ continuity of TCB-spline basis functions are restricted to a single knot line at the corresponding corner. Then, we merge all $C^0$ basis functions at each concave corner into a single $C^1$ basis function. The locally  merged basis functions are used in 
the subsequent surface reconstruction and shell analysis to achieve the global $C^1$ continuity.

\section{Kirchhoff-Love thin-shell theory \label{KL_theory}}
A detailed presentation of the shell theories based on the isogeometric concept can be found in~\cite{Kiendl:2011:phdthesis, Kiendl:2015:CMAME, Casquero:2020:CMAME}. In the KL shell theory, the following hypotheses hold:
\begin{itemize}
  \item [(a)] Shell cross-sections stay on a straight line and remain perpendicular to the mid-surface after any deformation.
  \item [(b)] Transverse normal stress is neglected.
\end{itemize}
Condition (a) implies that a linear strain distribution through the thickness and the transverse shear strains are neglected~\cite{Kiendl:2009:CMAME}. Thus, the kinematics of the shell structure can be described by its mid-surface and the thickness.

\subsection{ Mechanics of thin shells}
For the convenience of the subsequent description, ``~$\hat{\cdot}$~" and ``~${\cdot}$~" refer to the quantities in the undeformed and deformed geometries, respectively. Consider a shell model with thickness $t$, where the mid-surface $\hat{\bm{r}}(\xi^1, \xi^2)$ is parameterized by the coordinates $\xi^1$ and $\xi^2$. Then, any material point $\hat{\mpp}(\xi^1,\xi^2,\xi^3)$ in the reference geometry can be expressed as
\begin{equation}
 \hat{\mpp}(\xi^1,\xi^2,\xi^3) = \hat{\bm{r}}(\xi^1, \xi^2) + \xi^3\hat{\ma}_3(\xi^1, \xi^2), \quad -0.5t \leq \xi^3 \leq 0.5t,
\end{equation}
where $\hat{\ma}_3$ is the normal vector of the point $\hat{\bm{r}}(\xi^1, \xi^2)$ in the mid-surface. In the same way, any material point $\mpp(\xi^1,\xi^2,\xi^3)$ in the deformed geometry can be expressed as
\begin{equation}
  {\mpp}(\xi^1,\xi^2,\xi^3) = {\bm{r}}(\xi^1, \xi^2) + \xi^3\ma_3(\xi^1, \xi^2), \quad -0.5t \leq \xi^3 \leq 0.5t.
\end{equation}
Accordingly, we can obtain the displacement of a material point in the mid-surface as
\begin{equation}
  \muu(\xi^1, \xi^2) = \bm{r}(\xi^1, \xi^2) - \hat{\bm{r}}(\xi^1, \xi^2).
\end{equation}
Hereinafter, the Greek indices $\alpha,\beta$ are assumed to hold in the range $\{1, 2\}$. The tangent vectors of the mid-surfaces of undeformed and deformed geometries can be expressed as
\begin{equation}
  \hat{\ma}_\alpha = \frac{\partial \hat{\bm{r}}(\xi^1, \xi^2)}{\partial \xi^\alpha} \quad \textrm{and} \quad{\ma}_\alpha = \frac{\partial {\bm{r}(\xi^1, \xi^2)}}{\partial \xi^\alpha},
\end{equation}
respectively. Under hypothesis condition (a), the unit normal vectors of the mid-surfaces of undeformed and deformed geometries can be written as
\begin{equation}
  \hat{\ma}_3 = \frac{\hat{\ma}_1 \times \hat{\ma}_2}{||\hat{\ma}_1 \times \hat{\ma}_2||} \quad \textrm{and} \quad {\ma}_3 = \frac{{\ma}_1 \times {\ma}_2}{||{\ma}_1 \times {\ma}_2||},
\end{equation}
respectively. The covariant base vectors of undeformed and deformed geometries are defined as $(\hat{\ma}_1, \hat{\ma}_2, \hat{\ma}_3)$ and $(\ma_1, \ma_2, \ma_3)$, respectively. Furthermore, the covariant metric coefficients are computed by
\begin{equation}
  \hat{a}_{\alpha\beta}=\hat{\ma}_{\alpha}\cdot \hat{\ma}_{\beta}, \quad  {a}_{\alpha\beta}={\ma}_{\alpha}\cdot {\ma}_{\beta},
\end{equation}
where the dot symbol denotes the inner product of two vectors. By solving for the inverse of the covariant metric coefficients, the contravariant metric coefficients can be expressed as
\begin{equation}
  [\hat{a}^{\alpha\beta}] = [\hat{a}_{\alpha\beta}]^{-1}, \quad  [a^{\alpha\beta}]=[a^{\alpha\beta}]^{-1}.
\end{equation}
The coefficients of Green-Lagrange strain tensor are given by
\begin{equation}
  E_{\alpha\beta} = \varepsilon_{\alpha\beta} + \xi^3\kappa_{\alpha\beta},
\end{equation}
where the covariant membrane strain coefficients $\varepsilon_{\alpha\beta}$ and the coefficients of change-of-curvature  $\kappa_{\alpha\beta}$ are
$$\varepsilon_{\alpha\beta}=\frac{a_{\alpha\beta} - \hat{a}_{\alpha\beta}}{2} \quad \textrm{and} \quad \kappa_{\alpha\beta} = \frac{\partial \hat{\ma}_{\alpha}}{\partial \xi^{\beta}} \cdot {\hat \ma_3} - \frac{\partial {\ma}_{\alpha}}{\partial \xi^{\beta}} \cdot \ma_3,$$ respectively.

For notational brevity, we reformulate all stress-related variables as did in as~\cite{Kiendl:2015:CMAME}. In particular, the membrane stress $\bm{n}$, bending stress $\bm{m}$, membrane strain $\bm{{\varepsilon}}$ and change-of-curvature $\bm{{\kappa}}$ in Voigt notation~\cite{Voigt:1910} are as follows:
\begin{equation}
  \bm{n} = \begin{bmatrix}
    {n}^{11} \\
    {n}^{22} \\
    {n}^{12}
  \end{bmatrix},\quad
   \bm{{m}} = \begin{bmatrix}
    {m}^{11} \\
    {m}^{22} \\
    {m}^{12}
  \end{bmatrix},\quad
  \bm{{\varepsilon}} = \begin{bmatrix}
    {\varepsilon}^{11} \\
    {\varepsilon}^{22} \\
    2{\varepsilon}^{12}
  \end{bmatrix},\quad
   \bm{{\kappa}} = \begin{bmatrix}
    {\kappa}^{11} \\
    {\kappa}^{22} \\
    2{\kappa}^{12}
  \end{bmatrix}.
\end{equation}
Moreover, the membrane stress and bending stress are calculated by~\cite{Li:2018:CMAME}
\begin{equation} \label{equ:stresses}
  \bm{{n}} = t\bm{C\varepsilon},\quad
   \bm{{m}} = \frac{t^3}{12}\bm{C\kappa}.
\end{equation}
with
\begin{equation}
\bm{C} = \frac{E}{1-\nu^2}
\begin{bmatrix}
  (\hat{a}^{11})^2 &\nu\hat{a}^{11}\hat{a}^{22}+(1-\nu)(\hat{a}^{12})^2 & \hat{a}^{11}\hat{a}^{12} \\
  \vdots &(\hat{a}^{22})^2 &\hat{a}^{22}\hat{a}^{12} \\
   &\cdots &[(1-\nu)\hat{a}^{11}\hat{a}^{22} + (1+\nu)(\hat{a}^{12})^2]/2
 \end{bmatrix},
\end{equation}
where $E$ is the Young's modulus and $\nu$ is the Poisson's ratio.

Let $\delta$ denote the variation operator. Then  $\delta \meps$, $\delta \mkap$ and $\delta\bm{u}$ are the virtual membrane strain, virtual change-of-curvature and virtual displacement, respectively. Denote by $W_{int}$ and $W_{ext}$ the internal and external works, respectively. Based on the principle of virtual work~\cite{Wunderlich:2002}, internal virtual work $ \delta  W_{int}$ is equal to the external virtual work $\delta W_{ext}$ with
\begin{equation}
  \delta W_{int}=\int_{\hat{A}}({\mn} \cdot \delta \meps + {\mm} \cdot \delta \mkap)~{\rm{d}}\hat{A}=\delta W_{ext}=\int_{\hat{A}} \bm{f} \cdot \delta\bm{u}~{\rm{d}}\hat{A},
\end{equation}
where ${\rm{d}}\hat{A}=||\hat{\ma}_1 \times \hat{\ma}_2||~\rm{d}\xi^1\rm{d}\xi^2$, $\bm{f}$ is the external load.

\subsection{Isogeometric discretization}
Suppose the mid-surface is represented by TCB-splines with $n$ basis functions $\{B_i(\xi^1, \xi^2)\}, i = 1, \cdots, n$. The displacement field is defined by the same set of basis functions as follows:
\begin{equation}
  {\bm{u}}(\xi_1, \xi_2)=\sum_{i=1}^n B_i(\xi^1, \xi^2)\bm{u}_i,
\end{equation}
where $\bm{u}_i=(u_{3i-2},u_{3i-1},u_{3i})\in R^3$ is the control variables of the displacement field associated with the basis function $B_i(\xi^1, \xi^2)$. Since each $\bm{u}_i$ contains three components, the number of degrees of freedom is three times the number of basis functions.

For nonlinear deformations, the residual force vector $\bm{R} =[R_r]$ is calculated as
\begin{equation} \label{equ:nonlinear_residual_vector}
  {R}_r=\frac{\partial W_{ext}}{\partial u_r} - \frac{\partial W_{int}}{\partial u_r} = \int_{\hat{A}} \bm{f} \cdot \frac{\partial \bm{u}}{\partial u_r}~{\rm{d}}\hat{A} - \int_{\hat{A}} \left({\mn} \cdot \frac{\partial\meps}{\partial u_r} + {\mm} \cdot \frac{\partial \mkap}{\partial u_r}\right)~{\rm{d}}\hat{A}, \quad 1 \leq r\leq 3n.
\end{equation}
In this paper, we consider displacement-independent loads. In this case, we have
$$\frac{\partial^2 W_{ext}}{\partial u_r \partial u_s}=\int_{\hat{A}} \frac{\partial \bm{f}}{\partial u_s} \cdot \frac{\partial \bm{u}}{\partial u_r}~{\rm{d}}\hat{A} = 0, \quad 1 \leq r, s \leq 3n,$$
and the stiffness matrix $\bm{K} = [K_{r,s}]$ is obtained as
\begin{equation}\label{equ:nonlinear_stiffness matrix}
  {K}_{rs}=\frac{\partial^2 W_{int}}{\partial u_r \partial u_s} - \frac{\partial^2 W_{ext}}{\partial u_r \partial u_s} = \int_{\hat{A}} \frac{\partial {\mn}}{\partial u_s} \cdot \frac{\partial\meps}{\partial u_r} + {\mn} \cdot \frac{\partial^2\meps}{\partial u_r \partial u_s} + \frac{\partial {\mm}}{\partial u_s} \cdot \frac{\partial \mkap}{\partial u_r} + {\mm} \cdot \frac{\partial^2 \mkap}{\partial u_r \partial u_s}~{\rm{d}}\hat{A}.
\end{equation}
  The linearized system is given as
\begin{equation}
  \bm{K}\Delta\bm{u}=\bm{R},
\end{equation}
where the displacements are iteratively updated for the incremental displacement vector $\Delta\bm{u}$.

For small deformation problems, it is convenient to employ geometric linear analysis~\cite{Kiendl:2009:CMAME}.  In this case, the actual geometry is assumed to be equal to the reference geometry, and elements of the residual force vector $\bm{R}^{lin} =[R^{lin}_r]$ and stiffness matrices $\bm{K}^{lin} = [K^{lin}_{r,s}]$ are simplified as
\begin{equation} \label{equ:linear_element}
  {R}_r^{lin} = \int_{\hat{A}} \bm{f} \cdot \frac{\partial \bm{u}}{\partial u_r}~{\rm{d}}\hat{A}  \quad \textrm{and} \quad
  {K}_{rs}^{lin} = \int_{\hat{A}} \frac{\partial {\mn}}{\partial u_s} \cdot \frac{\partial\meps}{\partial u_r} + \frac{\partial {\mm}}{\partial u_s} \cdot \frac{\partial \mkap}{\partial u_r}~{\rm{d}}\hat{A}, \quad 1 \leq r, s \leq 3n,
\end{equation}
respectively. 
Similar to~\cite{Nguyen:2011:CMAME}, the strains can be obtained from the linear form of displacement as follows:
\begin{equation}\label{equ:linear_strain}
  \bm{\varepsilon}=\sum_{i=1}^n \bm{M}_i(\xi^1, \xi^2)\bm{u}_i,  \quad \bm{\kappa}=\sum_{i=1}^n \bm{D}_i(\xi^1, \xi^2)\bm{u}_i,
\end{equation}
where the membrane matrix $\bm{M}_i$ and bending matrix $\bm{D}_i$ are
\begin{equation}
 \bm{M}_i=[{B}_{i,1}\hat{\ma}_1, \quad {B}_{i,2}\hat{\ma}_2,  \quad {B}_{i,2}\hat{\ma}_1  + {B}_{i,1}\hat{\ma}_2 ]^T,\quad
 \bm{D}_i=[\bm{b}_{m11}^i, \quad \bm{b}_{m22}^i, \quad 2\bm{b}_{m12}^i]^T,\quad
\end{equation}
with
${B}_{i,\alpha} = \frac{B_i(\xi^1,\xi^2)}{\xi^{\alpha}}$ and
\begin{equation}
\begin{aligned}
 \bm{b}_{m\alpha\beta}^i &=\frac{1}{||\hat{\ma}_1 \times \hat{\ma}_2||} \left[ \frac{\partial \hat{\ma}_{\alpha}}{\partial \xi^{\beta}} \cdot \hat{\ma}_3  \frac{\partial B_i}{\partial \xi^{1}}\left(\hat{\ma}_2 \times \hat{\ma}_3\right) +  \frac{\partial B_i}{\partial \xi^{2}}\left(\hat{\ma}_3 \times \hat{\ma}_1\right) -  \frac{\partial B_i}{\partial \xi^{1}}\left(\hat{\ma}_2 \times  \frac{\partial \hat{\ma}_{\alpha}}{\partial \xi^{\beta}}\right) -  \frac{\partial B_i}{\partial \xi^{2}}\left(\frac{\partial \hat{\ma}_{\alpha}}{\partial \xi^{\beta}} \times \hat{\ma}_1\right) \right] -  \frac{\partial^2 B_i}{\partial \xi^{\alpha\beta}}\hat{\ma}_3. \\
\end{aligned}
\end{equation}
Bringing Eq.~(\ref{equ:stresses}) and Eq.~(\ref{equ:linear_strain}) into Eq.~(\ref{equ:linear_element}) yields the global linear deformation equation
\begin{equation}
  \bm{K}^{lin}\bm{u}=\bm{R}^{lin}.
\end{equation}

\section{TCB-spline-based reparameterization\label{surface_fitting}}

\begin{figure}
  \centering
\graphicspath{{figures/}}
  \begin{subfigure}[b]{0.245\textwidth} \centering \includegraphics[width=\textwidth]{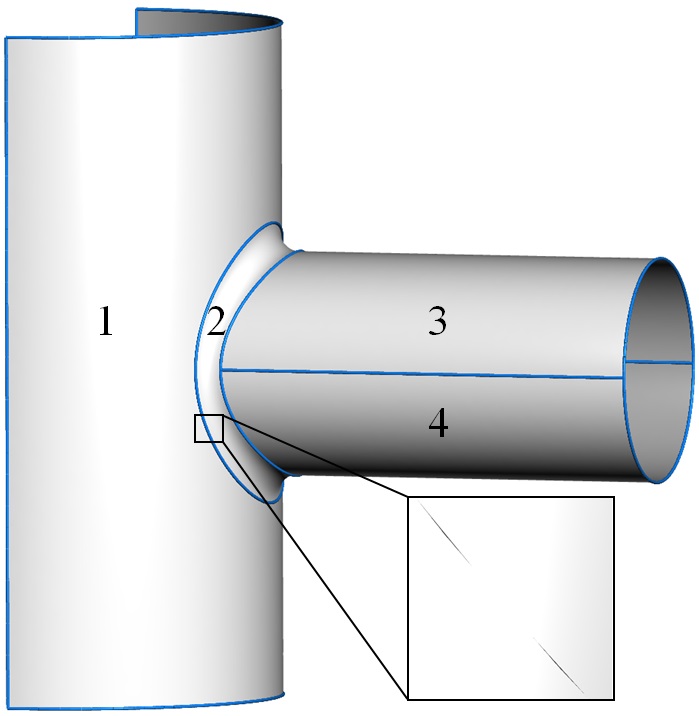} \caption{Given CAD model}  \end{subfigure} ~
  \begin{subfigure}[b]{0.245\textwidth} \centering \includegraphics[width=\textwidth]{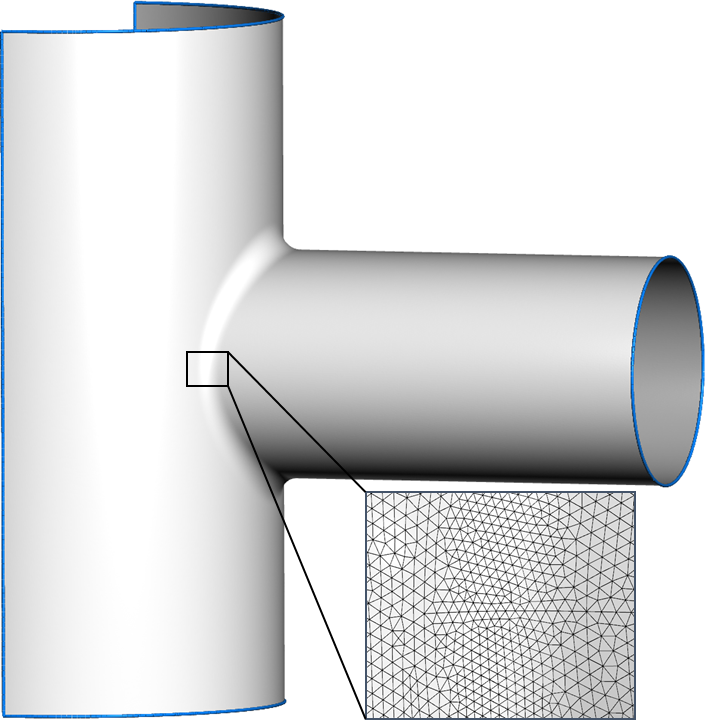} \caption{Surface triangulation}  \end{subfigure} ~
  \begin{subfigure}[b]{0.47\textwidth} \centering \includegraphics[width=\textwidth]{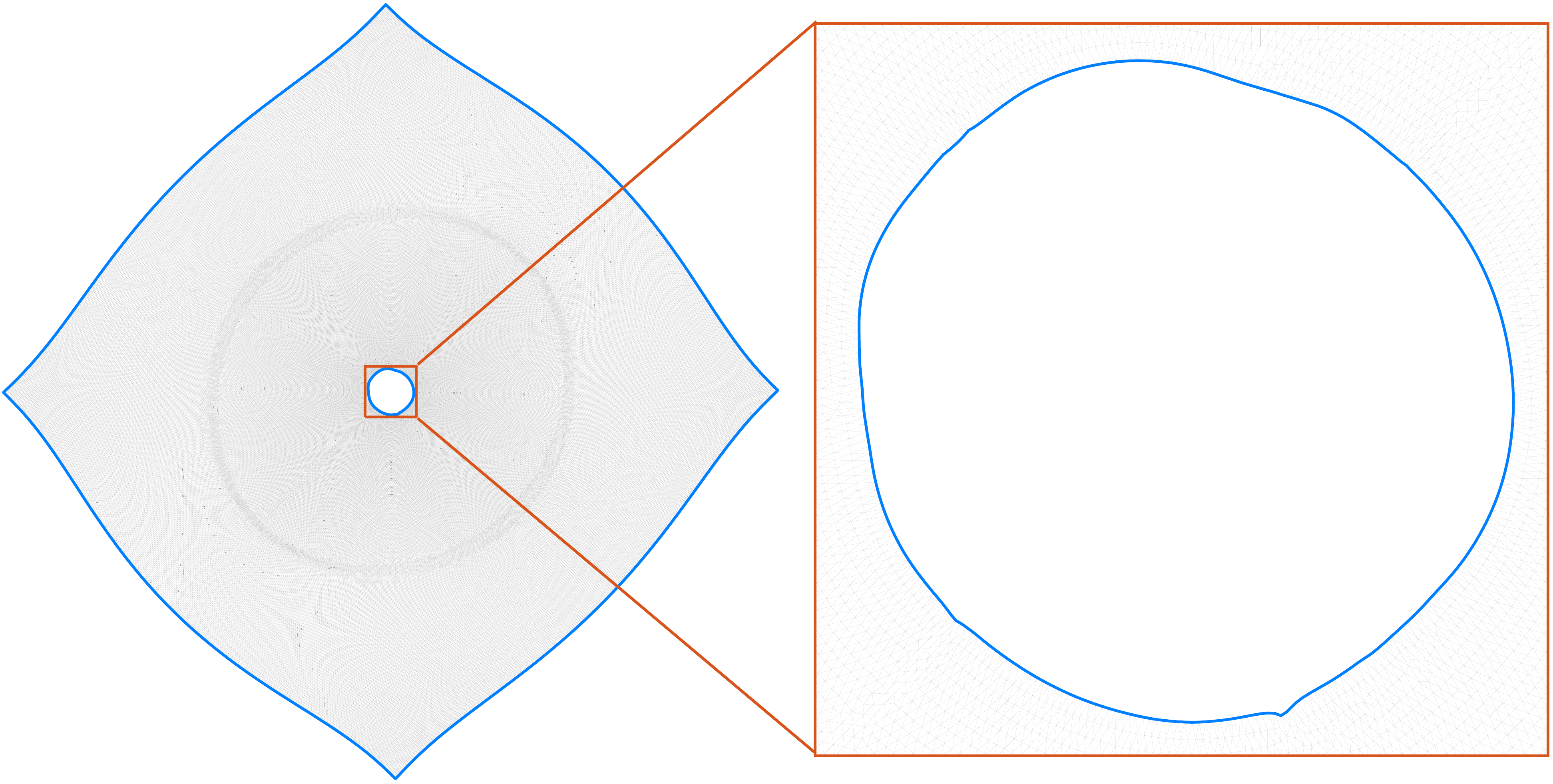} \caption{Surface parameterization}  \end{subfigure}
  \begin{subfigure}[b]{0.47\textwidth} \centering \includegraphics[width=\textwidth]{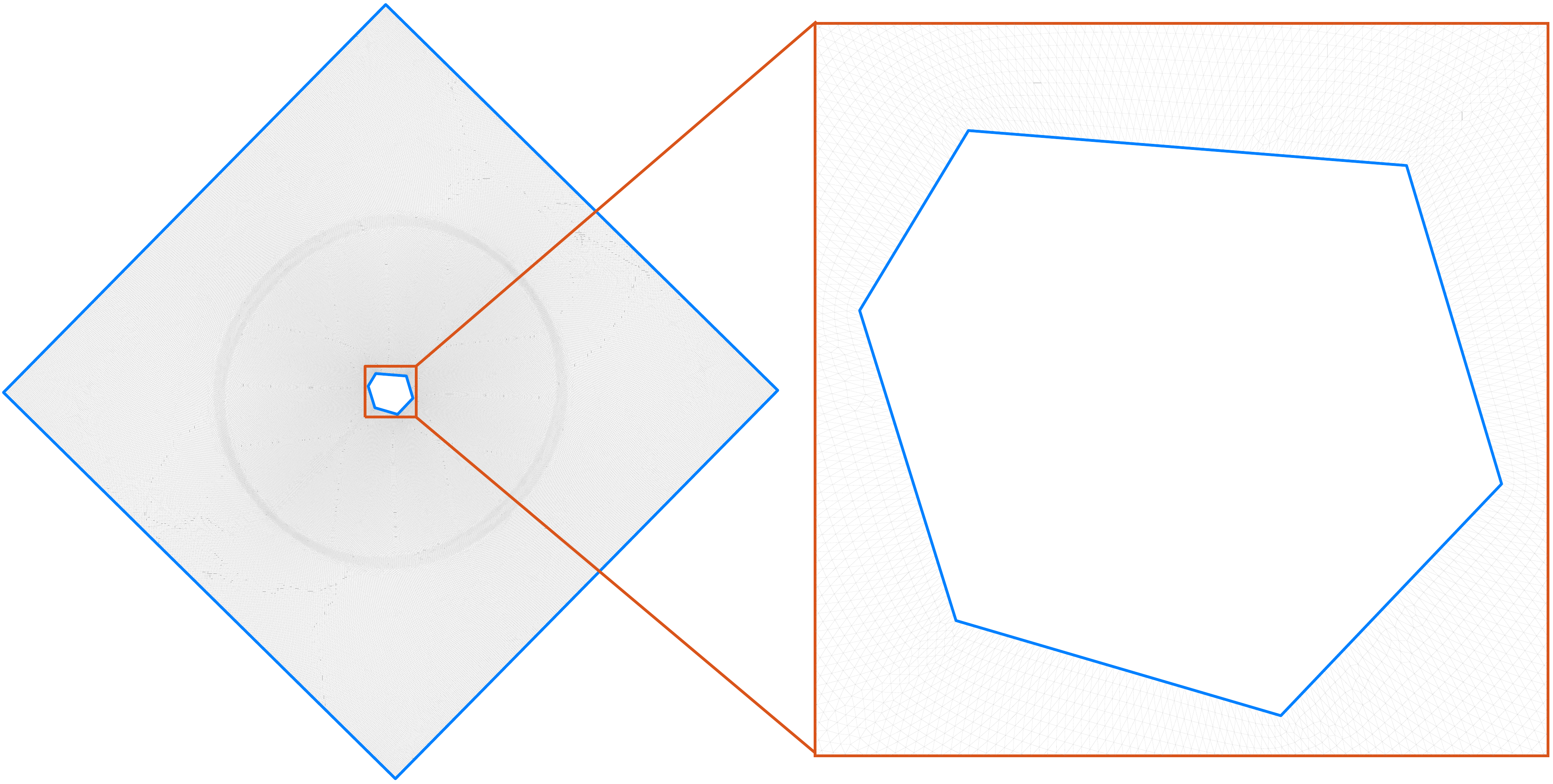}  \caption{The generation of parametric domain} \end{subfigure} ~
  \begin{subfigure}[b]{0.245\textwidth} \centering \includegraphics[width=\textwidth]{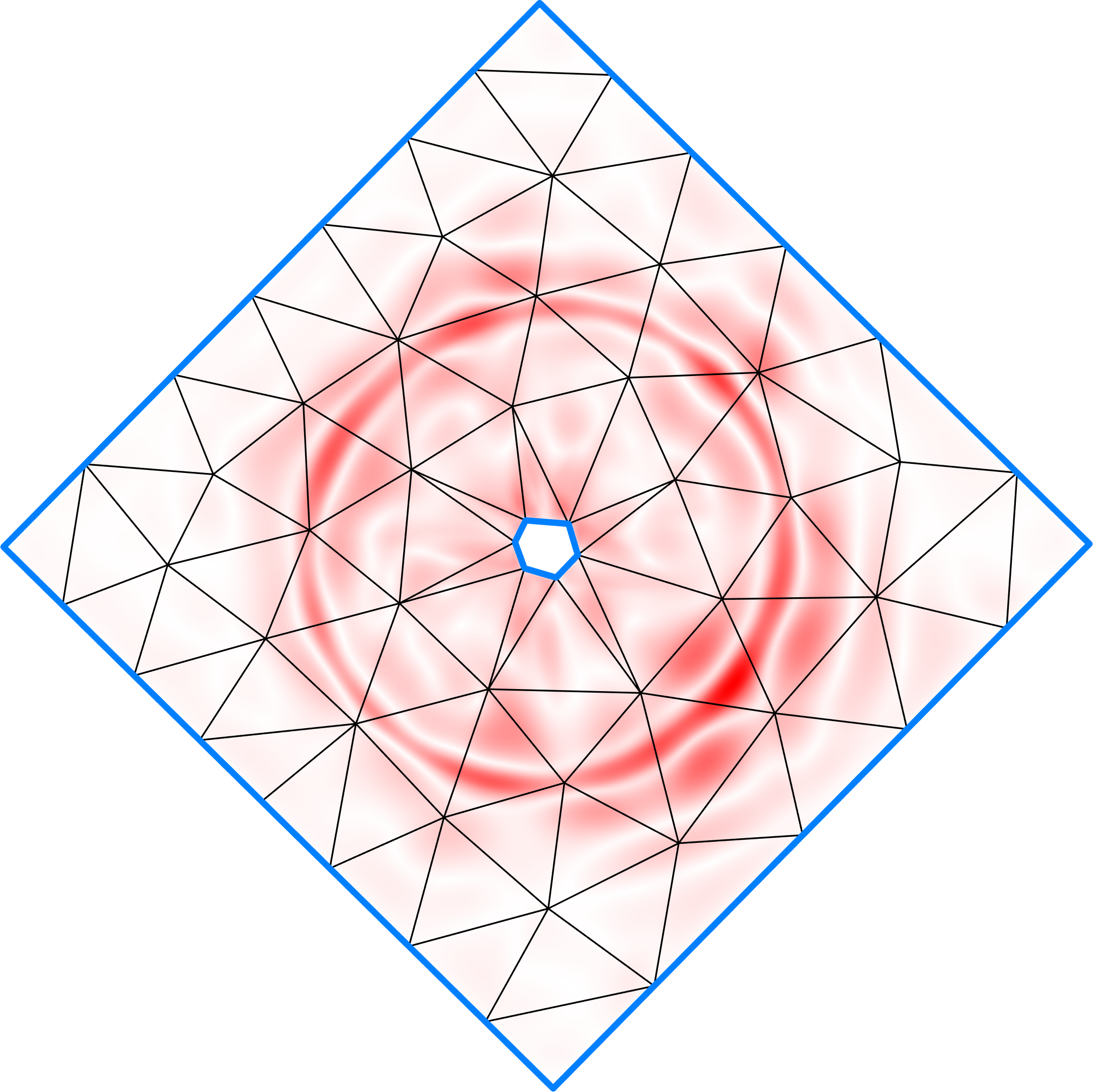}  \caption{Initial knot mesh} \end{subfigure} ~
  \begin{subfigure}[b]{0.245\textwidth} \centering \includegraphics[width=\textwidth]{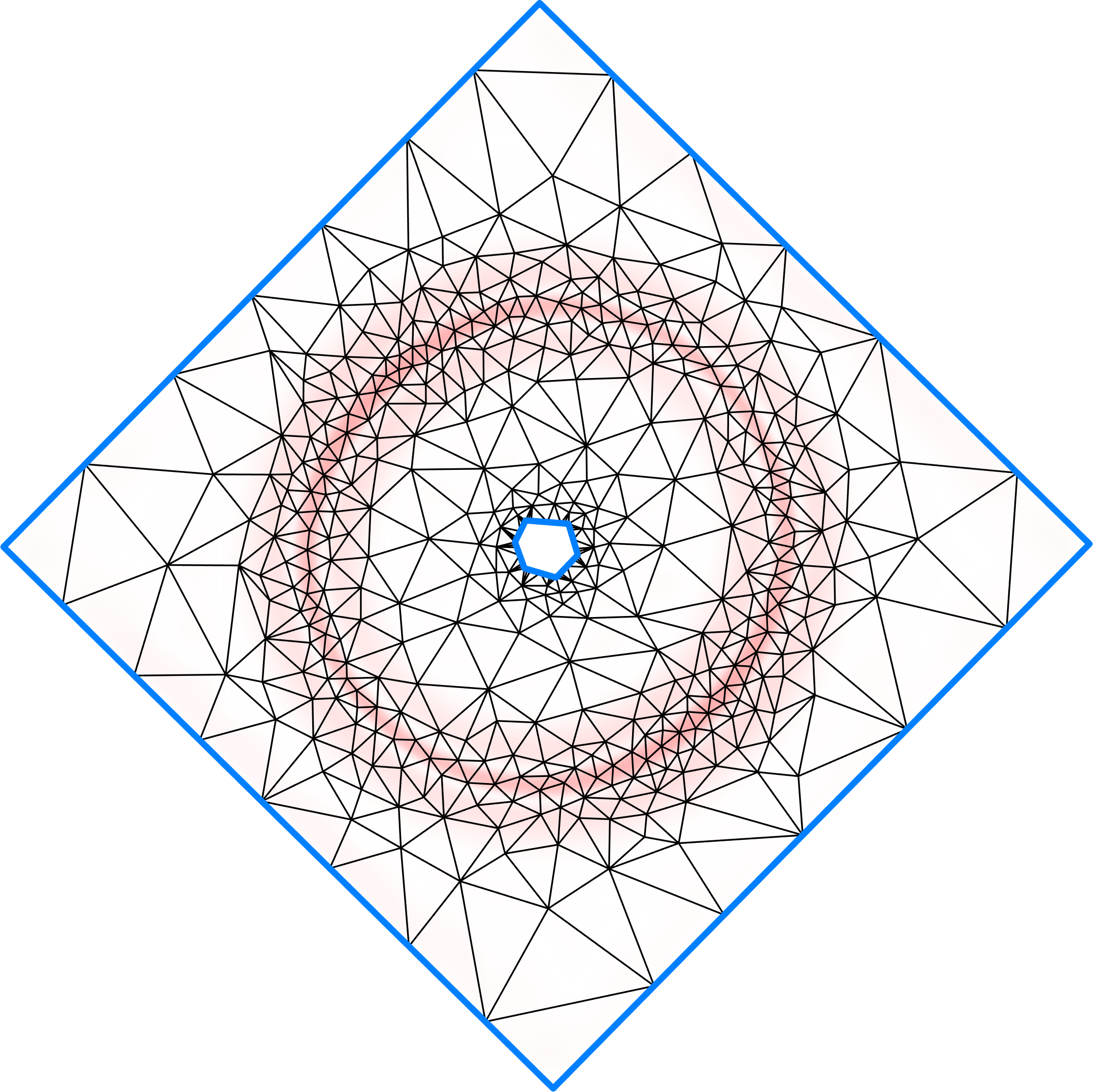}  \caption{Refined knot mesh} \end{subfigure}
 \begin{subfigure}[b]{0.245\textwidth} \centering \includegraphics[width=\textwidth]{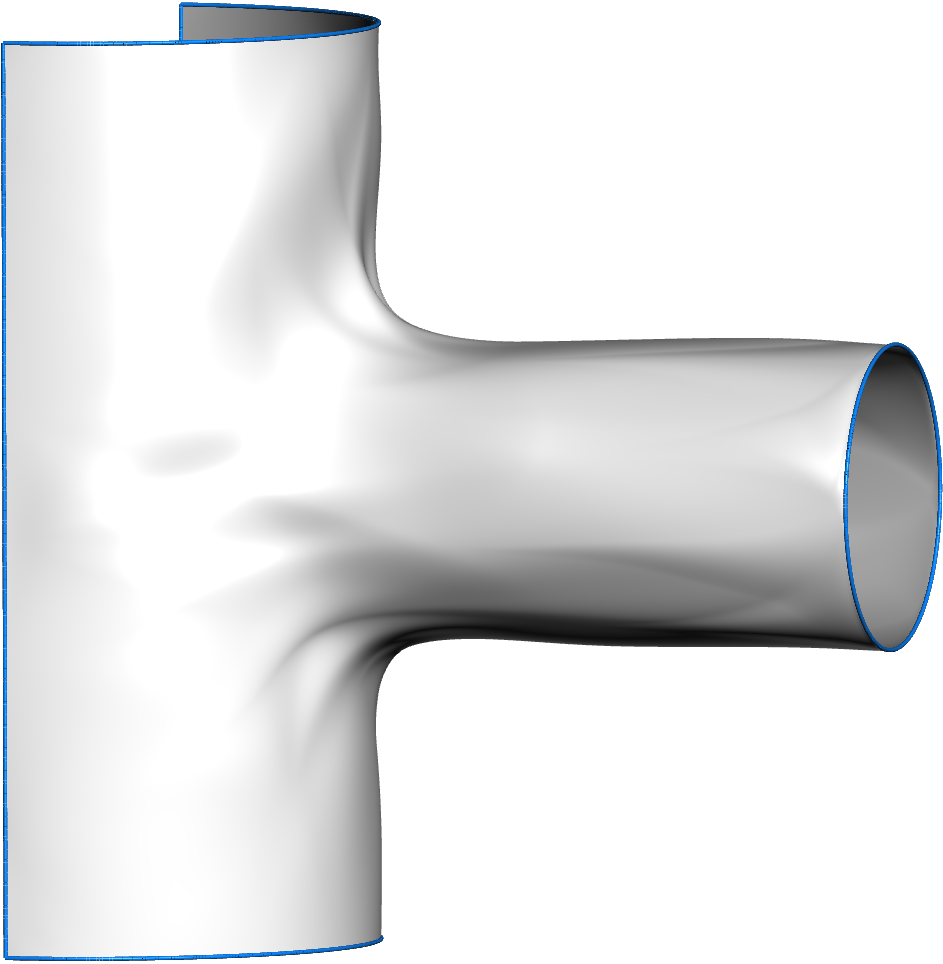}  \caption{Initial fitting result} \end{subfigure} \hspace{1.35cm}
 \begin{subfigure}[b]{0.245\textwidth} \centering \includegraphics[width=\textwidth]{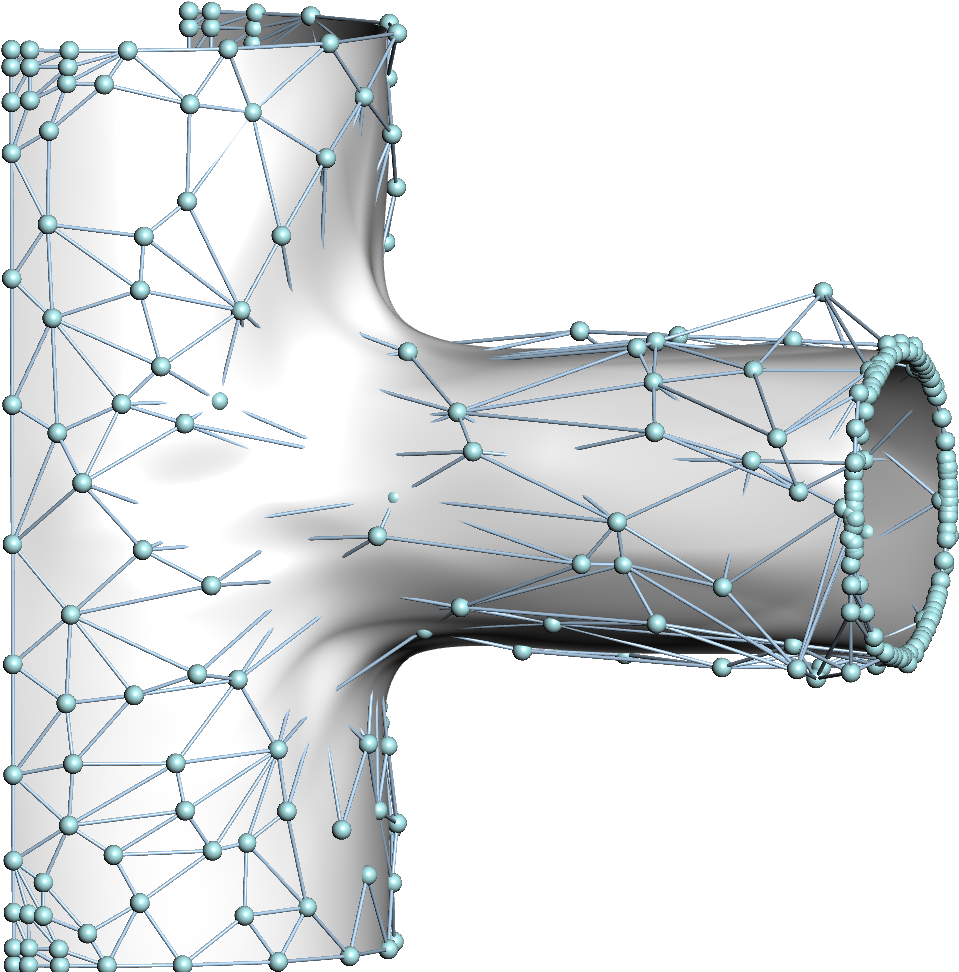}  \caption{Control net} \end{subfigure} \hspace{1.35cm}
 \begin{subfigure}[b]{0.245\textwidth} \centering \includegraphics[width=\textwidth]{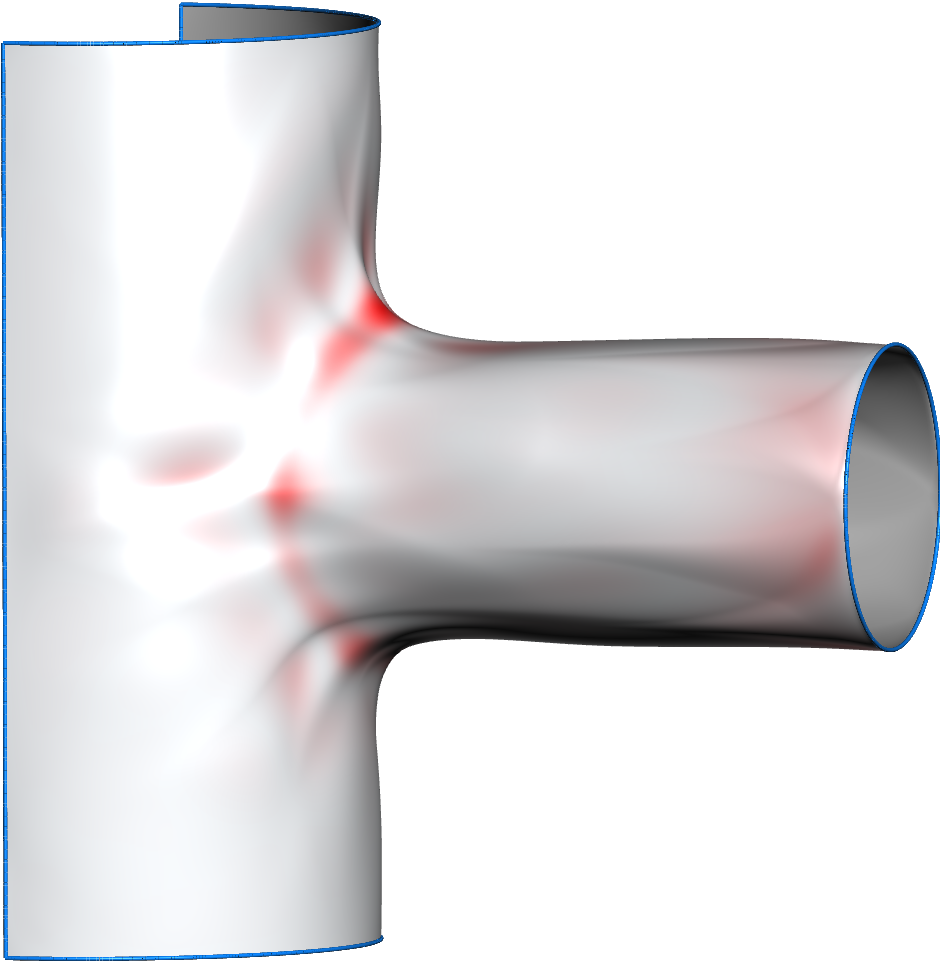}  \caption{Fitting error} \end{subfigure} \hspace{0.4cm}

\hspace{0.3cm}
 \begin{subfigure}[b]{0.245\textwidth} \centering \includegraphics[width=\textwidth]{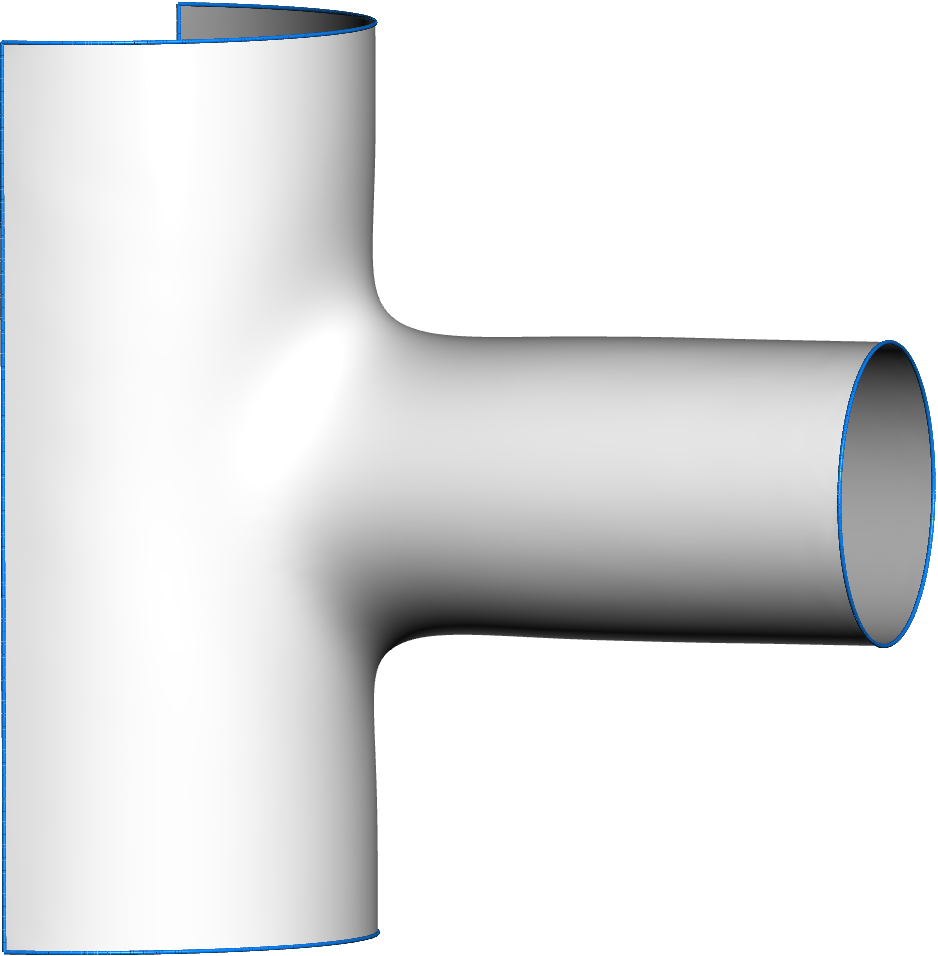}  \caption{Final fitting result} \end{subfigure} \hspace{1.35cm}
 \begin{subfigure}[b]{0.245\textwidth} \centering \includegraphics[width=\textwidth]{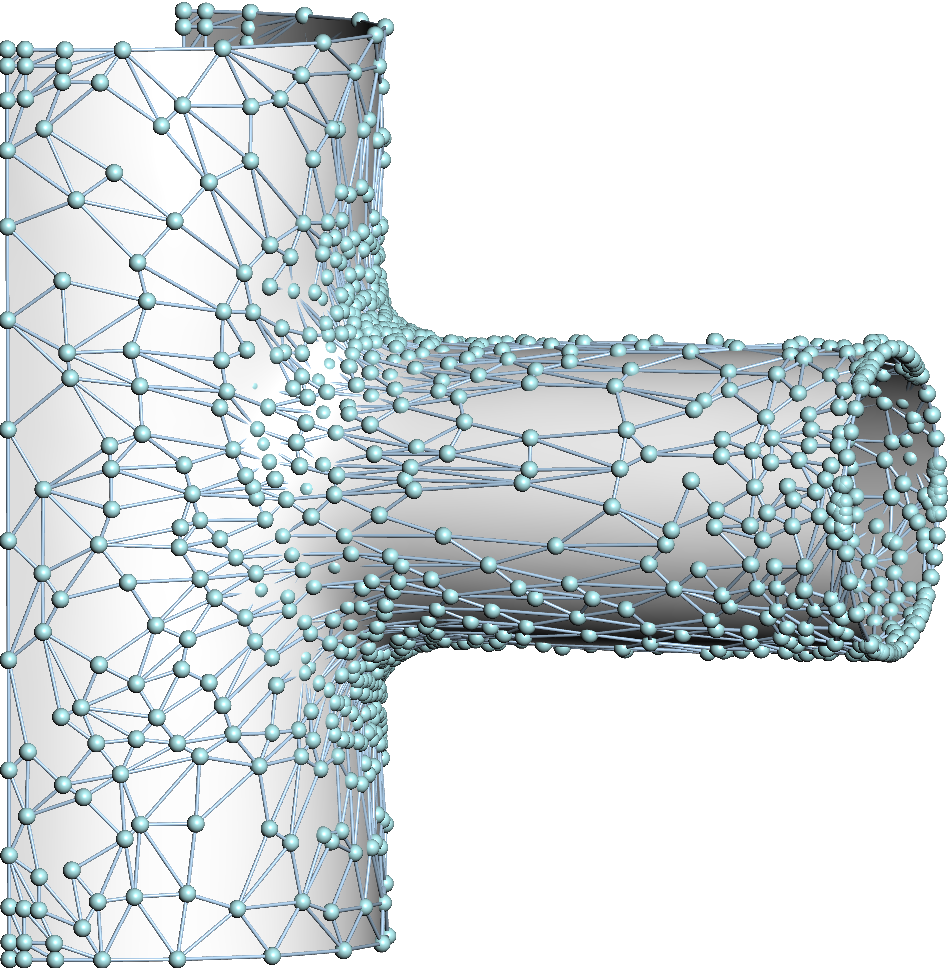}  \caption{Control net} \end{subfigure} \hspace{1.35cm}
 \begin{subfigure}[b]{0.3\textwidth} \centering \includegraphics[width=\textwidth]{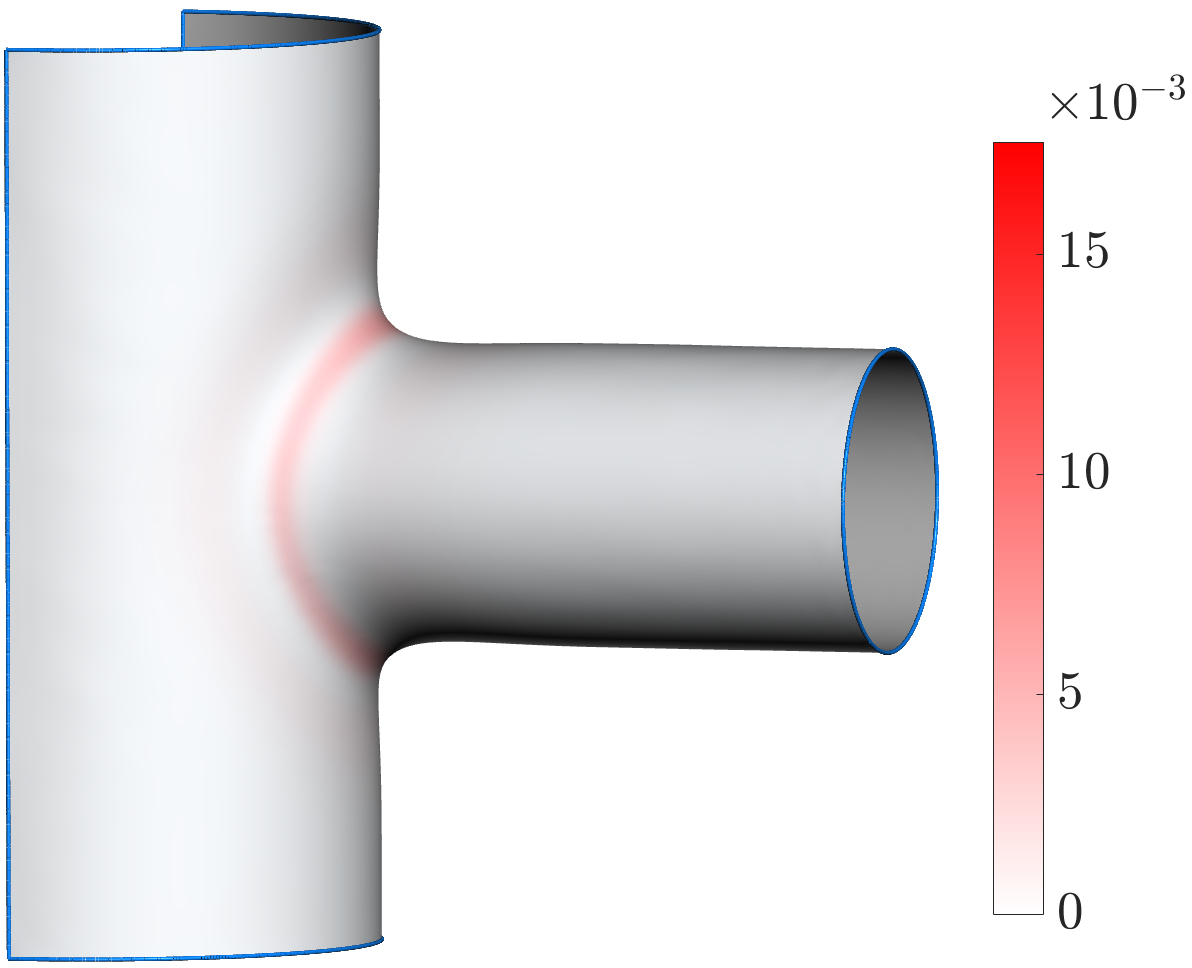}  \caption{Fitting error} \end{subfigure} 
\caption{Fitting pipeline. The given model with four trimmed patches (boundaries marked in blue) and gaps between patches in (a) is triangulated (b); mapping the surface mesh in (b) to the planar mesh in (c); the parametric domain boundary (marked by blue segment lines) obtained by simplifying the boundary of (c) and mapping the mesh into the parametric domain in (d); the single patch and no gap fitting result represented by the TCB-splines in (g), the control net in (h) and the fitting error in (i) generated from the evenly distributed knots corresponds to the triangulation and normalized fitting error distribution in (e); the final fitting result represented by the TCB-splines in (j), the control net in (k) and the fitting error in (l) generated from the refinement knots corresponds to the triangulation and normalized fitting error distribution in (f). }\label{fig:fit_algorithm}
\end{figure}

In this section, we remodel the mid-surface of a thin shell consisting of multiple (trimmed) NURBS patches using the TCB-spline-based linear least-squares fitting method. The remodeled surface is a single piece of TCB-spline surface with at least $C^1$ continuity everywhere, making it suitable for KL shell analysis. We will first give an overview of our fitting algorithm and then describe the details of each step. 

\subsection{Algorithm overview}
Spline surface fitting is a fundamental problem in many application fields, such as computer graphics and computer-aided design. In our applications, we attempt to find a TCB-spline surface that well approximates the mid-surface of the input shell model. In our initial input, the mid-surface may consist of multiple trimmed NURBS patches with undesirable gaps and overlaps. We assume that the mid-surface is topologically equivalent to a disk with finite holes if the gaps and overlaps between patches are not considered; see the pipe junction model consisting of four trimmed NURBS patches in Fig.~\ref{fig:fit_algorithm} (a).  Our output is a single patch, watertight, conforming, and global $C^1$-continuous TCB-spline surface suited directly to the analysis.

As a pre-processing, the input mid-surface is repaired using commercial software such as CADfix~\cite{Cadfix:ITI} if there are gaps or overlaps between patch interfaces. The fixed surface is then converted into a valid and conforming triangular mesh $\mathcal{M}_0$ using Gmsh software~\cite{Geuzaine:2009:IJNME}; see an example of triangular mesh in Fig.~\ref{fig:fit_algorithm}(b). Finally, our surface fitting is performed on this triangular mesh, whose vertices are sampled from the input mid-surface.

Traditional spline-based surface fitting starts with estimating parameters, i.e., determining the parameter values of the vertices on the mesh. This process is also referred to as mesh parameterization. In particular, the triangular mesh $\mathcal{M}_0$ is mapped onto the specified parametric domain so that each vertex $\mx_i\in \mathcal{M}_0$ is associated with a parameter point $\mpp_i$ in the 2D parametric domain. The pre-specified parametric domain is typically square-shaped~\cite{Kiss:2014:GM,Zhang:2016:CG,Zhang:2017:cagd}. Subsequently, a set of knots is placed onto the parametric domain to compute a set of spline basis functions. Assume $\mss(\mt)$ be the reconstructed TCB-spline surface defined over the knots in the form of Eq.~(\ref{eq:tcbsurface}). We then use the euclidean distance $e_i = \|\mx_i - \mss(\mpp_i) \|$, referred to as the fitting error of $\mx_i$, to measure the distance between the reconstruction and the input mesh. Then control points $\{\mc_j\}$ can be found by minimizing the fitting error between the triangular mesh and the spline surfaces in the linear least-squares sense as follows:
\begin{equation} \label{equ:ls}
     \mathop{\min}\limits_{\mc_j} {\sum_{\mx_i \in \mathcal{M}_0} e_i = \mathop{\min}\limits_{\mc_j}} {\sum_{\mx_i \in \mathcal{M}_0} \|~\textbf{\emph{x}}_i - \sum_{j} B_j(\mpp_i) \mc_j\|}.
\end{equation}
Our fitting method follows the traditional three-step procedure mentioned above, i.e., parameterization, knot placement, and least-squares minimization. 

For spline fitting, the parameterization is required to be at least bijective. Moreover, the parameterization is desired to have low angle and area distortion, which is potentially crucial for high-quality fitting in the subsequent steps~\cite{Zhang:2016:CG}. However, the triangular meshes discrete from the given mid-surfaces may not be four-sided shapes in nature in the application of this paper. To avoid large angle and area distortions in the parameterization, it may be challenging to artificially specify a parametric domain for those complex models. Note that TCB-splines are capable of accommodating general polygonal domains. Benefiting from this flexibility of TCB-splines, it is natural to consider free-boundary parameterizations of triangular meshes~\cite{Smith:2015:TOG}, which minimize the local geometric distortion of the mapping without constraining the overall shape of the 2D parametric domain; see Fig.~\ref{fig:fit_algorithm}(c). We can take the 2D polygonal region formed by the free boundary parameterization as the parametric domain for the TCB-spline definition. However, this straightforward approach will introduce redundant knots along the fragmentized boundary in the subsequent knot placement stage, leading to overfitting. We, therefore, further simplify the polygonal region to reduce the edge number and update the parameterization accordingly while preserving the domain shapes as much as possible; see Fig.~\ref{fig:fit_algorithm}(d). In Section~\ref{parameterization}, the parameterization and parametric domain generation details are described.

Knot distribution is also crucial for recovering the underlying shapes of the triangular meshes. Unreasonable knot distribution may result in reconstructed surfaces with shapes significantly different from the input. Since it is usually unknown how many knots are needed to recover the underlying shapes, we use a fitting error-driven knot placement strategy to increase the number of knots, adjust their distributions, and iteratively solve the linear least-squares  problem to progressively improve the reconstruction quality progressively. We first place evenly spaced knots on the generated polygonal domain and construct TCB-spline basis functions. Each basis function can be associated with a control point. We then optimize the control point positions by solving the above linear least-squares problem; see Fig.~\ref{fig:fit_algorithm}(e,~g-i) for the fitting surface obtained based on the evenly spaced knots. The details of surface fitting are given in Section~\ref{surface_remodeling}. Then more knots are added to the regions with larger fitting errors. The adaptive knot placement algorithm is described in Section~\ref{sec:knot_refinement}. We obtain a quality-improved reconstruction by solving the linear least-squares fitting problem over the refined knots. We repeat the adaptive knot refinement and surface fitting until a required number of iterations is reached, or the fitting errors are smaller than a prescribed threshold. An overview of our algorithm pipeline is given in Fig.~\ref{fig:fit_algorithm}.

\subsection{Mesh parameterization \label{parameterization}}
 The traditional spline fitting method computes the parameterization on a pre-specified simple domain. In contrast, we parameterize the triangular mesh onto a general polygonal domain adapted to the given mesh. The parameterization is bijective and naturally has low angle and area distortion. Our method consists of three steps: free-boundary parameterization, parametric domain generation, and reparametrization.

\textbf{Free-boundary parameterization}
The quality of the mesh parameterization is important for spline fitting. For example, if the mesh parameterization is not bijective, the resultant fitting surface will fold. Large distortions in mesh parameterization make the subsequent knot distribution more challenging, and it becomes difficult to recover the geometric details of the input surface. Most mesh parameterizations are computed by minimizing the distortion in terms of the deviation of angles, areas, or some combination thereof~\cite{Smith:2015:TOG}. For a comprehensive overview of triangular mesh parameterization, we refer the reader to~\cite{Fu:2021:CVM}. Constraining the boundary reduces the degree of freedom in the minimization,  thus producing more distortion in the parameterization than necessary. Free-boundary parameterization methods do not constrain the boundary to some specified shape. By allowing the optimization to modify the boundary, they better reduce distortions. In this paper, we adopt the simplicial complex augmentation framework (SCAF for short) method~\cite{Jiang:2017:TOG}, which has proven to be efficient in obtaining bijective and low-distortion parameterization results for complex models. Then, the triangular mesh $\mathcal{M}_0$ is flattened to a 2D discrete mesh, denoted by $\mathcal{M}_1$; see Fig.~\ref{fig:fit_algorithm}(c). 
When there is no ambiguity, we also use $\mathcal{M}_1$ to represent the domain covered by the planar triangular mesh $\mathcal{M}_1$. Moreover, we refer to the mapping from $\mathcal{M}_0$ to $\mathcal{M}_1$ as $\phi_0$.

\begin{figure}
  \centering
\graphicspath{{figures/}}
   \begin{subfigure}{0.196\textwidth} \centering \includegraphics[width=\textwidth]{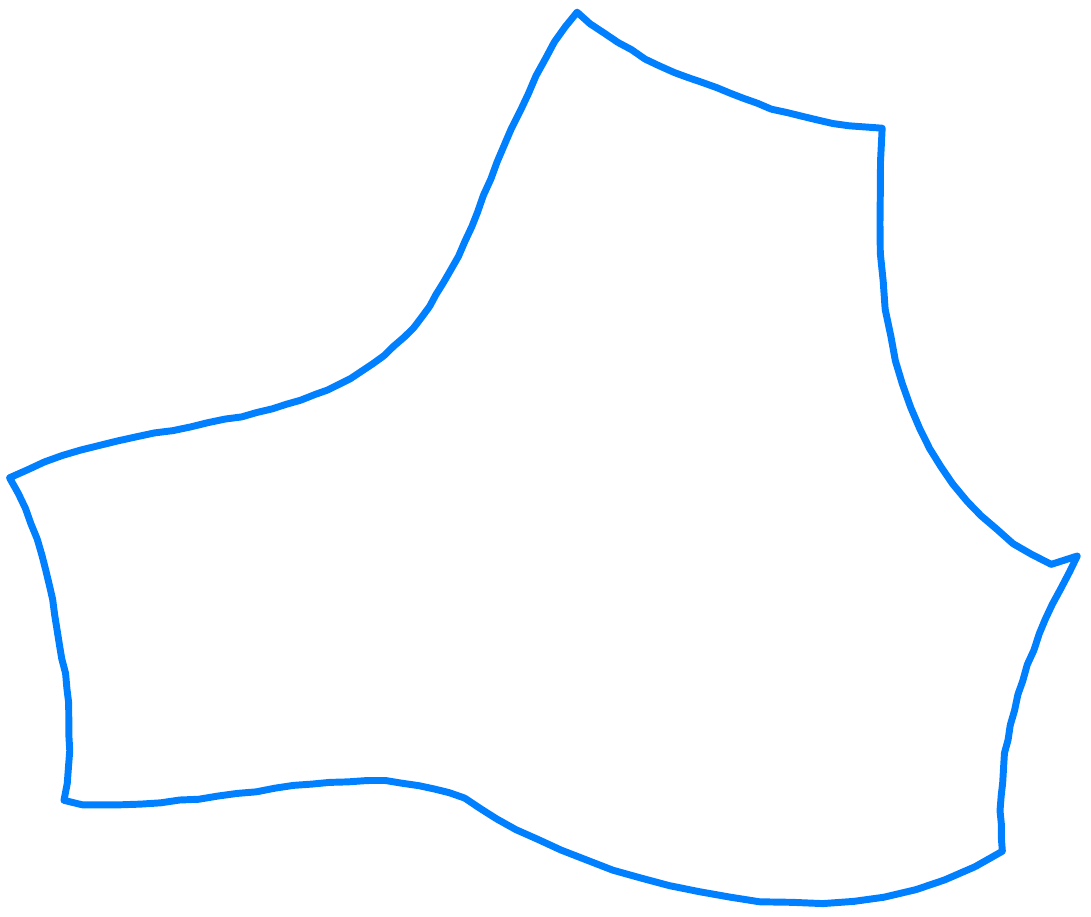}  \caption{}  \end{subfigure} 
   \begin{subfigure}{0.196\textwidth} \centering \includegraphics[width=\textwidth]{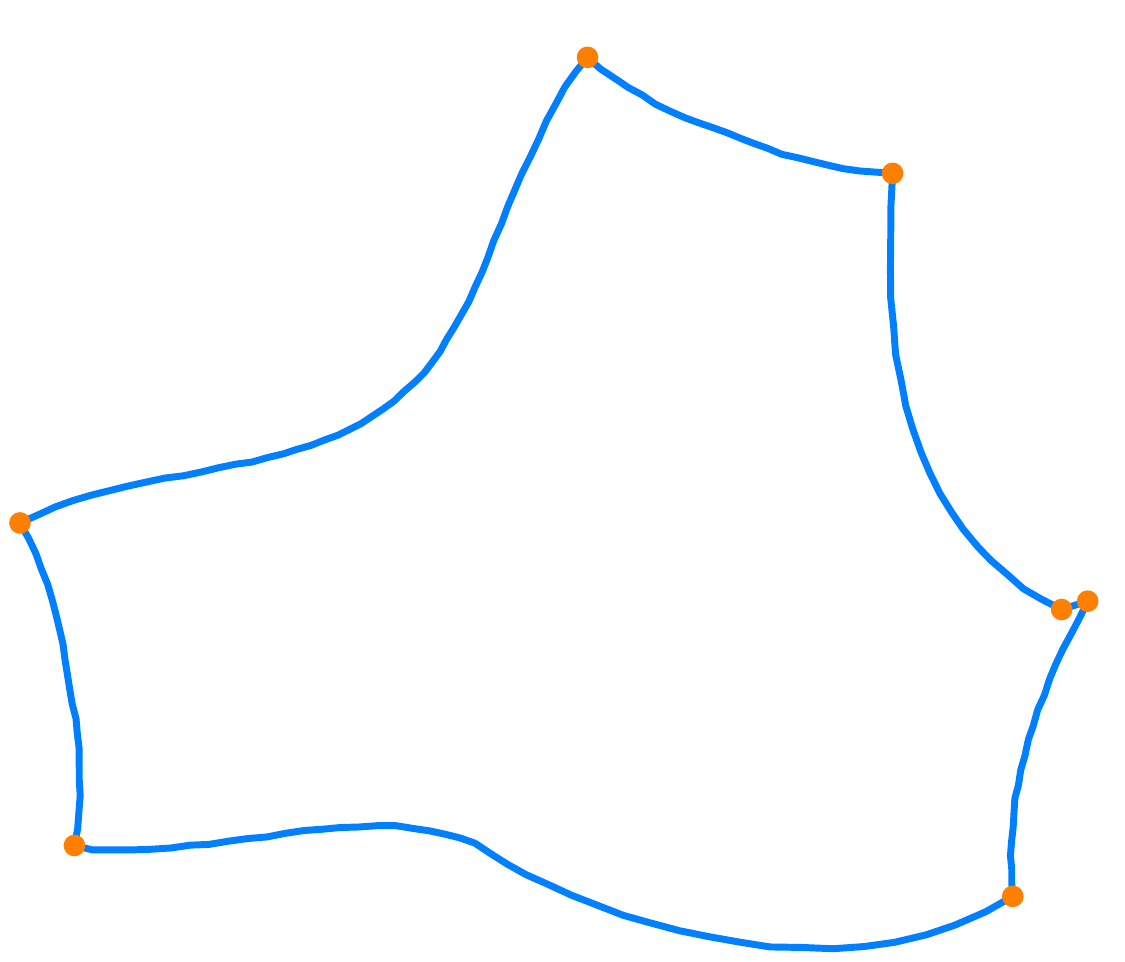} \caption{}  \end{subfigure} 
   \begin{subfigure}{0.196\textwidth} \centering \includegraphics[width=\textwidth]{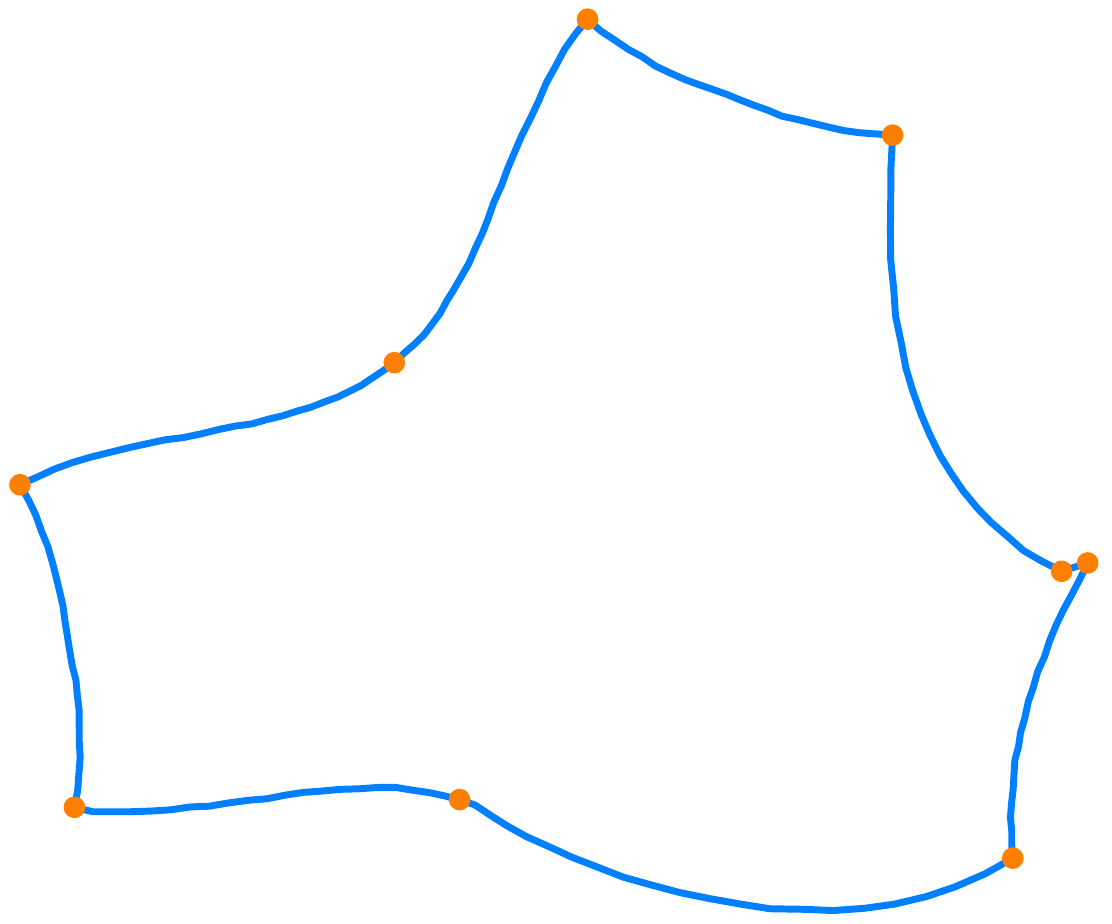}  \caption{}  \end{subfigure} 
   \begin{subfigure}{0.196\textwidth} \centering \includegraphics[width=\textwidth]{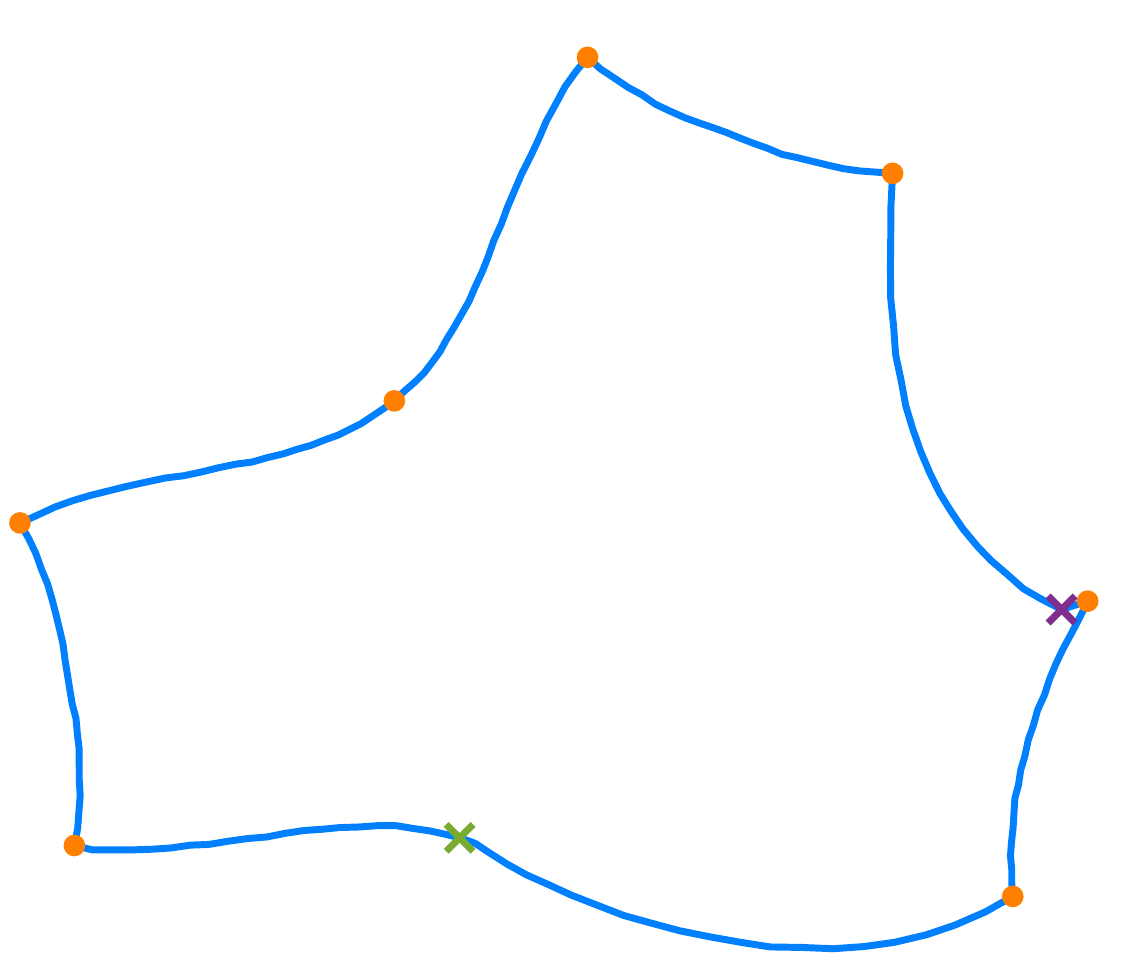} \caption{}    \end{subfigure} 
   \begin{subfigure}{0.196\textwidth} \centering \includegraphics[width=\textwidth]{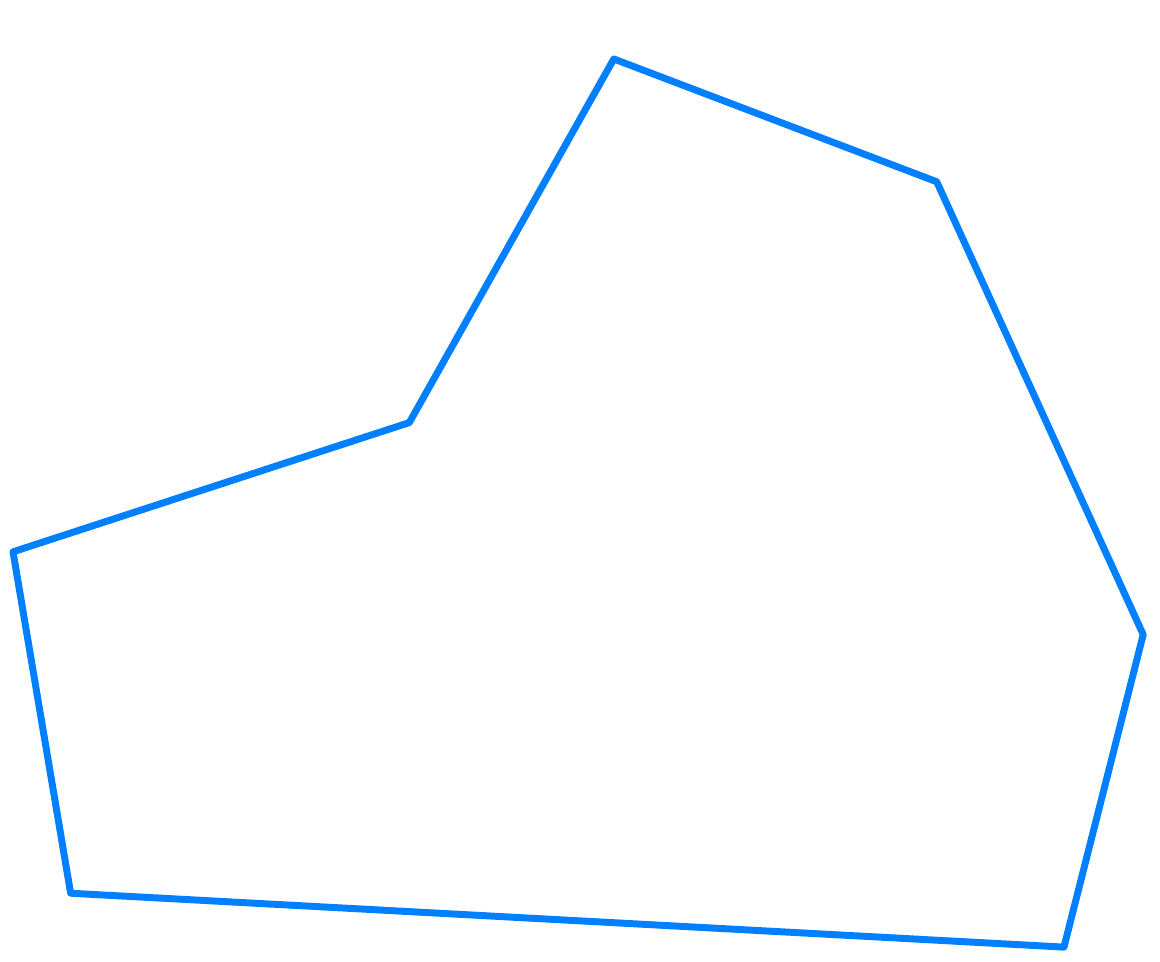}  \caption{}   \end{subfigure} 
\caption{A toy example of parametric domain generation. (a) $\mathcal{M}_1$ with a single boundary consisting of 173 vertices; (b) corners (solid orange dots); (c) more corners selected by step 2; (d) redundant corners selected by step 3 (marked by green crosses) and step 4 (marked by purple crosses) respectively; and (e) the final parametric domain with 7 corners.}\label{fig:para_generation}
\end{figure}

\textbf{Parametric domain generation.} Similar to the traditional B-spline surface, it is required to place at least $k$ knots on each edge of a parametric domain boundary for degree-$k$ TCB-splines to guarantee the partition of unity of basis functions~\cite{Cao:2019:CMAME}. Taking the 2D domain $\mathcal{M}_1$ directly as a parametric domain will introduce too many redundant knots along the boundary. Therefore, there will be many more control points than boundary vertices, leading to overfitting during the surface fitting. To avoid overfitting, we generate a polygonal parametric domain $\Omega$ that has a similar shape to the 2D domain $\mathcal{M}_1$, but with many fewer vertices at the boundary. Without loss of generality, we assume that the 2D domain $\mathcal{M}_1$ is bounded by a single boundary $Q$ defined by $n+1$ counter-clockwise order vertices $Q = \{\mq_1,\mq_2,...,\mq_n\}$. We refer to the polyline $\mq_i,\mq_{i+1},\cdots,\mq_{j-1},\mq_j$ as the boundary segment $[\mq_i,\mq_j]$. The subscripts are interpreted modulo $n+1$, i.e., vertex $\mq_{n+1}$ corresponds to vertex $\mq_{0}$. We define the discrete change-of-curvature on each line segment $\mq_i\mq_{i+1}$ as 
$\kappa_i=\frac{1}{2}||\mq_i\mq_{i+1}||(|\kappa(\mq_i)| + |\kappa(\mq_{i+1})|)$, where $|\kappa (\mq_i)|=\frac{2\theta_i}{||\mq_{i-1}\mq_{i}|| + ||\mq_i\mq_{i+1}||}$ is the absolute value of the discrete curvature at $\mq_i$ and $\theta_i= |\pi - \angle\mq_{i-1}\mq_{i}\mq_{i+1}|$ is the unsigned turning angle of the boundary $Q$ at $\mq_i$ ~\cite{Grinspun:2008:SIGGRAPH}.  The accumulated change-of-curvature of the boundary segment $[\mq_i,\mq_j]$ is then defined as $\kappa[\mq_i,\mq_j] = \sum_{k=i}^j \kappa_k$. In the following, we select a small number of the vertices as the corners of $\Omega$ by four steps:
 \begin{description}
  \item \textbf{Step~1.} For each vertex $\mq_{i}$ of $Q$, if $\theta_i>\delta_1$, then $\mq_{i}$ is included in the corner list of $\Omega$; see Fig.~\ref{fig:para_generation}(b).
  
  \item \textbf{Step~2.} For each two consecutive corners $\mq_i$ and $\mq_j$ selected by Step 1. If $\kappa[\mq_i,\mq_j] < \gamma $, then no more vertices between $\mq_i$ and $\mq_j $ are selected as corners. Otherwise, include the vertex $\mq_s$ between $\mq_i$ and $\mq_j$ which minimizes $|\kappa[\mq_i,\mq_s]-\kappa[\mq_s,\mq_j]|$ in the corner list of $\Omega$.  Two new boundary segments $[\mq_i,\mq_s]$ and $[\mq_s,\mq_j]$ are then recursively handled; see Fig.~\ref{fig:para_generation}(c).
  
  \item \textbf{Step~3.} For each two consecutive corners $\mq_i$ and $\mq_j$ obtained by Step 2, if the boundary segment $[\mq_i,\mq_j]$ contains less than $m$ vertices and $\theta_s$ is the smaller angle between $\alpha_i$ and $\alpha_j$, then $\mpp_s$ is removed from the corner list; see Fig.~\ref{fig:para_generation}(d). 
  
  \item \textbf{Step~4.} For each three consecutive corners $\mpp_i$, $\mpp_j$ and $\mpp_k$ obtained by Step 3, if $\mpp_i$, $\mpp_j$ and $\mpp_k$ are approximately co-linear, i.e., $|\pi - \angle\mpp_i\mpp_j\mpp_k| < \delta_2$, then $\mpp_j$ is removed from the corner list; see Fig.~\ref{fig:para_generation}(d).  
\end{description}
Intuitively, the vertices with large turning angles are selected as the corners of the parametric domain $\Omega$ in Step 1. The boundary segment formed by two consecutive corners chosen in Step 1 may have small oscillations. In Step 2, more vertices of the domain boundary are included in the corners according to the discrete curvature distribution to better preserve the shape of  $\mathcal{M}_1$. Steps 3-4 remove redundant corners that do not affect the shape of the parametric domain. We use four user-specified parameters $\delta_1$,  $\gamma$, $m$, and $\delta_2$ to control the generation of the parametric domain. Reducing the number of corners in the parametric domain and avoiding redundant knots in the surface fitting is desirable in our application. In all of our experiments, we use the parameter
$\delta_1 = 10^{\circ}, \gamma = \pi/2$, $m=3$, and $\delta_2=5^{\circ}$, which generates a parametric domain $\Omega$ well approximating the original domain $\mathcal{M}_1$ with a small number of vertices; see Fig.~\ref{fig:para_generation}(e).

\textbf{Reparameterization.} We now produce a bijective mapping $\phi_1$ from $\mathcal{M}_1$ to $\Omega$. The parameterization from the 3D mesh $\mathcal{M}_0$ to the 2D parametric domain $\Omega$ can be computed as the composition of $\phi_0$ and $\phi_1$. To guarantee a bijective mapping between $\mathcal{M}_1$ and $\Omega$, we first have to ensure a bijective mapping between the boundaries of $\mathcal{M}_1$ and $\Omega$. To this end, the boundary vertices of $\mathcal{M}_1$ between two consecutive corners $\mq_i$ and $\mq_{j}$ of $\Omega$ are mapped to the segment $\mq_i\mq_{j}$ using the chord length parameterization method~\cite{Farin:2002}. Then, the interior vertices of $\mathcal{M}_1$ are mapped to the interior of $\Omega$ using mean value mapping~\cite{Schneider:2013:CGF}, as done in~\cite{Wang:2022:CMAME}; see Fig.~\ref{fig:fit_algorithm}(c). It has been proved that the mean value mapping is bijective if the source and target polygons are sufficiently close. In the context of this work, $\mathcal{M}_1$ and $\Omega$ have almost the same shape; therefore, the mean value mapping $\phi_1$ between $\mathcal{M}_1$ and $\Omega$ is close to the identity mapping, which is bijective and has low distortion. Although there is no guarantee that $\phi_1$ is bijective,  numerically, we never find overlaps in the parameterization in our experiments. If, in any case,  portions of parameterization $\phi_1$ overlap, we can create sufficiently many intermediate domains between $\Omega$ and $\mathcal{M}_1$ as done in~\cite{Wang:2022:CMAME} and successively map from one to another to get a bijective composite mapping numerically. 

\subsection{Linear least-squares surface fitting \label{surface_remodeling}}
Given the parametric domain, we need to place a set of knots before defining the TCB-spline basis functions. Generally speaking, the knot distribution should be adapted to the geometry to achieve a more compact surface representation~\cite{Zhang:2017:cagd}. However, there is no effective and intuitive way to construct a knot distribution in the spline-based fitting. For simplicity, we first generate a set of evenly distributed knots over $\Omega$ as described in Section~\ref{TCB-surface}; see Fig.~\ref{fig:fit_algorithm}(e) for an example of an initial evenly spaced knot distribution. We will defer the discussion of knot refinement until Section~\ref{sec:knot_refinement}.

Upon the parametric domain and knot set, we generate the degree-$k$ t-config family using LTP in Section~\ref{sec:LTP} and define the degree-$k$ TCB-spline basis functions. In most fitting processes, all control points of the splines are computed at once by minimizing the fitting error in the least-squares sense as formulated in Eq.~(\ref{equ:ls}). In contrast, we first perform the boundary curve fitting and then complete the entire surface fitting with fixed boundary control points to better capture the boundary features of the model.

The set of vertices of $\mathcal{M}_0$ is the union of the set of boundary vertices $\mathcal{M}_0^{B}$ and the set of inner vertices  $\mathcal{M}_0^I$. Then, we obtain the boundary control points $\mc_I^B$ with $I\in\mathcal{I}_k^b$ by minimizing the following linear least-squares energy
\begin{equation}E_B(\{\mc_I^B|I\in\mathcal{I}_k^b\})={\sum_{\mx_i \in \mathcal{M}_0^{B}} ||~\mx_i - \sum_{I\in\mathcal{I}_k^b} B_I(\mpp_i) \mc_I||},
\end{equation}
where $\mpp_i$ is the parameter point of $\mx_i$, $\mathcal{I}_k^b\subset \mathcal{I}_k$ has all knots on the parametric domain boundary. In the following, we consider the computation of inner control points. A common problem in surface fitting is that the resulting surface may have unwanted fluctuations. Reconstructed surfaces can be smoothed by augmenting the linear least-squares energy for the computation of inner control points with an additional fairness term. Here, we adopt the often-used thin plate energy functional as the fairness term: 
 \begin{equation}\label{equ:fitting_energy}
   E_{fair}(\mss) = \int ||\mss''(\mt)||^2d\mt. \end{equation}
Accordingly, we compute the inner control points of the TCB-spline surface by solving the following weighted linear least-squares problem:
$$E_I(\{\mc_I^I|I\in\mathcal{I}_k\setminus\mathcal{I}_k^b\}) = \sum_{\mx_i \in \mathcal{M}_0^{I}} w_i|||~\mx_i - \mss(\mpp_i)|| + \lambda E_{fair}(\mss),$$ 
where $w_i$ is the point-wise fitting weight for $\mx_i$, $\mpp_i$ is the parameter value of $\mx_i$, $\lambda$ is the fairness weight. A large value of $\lambda$ increases the fairness of the approximating surface while simultaneously increasing the fitting error.  In all our examples, the fairness weight is set to $\lambda=0.01$. We set a uniform point-wise weight for all vertices in our initial fitting iteration, i.e.,  $w_i = 1$. We initially assign a uniform point-wise weight to all vertices, with $w_i = 1$, until the surface fitting process reaches the desired number of iterations or the maximum fitting error becomes smaller than a predefined threshold. However, when using uniform weights, the fitting error in the transition area (e.g., region 2 in Fig.~\ref{fig:fit_algorithm}(a)) may be significantly greater compared to other regions. To achieve a smoother result in the transition area and further reduce the fitting errors in the remaining regions, we introduce an additional step of fitting iteration by setting non-uniform weights. Specifically, we set $w_i = 1,000$ if the corresponding point $\mx_i$ is outside the transition area and has one of the top $5\%$ largest fitting errors among the points in the non-transition area.

\subsection{Adaptive knot refinement \label{sec:knot_refinement}}

\begin{figure}
  \centering
\graphicspath{{figures/}}
   \begin{subfigure}{0.222\textwidth} \centering \includegraphics[width=\textwidth]{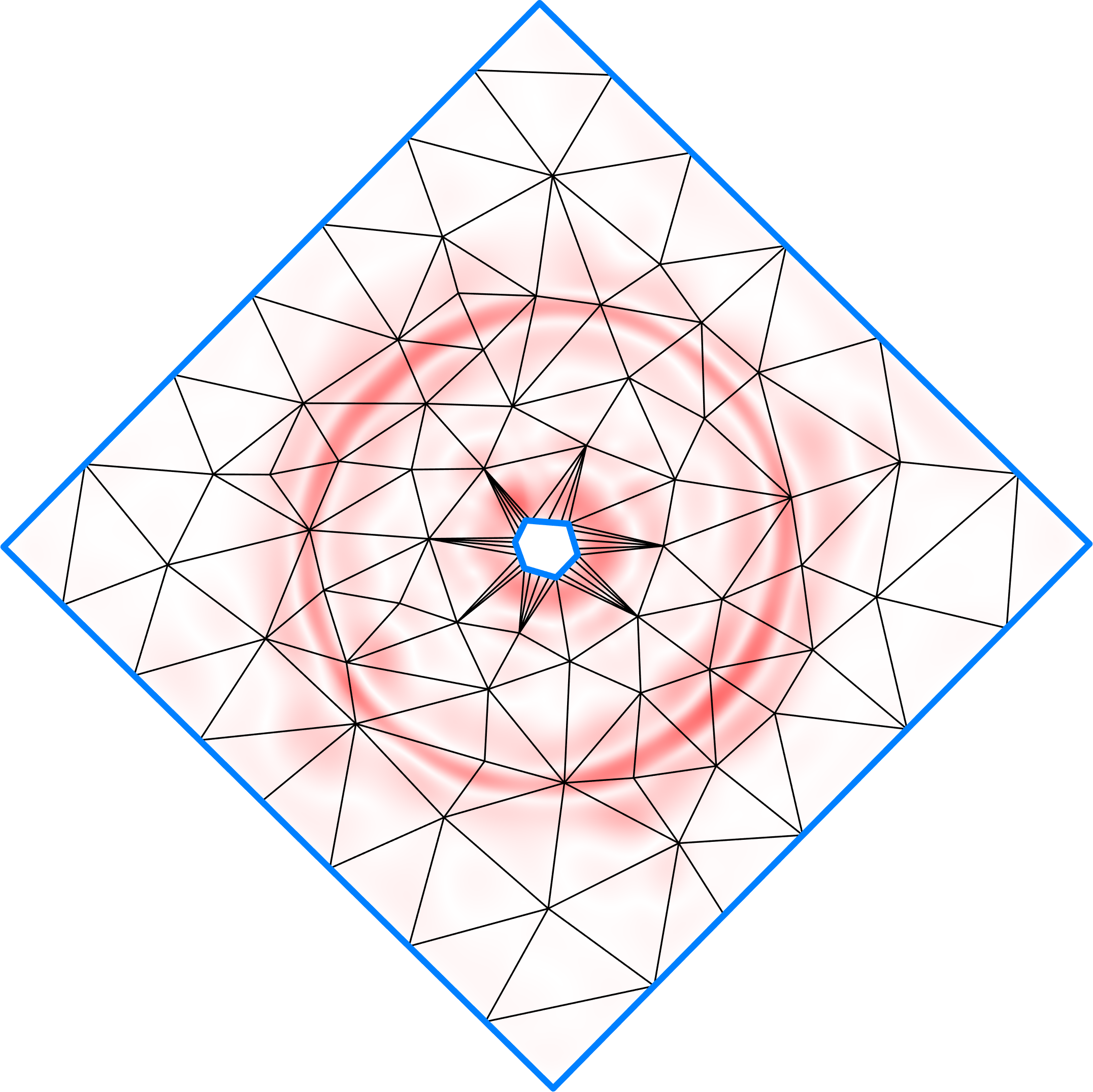}  \caption{3rd}  \end{subfigure} ~
   \begin{subfigure}{0.222\textwidth} \centering \includegraphics[width=\textwidth]{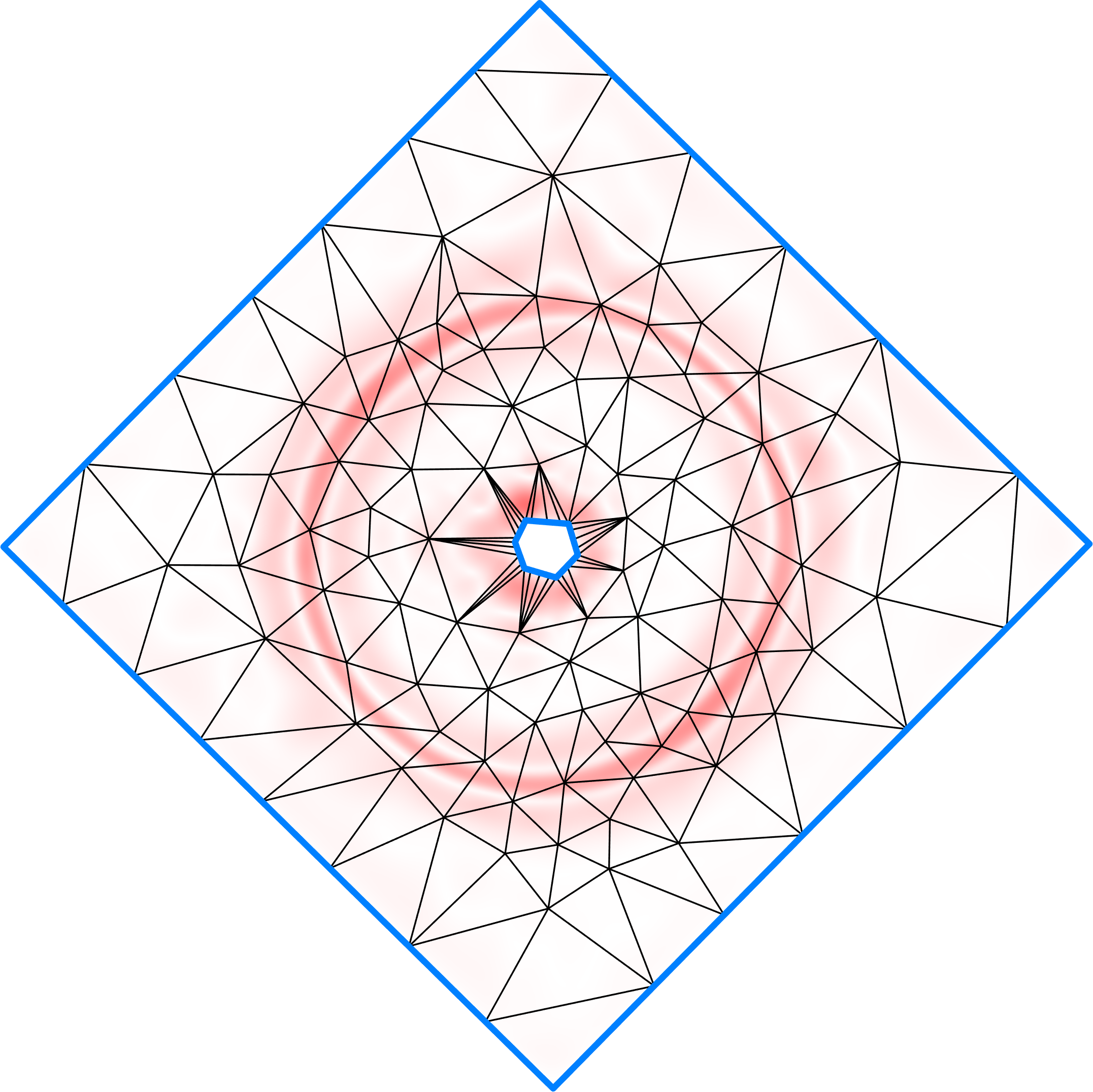} \caption{6th}  \end{subfigure} ~
   \begin{subfigure}{0.222\textwidth} \centering \includegraphics[width=\textwidth]{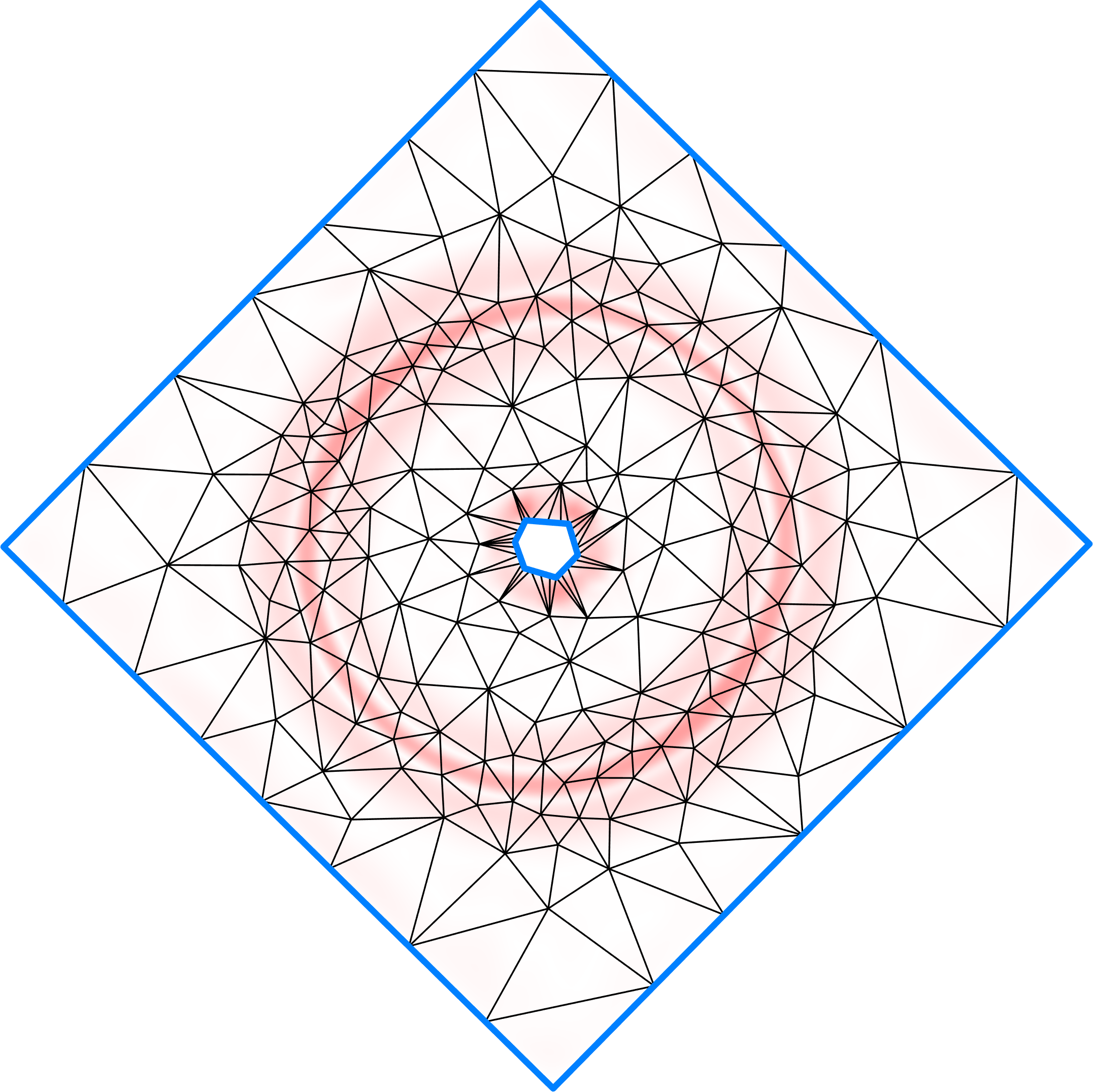} \caption{9th}  \end{subfigure} ~
   \begin{subfigure}{0.222\textwidth} \centering \includegraphics[width=\textwidth]{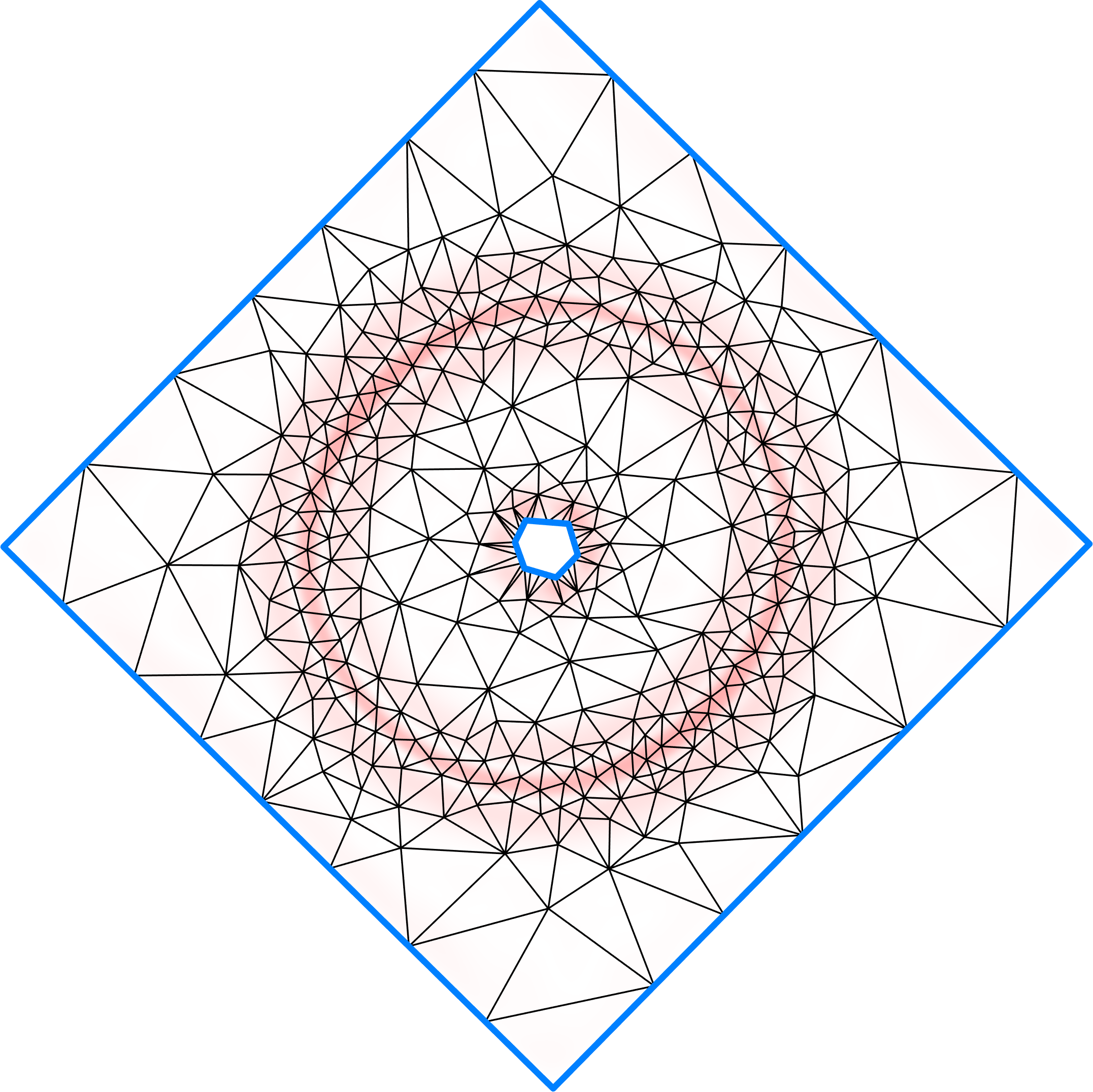}  \caption{12th}  \end{subfigure} ~
   \begin{subfigure}{0.05\textwidth} \centering \includegraphics[width=\textwidth]{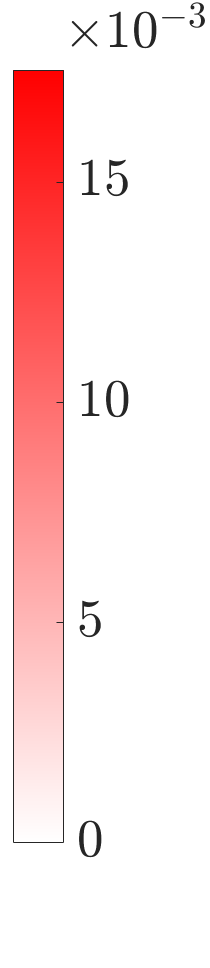}    \end{subfigure} ~

    \begin{subfigure}{0.22\textwidth} \centering \includegraphics[width=\textwidth]{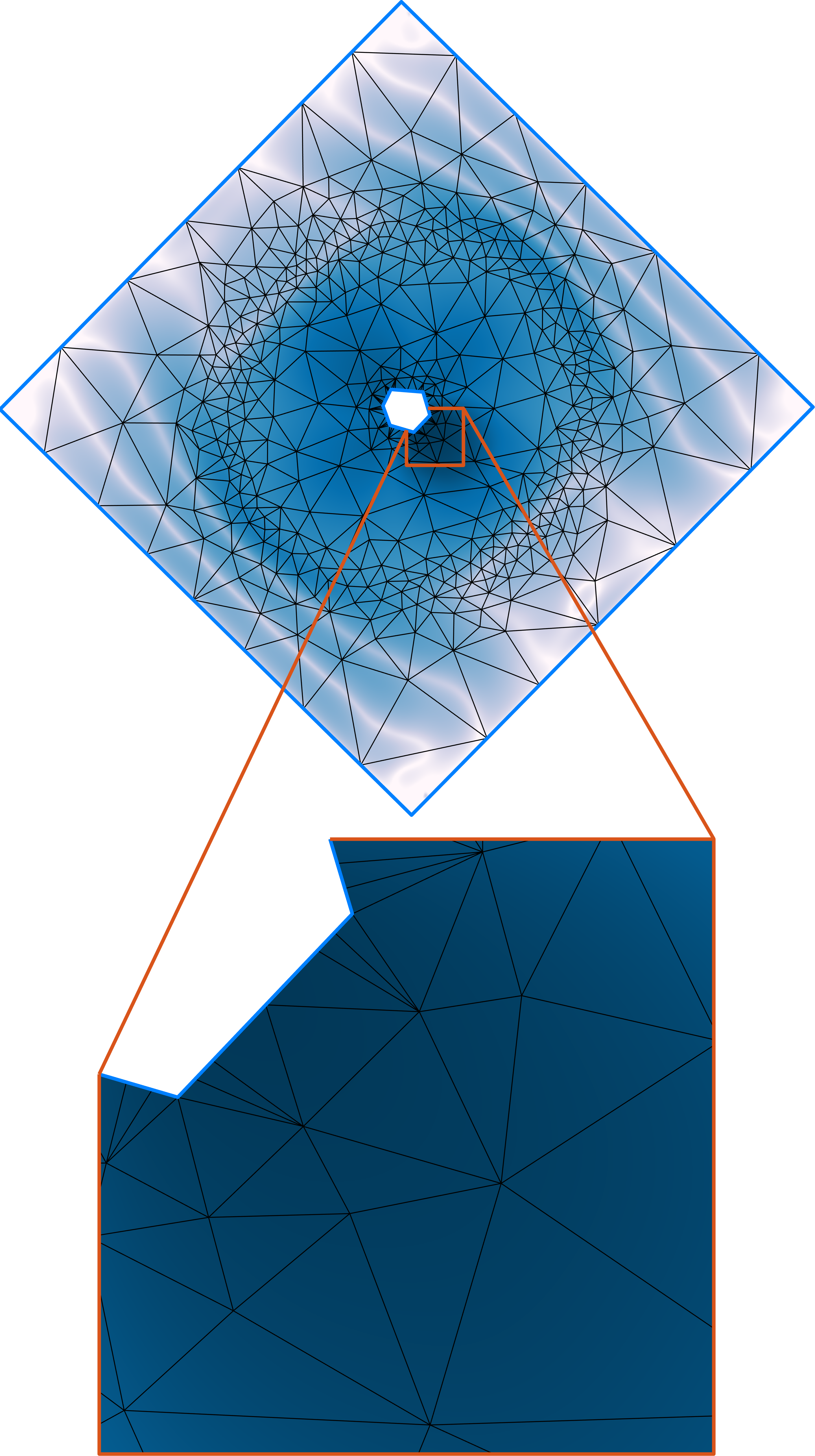}  \caption{1st}  \end{subfigure} ~
   \begin{subfigure}{0.22\textwidth} \centering \includegraphics[width=\textwidth]{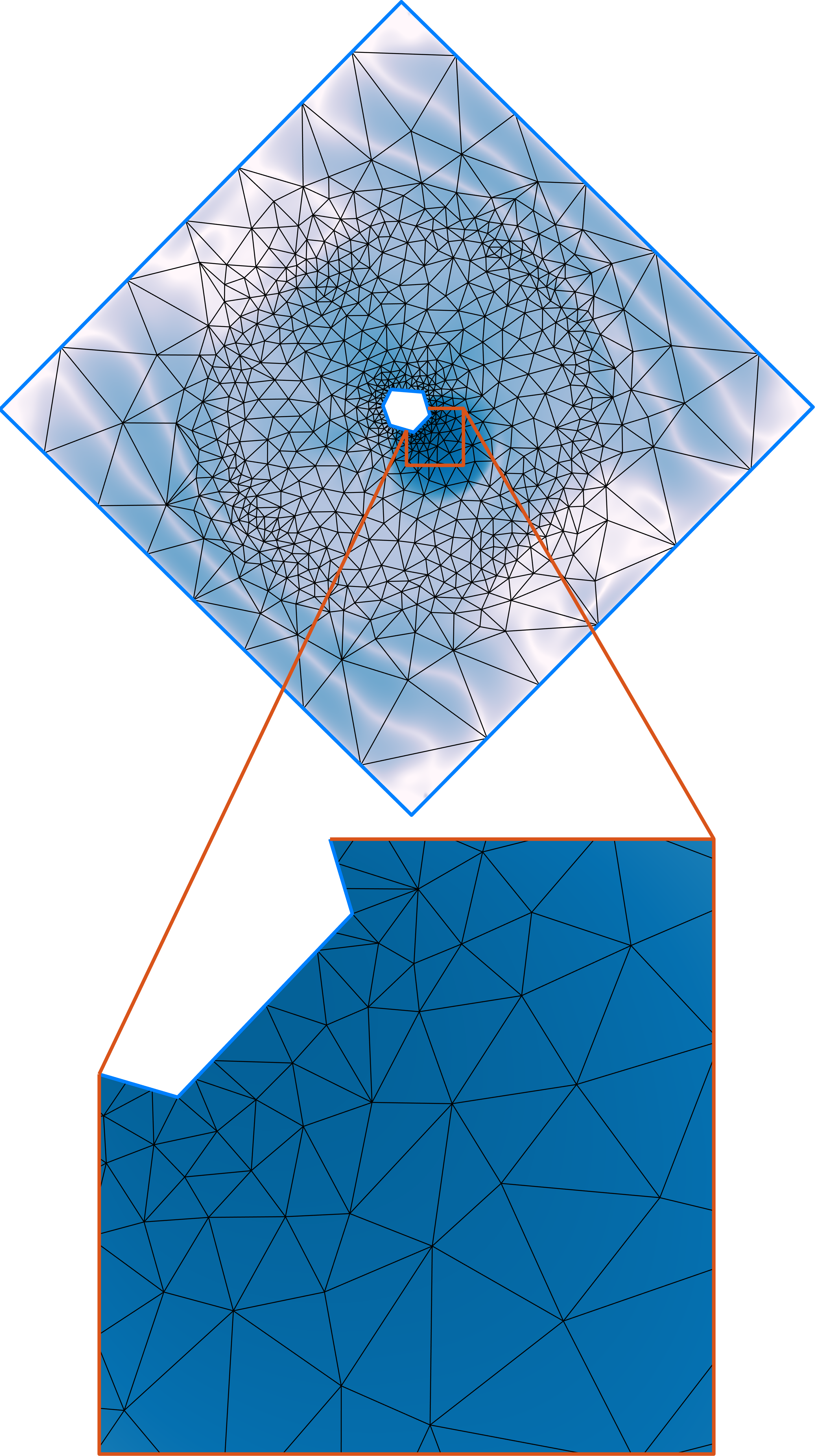} \caption{2nd}  \end{subfigure} ~
   \begin{subfigure}{0.22\textwidth} \centering \includegraphics[width=\textwidth]{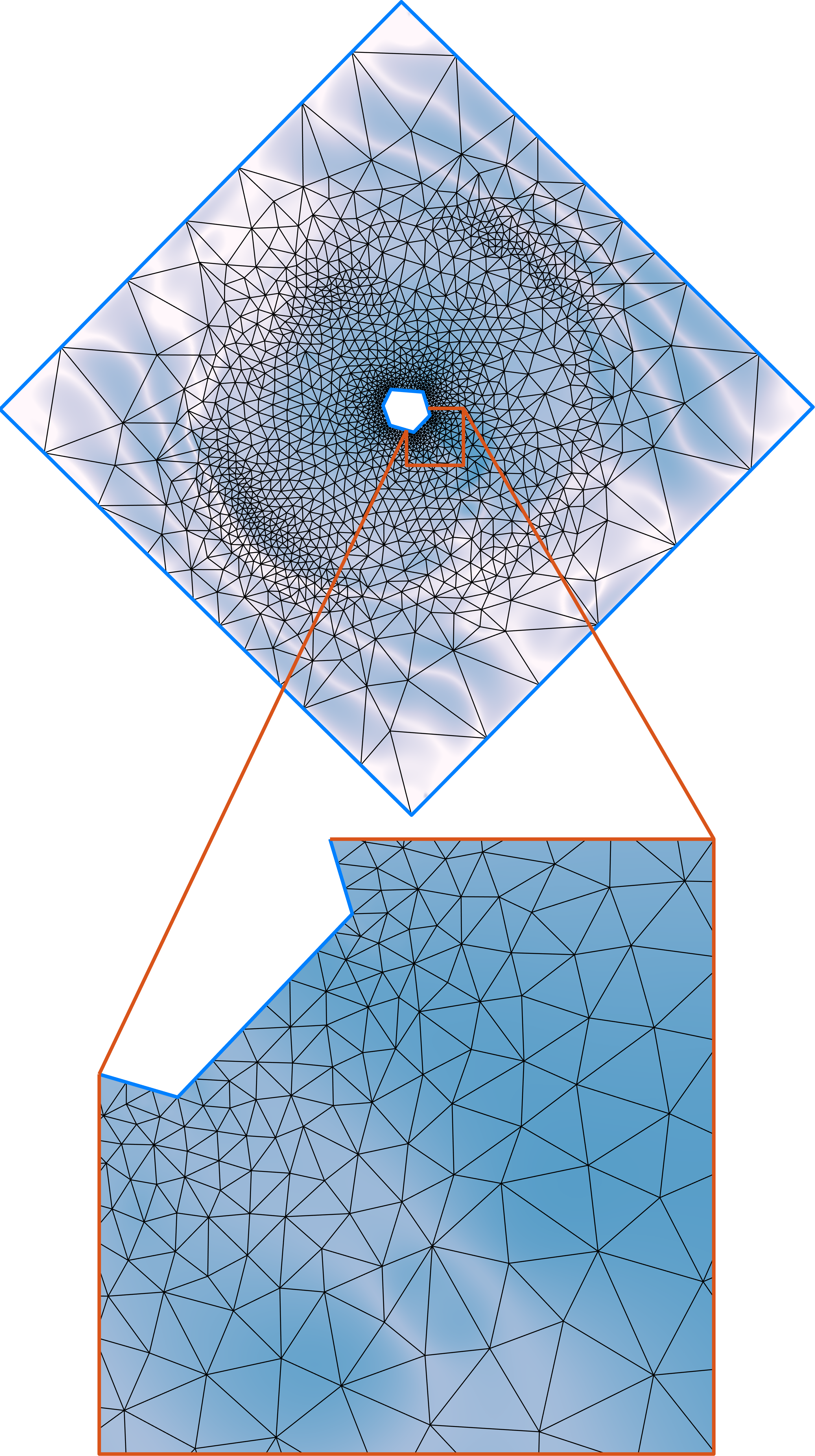}  \caption{3rd}  \end{subfigure} ~
   \begin{subfigure}{0.22\textwidth} \centering \includegraphics[width=\textwidth]{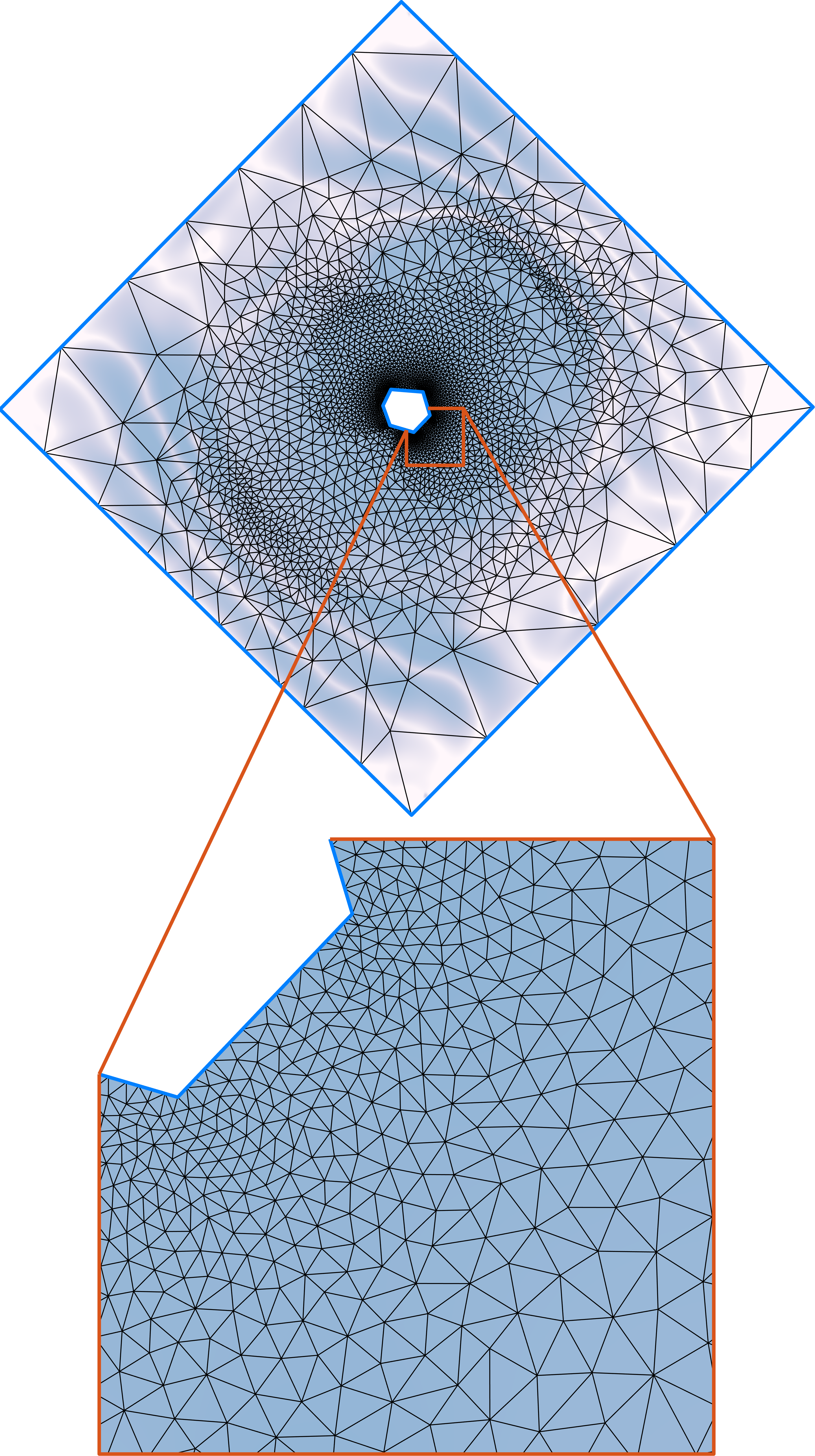}  \caption{4th}  \end{subfigure} ~
   \begin{subfigure}{0.04\textwidth} \centering \includegraphics[width=\textwidth]{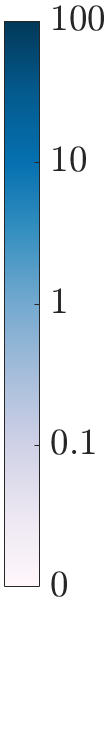}    \end{subfigure} ~
\caption{The knot refinement process for the pipe junction. The knot distribution and the color-coded normalized fitting error of the refinement process at the (a) 3rd, (b) 6th, (c) 9th and (d) 12th mesh refinement steps for the fitting phase; (e-g) the knot distribution and the color-coded displacement error at the each refinement step for the analysis phase. In (e-g), the color bar uses a nonlinear mapping in order to better represent small displacement error variations. }\label{fig:knot_refinement}
\end{figure}

In our method, knot refinement is applied to improve the quality of the remodeled surface in the fitting phase and to increase the degree of freedom in the analysis phase. In the surface fitting phase, knot refinement is adaptively controlled by the fitting error distribution. Let $\mathcal{T}_s$ denote the knot mesh for the $s$-th iteration of fitting, where $s=1$ means the initial fitting iteration. In the initial fitting step, the knot mesh $\mathcal{T}_1$ is simply an evenly distributed triangular mesh. After each fitting step, we can measure the fitting error $e_j$ of each vertex $\mx_j$.
The knot mesh $\mathcal{T}_s$ is adaptively refined to be $\mathcal{T}_{s+1}$ according to the fitting errors obtained from the $s$-th fitting iteration by the following four steps:
\begin{description}
\item[Step 1.] Compute the fitting error $\rho_{s,i}$ of each face $T_{s,i}$ of the knot mesh $\mathcal{T}_s$ as $\rho_{s,i}=|\textrm{Area}|(T_{s,i})\bar{\rho}_{s,i}$, where $\textrm{Area}(T_{s,i})$ is the area of triangle $T_{s,i}$, and $\bar{\rho}_{s,i}$ is the average of the fitting error of the vertices whose parameter points are in $T_{s,i}$.    

\item[Step 2.] Collect the circumcenters of triangles $T_{s,i}$ with error $\rho_{s,i} >{\alpha^{\star} \mathop{\max}\limits_{i}\{\rho_{s,i}\}}$, and denote by $V^s$. In our experiments we empirically choose the constant $\alpha^{\star}$ as $\alpha^{\star}=0.1$.

\item[Step 3.] $\mathcal{T}_{s+1}$ is obtained by inserting the points in $V^s$ into the knot mesh $\mathcal{T}_s$, where the newly added knots whose Voronoi cells intersect with the parametric domain boundary are projected onto the corresponding boundary.

\item[Step 4.] The vertex distribution of $\mathcal{T}_{s+1}$ is further slightly optimized by fixing the knots already existing in $\mathcal{T}_s$ and adjusting the newly added knots using two steps of Lloyd's relaxation, where knots on the domain boundaries are restricted to move along the boundary.
\end{description}

We can then resolve the fitting problem on the refined knot set, leading to a reconstruction result with improved quality. Knot mesh refinement and surface fitting are repeated until a required
number of iterations is reached, or the maximum fitting error is less than a pre-specified threshold. It is also worth pointing out that we only need to update the computation of the TCB-spline basis involving newly added knots in each fitting iteration to improve the computational efficiency, owing to their local support property. Fig.~\ref{fig:fit_algorithm}(e-f) shows the initial knot mesh and the refined mesh, respectively, where we can observe that the triangles are locally refined in the regions with large fitting errors. The quality of the remodeled geometry on the refined knot mesh is significantly improved, as shown in Fig.~\ref{fig:fit_algorithm}(g-l). Furthermore, the knot meshes at selected intermediate iteration steps are illustrated in Fig.~\ref{fig:knot_refinement}(a-d).

In the shell analysis phase, the knot mesh is refined similarly to the surface fitting phase.  Since the exact displacement of a given model is usually unknown, the result obtained on a sufficiently refined knot mesh is used as the ``exact solution''. We therefore measure the error of displacement $d_j$ for each vertex $\mx_j$ of the triangular mesh with respect to the exact displacement of the shells. The knot mesh $\mathcal{T}_s$ is adaptively refined to be $\mathcal{T}_{s+1}$ according to the displacement errors obtained from the $s$-th analysis iteration by the above four steps, where Steps 1-2 are slightly modified as follows:
\begin{description}
\item[Step $1^*$.] 
For each face $T_{s,i}$ of the knot mesh $\mathcal{T}_s$, compute the average of and maximum displacement error of vertices whose parameter points are in $T_{s,i}$, and denote them by $\bar{\delta}_{s,i}$ and $\hat{\delta}_{s,i}$, respectively.    
\item[Step $2^*$.] {Find the triangles in $\mathcal{T}_s$, each with the average displacement error  $\bar{\delta}_{s,i} >\eta_1 \mathop{\min}\limits_{j}\bar{\delta}_{s,j} + (1-\eta_1)\mathop{\max}\limits_{j}\bar{\delta}_{s,j}$ or the maximum displacement error
$\hat{\delta}_{s,i} >\eta_2 \mathop{\min}\limits_{j}\hat{\delta}_{s,j} + (1-\eta_2)\mathop{\max}\limits_{j}\hat{\delta}_{s,j}$ where $\eta_1$ and $\eta_2$ are parameters for adjusting the adaptivity of the refinement in different problems.} $ V^s$ denotes the collection of the circumcenters of selected triangles.
\end{description}
Again, we update the geometric representation on the refined knot mesh by locally updating the TCB-spline involving newly added knots and solving the weighted linear least-square problem to fit the surface obtained in Section~\ref{surface_remodeling} using uniform weights. {If the maximum displacement error is less than the prescribed tolerance or the number of iterations exceeds the specified number of steps, the knot mesh refinement for the analysis phase is terminated. Fig~\ref{fig:knot_refinement}(e-h) show the knot meshes and the color-coded displacement error on the parametric domain of the pipe junction during the process of adaptive knot refinement, where the definition of the analysis problem is stated in the Section~\ref{Pipe_junction_problem}.}

\begin{remark}
It is generally a challenging task for automatic adaptive refinement for IGA, as discussed in~\cite{Wu:2015:CMAME}. Here, we only give a straightforward refinement method for solving the shell problem to demonstrate the flexibility of our TCB-splines in adaptive refinement. In the future, more sophisticated adaptive refinement methods~\cite{Coradello:2020:CMAME}, such as displacement gradient information-based or posterior error estimates-based techniques, are expected to be incorporated into our approach.
\end{remark}

\begin{figure}
  \centering
\graphicspath{{figures/}}
   \begin{subfigure}[b]{0.33\textwidth} \centering \includegraphics[width=\textwidth]{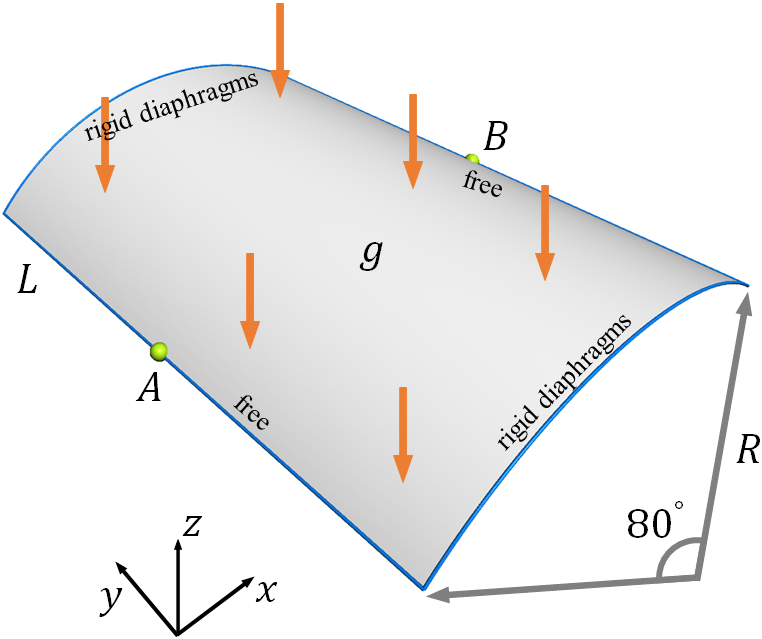}  \caption{}  \end{subfigure} ~
   \begin{subfigure}[b]{0.28\textwidth} \centering \includegraphics[width=\textwidth]{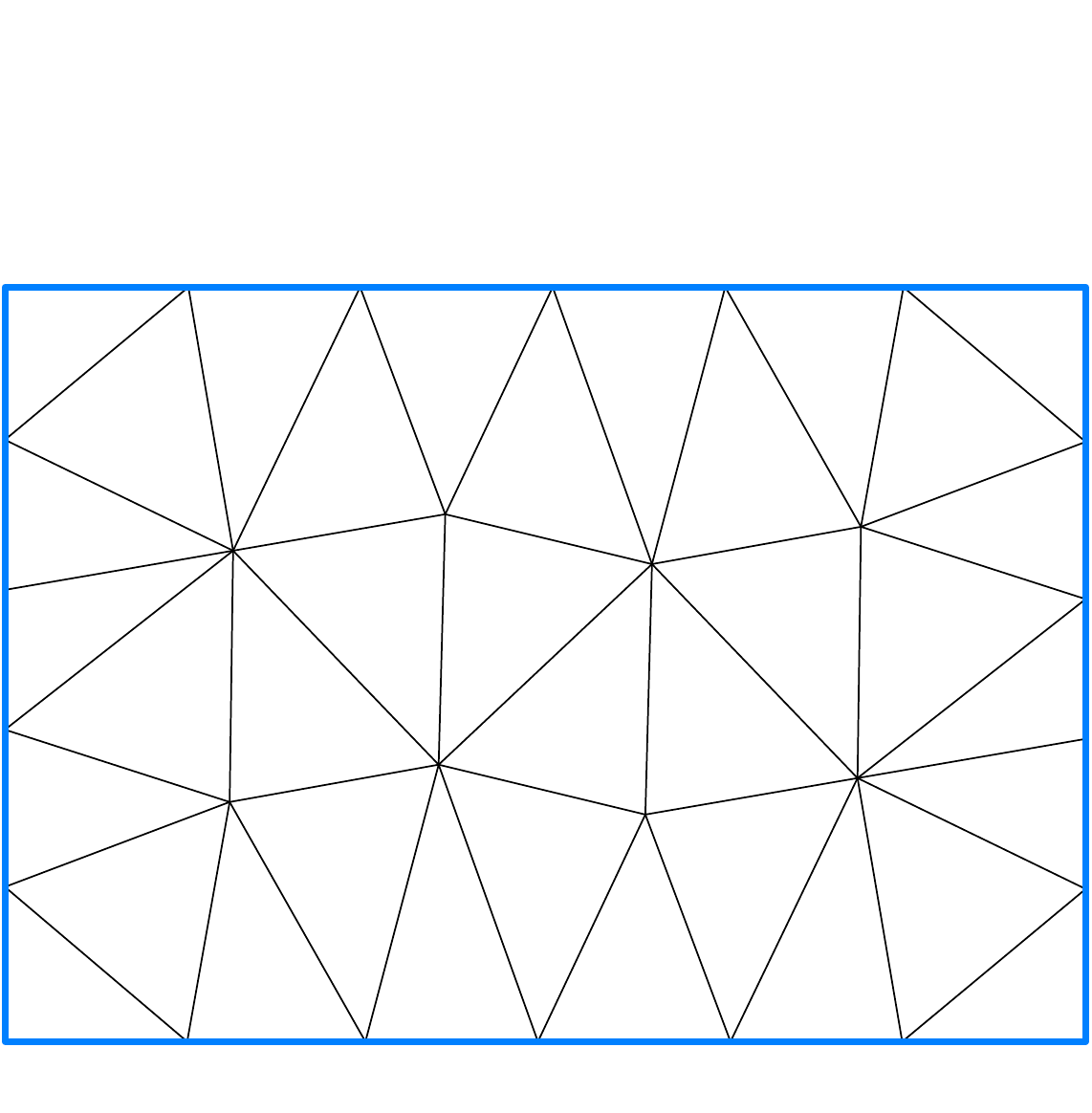} \caption{}  \end{subfigure} ~
   \begin{subfigure}[b]{0.33\textwidth} \centering \includegraphics[width=\textwidth]{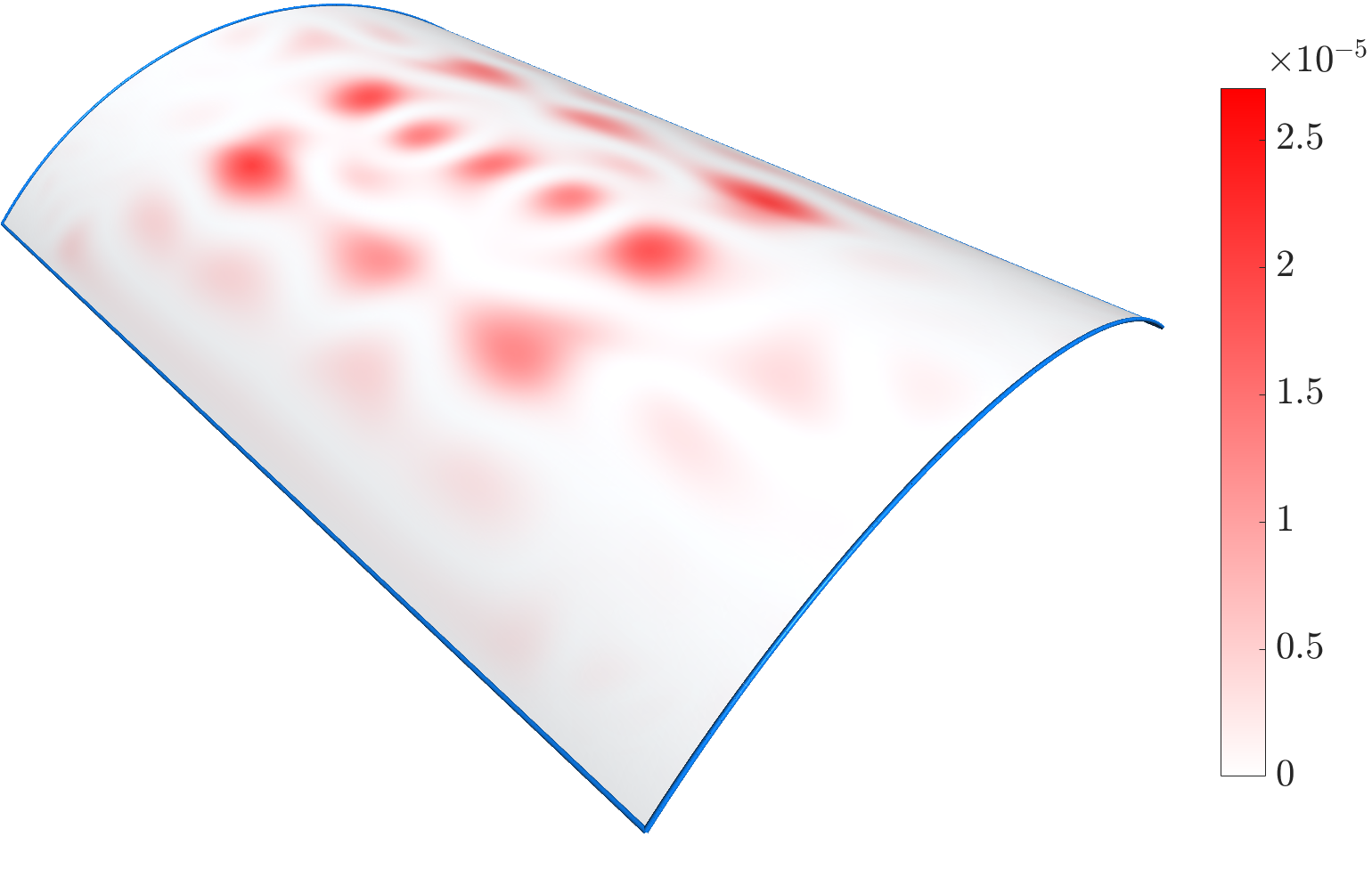}  \caption{}  \end{subfigure} ~

   \begin{subfigure}[b]{0.3\textwidth} \centering \includegraphics[width=\textwidth]{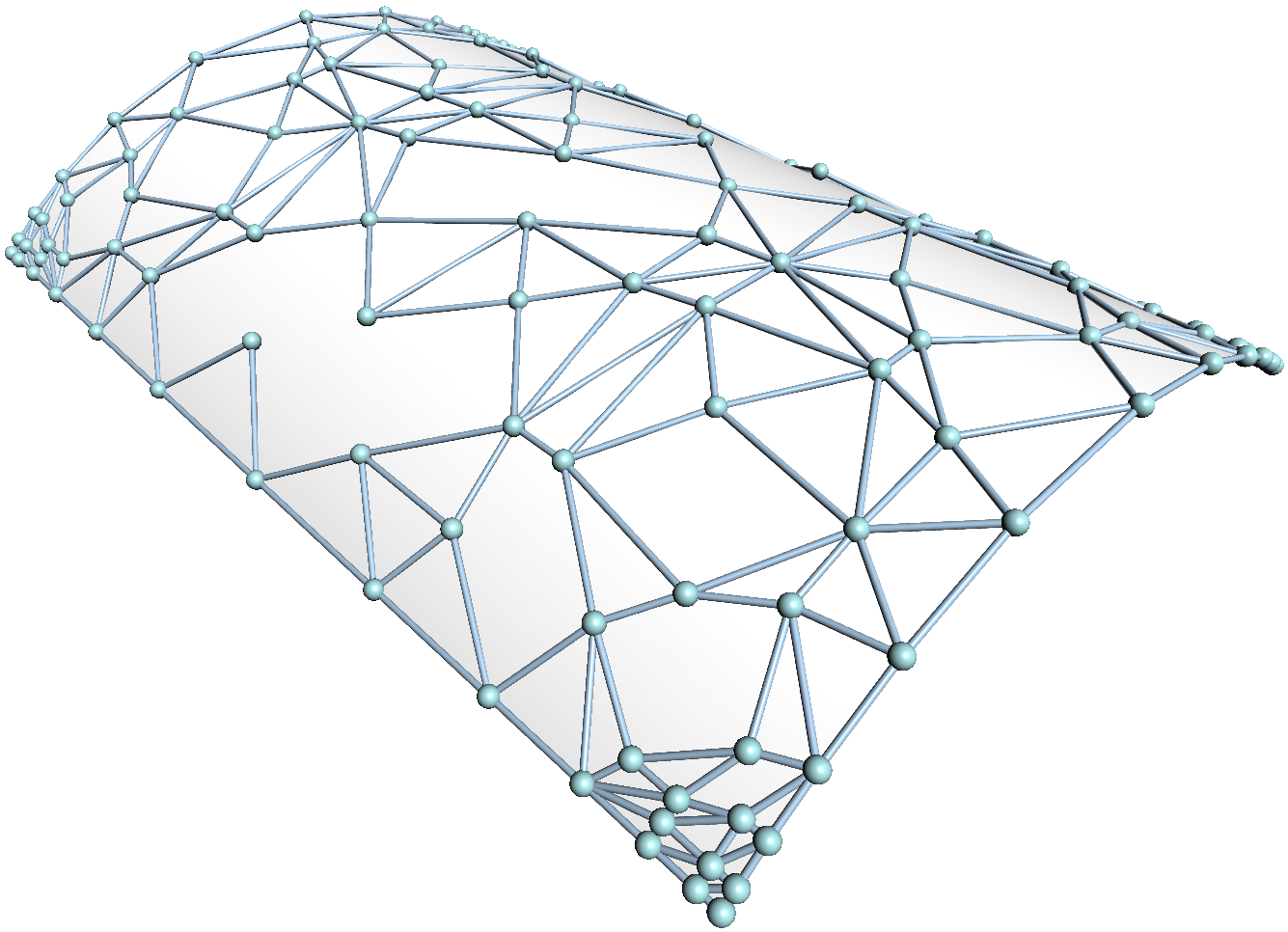}  \caption{}  \end{subfigure} ~
   \begin{subfigure}[b]{0.35\textwidth} \centering \includegraphics[width=\textwidth]{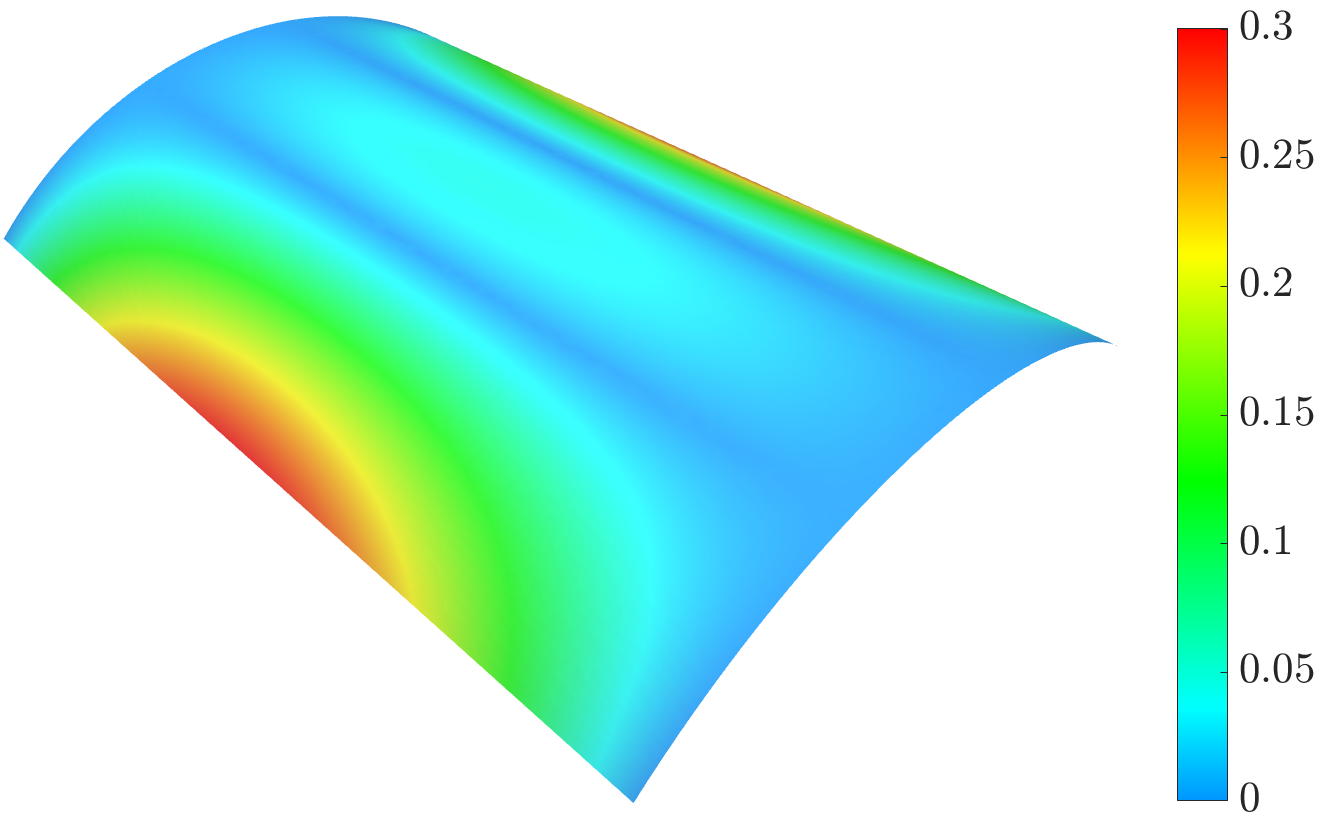}  \caption{}  \end{subfigure} ~
   \begin{subfigure}[b]{0.28\textwidth} \centering \includegraphics[width=\textwidth]{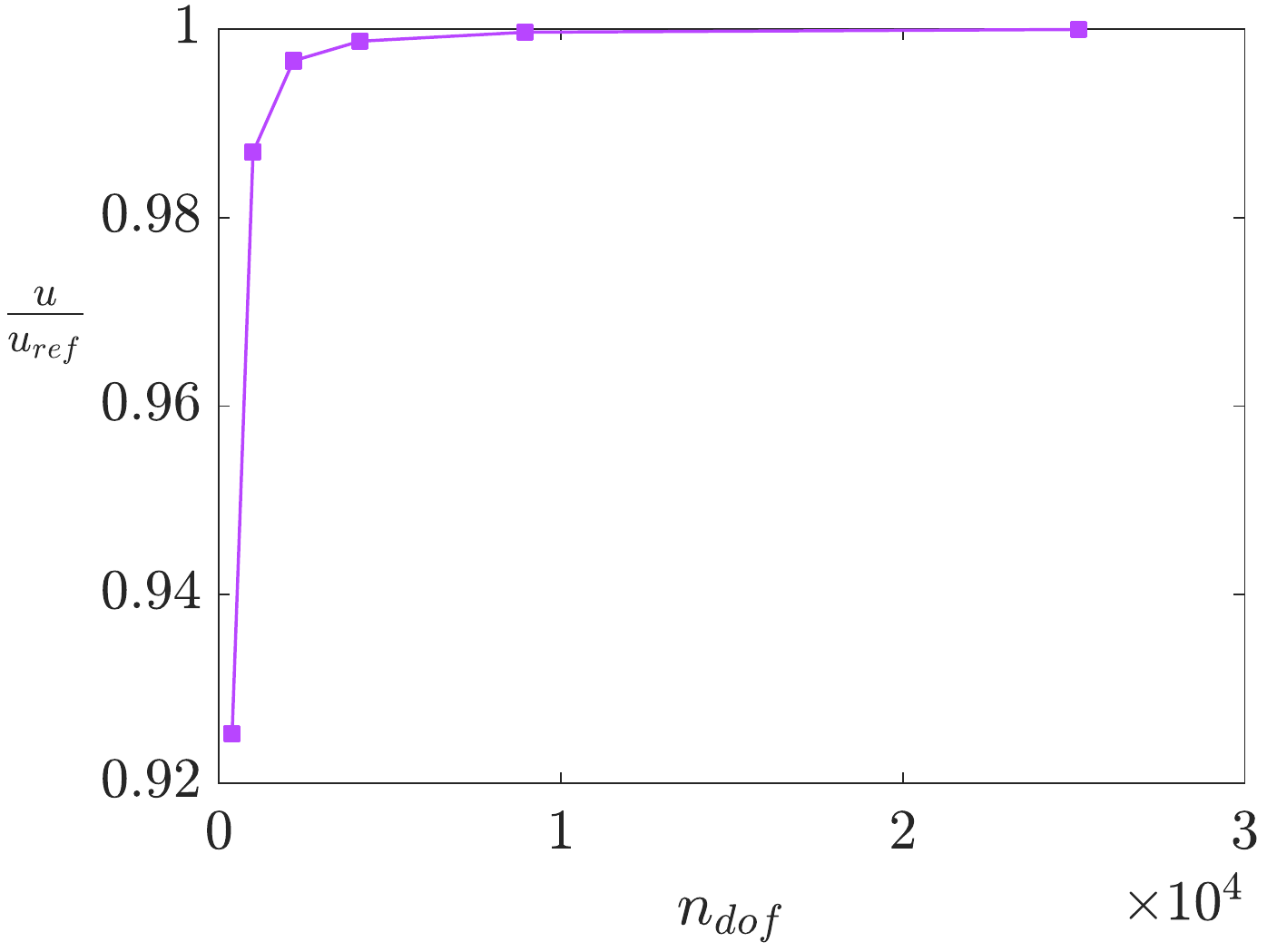}  \caption{}  \end{subfigure}
\caption{Scordelis-Lo roof. (a) The geometry for the Scordelis-Lo roof; (b-d) the knot mesh, fitting error and control net for the fitting result; (e) the final analyzed displacement magnitude; and (f) the convergence of normalized displacement.}\label{fig:scordelis-lo_roof}

\end{figure}

\section{Numerical examples \label{numerical_examples}}
In this section, the proposed framework is applied to solve thin-shell problems containing linear and nonlinear deformations. Then, the performance and flexibility of our TCB-spline-based representation and analysis framework are further exemplified using two geometrically nonlinear KL shells consisting of multiple trimmed NURBS patches. We use cubic TCB-spline basis functions for modeling and analysis in all examples. Note that TCB-splines are polynomials on each triangular cell obtained by triangulating the entire parametric domain restricted to knot lines~\cite{Cao:2019:CMAME}. The one-point rule for triangles is applied to each cell for the integration involved in the surface fitting, giving the exact integral value for the second-order derivative of cubic polynomials. For the integration in analysis, we use the $25$-point rule~\cite{Dunavant:1985:IJNME} on each cell to achieve an up to degree $10$ approximation of the integration of rational polynomials. To enhance the visual representation of our fitting results, we utilize the normalized fitting errors in all examples. Specifically, for each vertex $\mx_i$ of $\mathcal{M}_0$, we denote its normalized fitting error as $\tilde{e}_i=\frac{e_i}{d}$, where $e_i$ represents the fitting error of the vertex $\mx_i$, and $d$ corresponds to the diagonal length of the bounding box that encompasses the mid-surface. This normalization allows for a more meaningful comparison of fitting errors across models, taking into account the size of the surface being analyzed.

\subsection{Linear problem}
To test the accuracy under complex strain states, we first apply our framework to the two geometrically linear problems, namely the Scordelis-Lo roof and the pinched cylindrical shell with diaphragms. Small displacements occur in all problems, so geometrically linear computations are performed. Note that the shells in these problems have relatively simple shapes and natural parametric domains. For simplicity, we remodel these shapes over manually specified parametric domains as a single patch of TCB-splines without regarding shape symmetry.

\subsubsection{Scordelis-Lo roof}
The Scordelis-Lo roof is a cylindrical shell section supported by rigid diaphragms at its ends and subjected to the uniform gravity load of $g = 90.0~\rm{N}$ per unit area; see Fig.~\ref{fig:scordelis-lo_roof}(a). The geometry parameters are the radius $R = 25.0~\rm{mm}$, the length $L=50.0~\rm{mm}$, the thickness $t=0.25~\rm{m}$. The material properties are the Poisson's ratio $\nu = 0.0$ and Young's modulus $E=4.32\times 10^8~\rm{N/{mm^2}}$. The reference value $u_{ref}=0.3006$  of the vertical displacement at the midpoint (i.e., points $A$ and $B$ in Fig.~\ref{fig:scordelis-lo_roof}(a)) is given in ~\cite{Kiendl:2009:CMAME,Nguyen:2011:CMAME}.

{ Since the roof model is four-sided with a simple shape, we can unfold the model to a rectangular domain. To remodel this geometry, we first define the TCB-spline basis functions over the rectangular domain. }  Fig.~\ref{fig:scordelis-lo_roof}(b-d) show the evenly distributed knot mesh, the normalzied fitting error, and the control net for the remodeled roof.  {As we have seen, for this simple example, we have obtained a good fitting result without adaptive refinement. Therefore, in the analysis phase, our adaptive refinement is constructed directly from the knot mesh in Fig.~\ref{fig:scordelis-lo_roof}(b). In the adaptive refinement, we set the parameters $\eta_1=0.8$ and $\eta_2=0.8$. After several iterations of knot refinement, we obtain a converged result where the color-coded vertical displacement is shown in Fig.~\ref{fig:scordelis-lo_roof}(e). Moreover, the vertical displacements at the point $A$ converge to the reference value $u_{ref}=0.3006$; see the plot of the normalized displacement versus the number of degrees of freedom ($n_{dof}$) in Fig.~\ref{fig:scordelis-lo_roof}(f).}

\begin{figure}
  \centering
 \graphicspath{{figures/}}
   \begin{subfigure}[b]{0.27\textwidth} \centering \includegraphics[width=\textwidth]{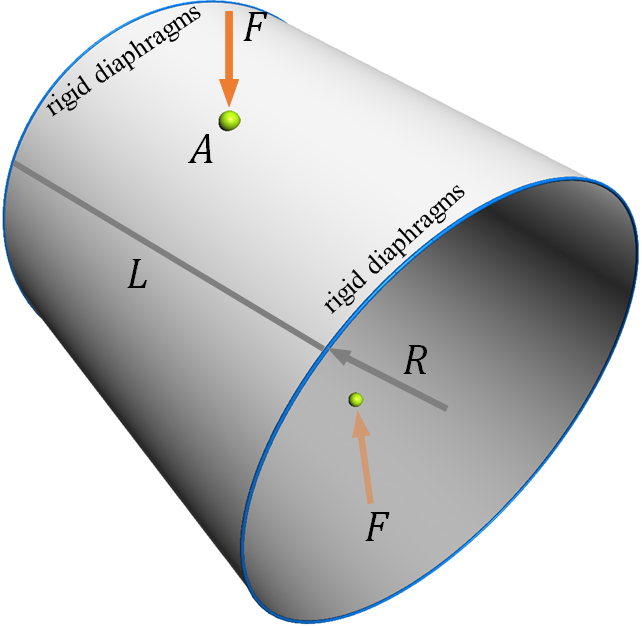} \caption{}  \end{subfigure} ~
   \begin{subfigure}[b]{0.37\textwidth} \centering \includegraphics[width=\textwidth]{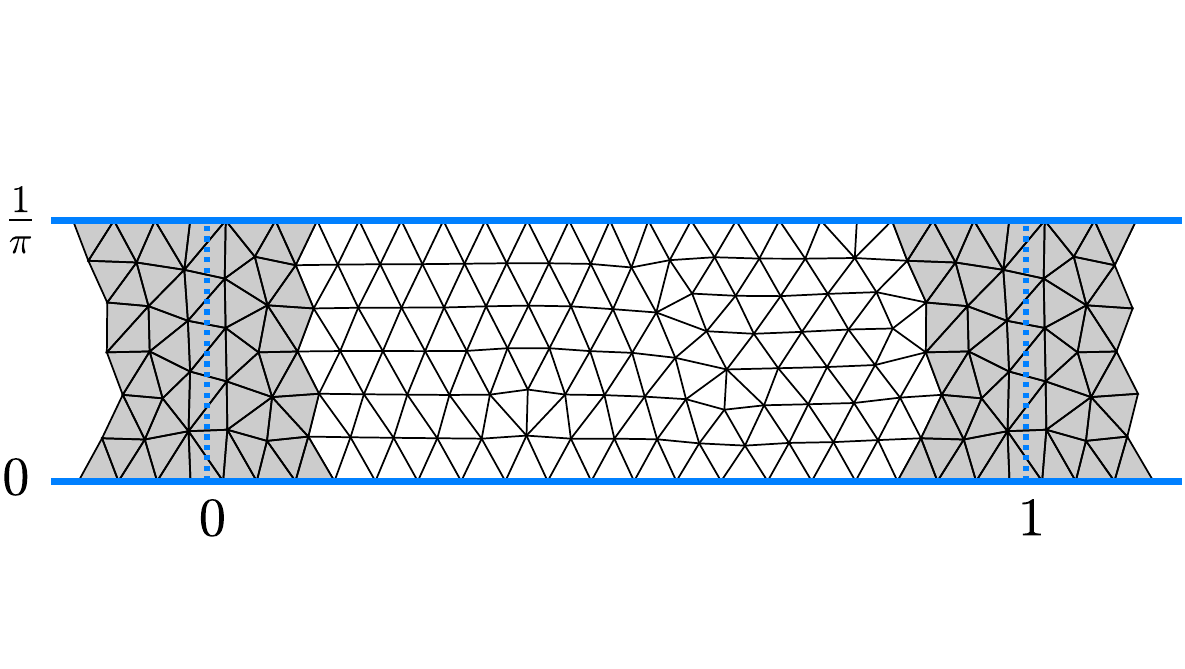}  \caption{}  \end{subfigure} ~
   \begin{subfigure}[b]{0.33\textwidth} \centering \includegraphics[width=\textwidth]{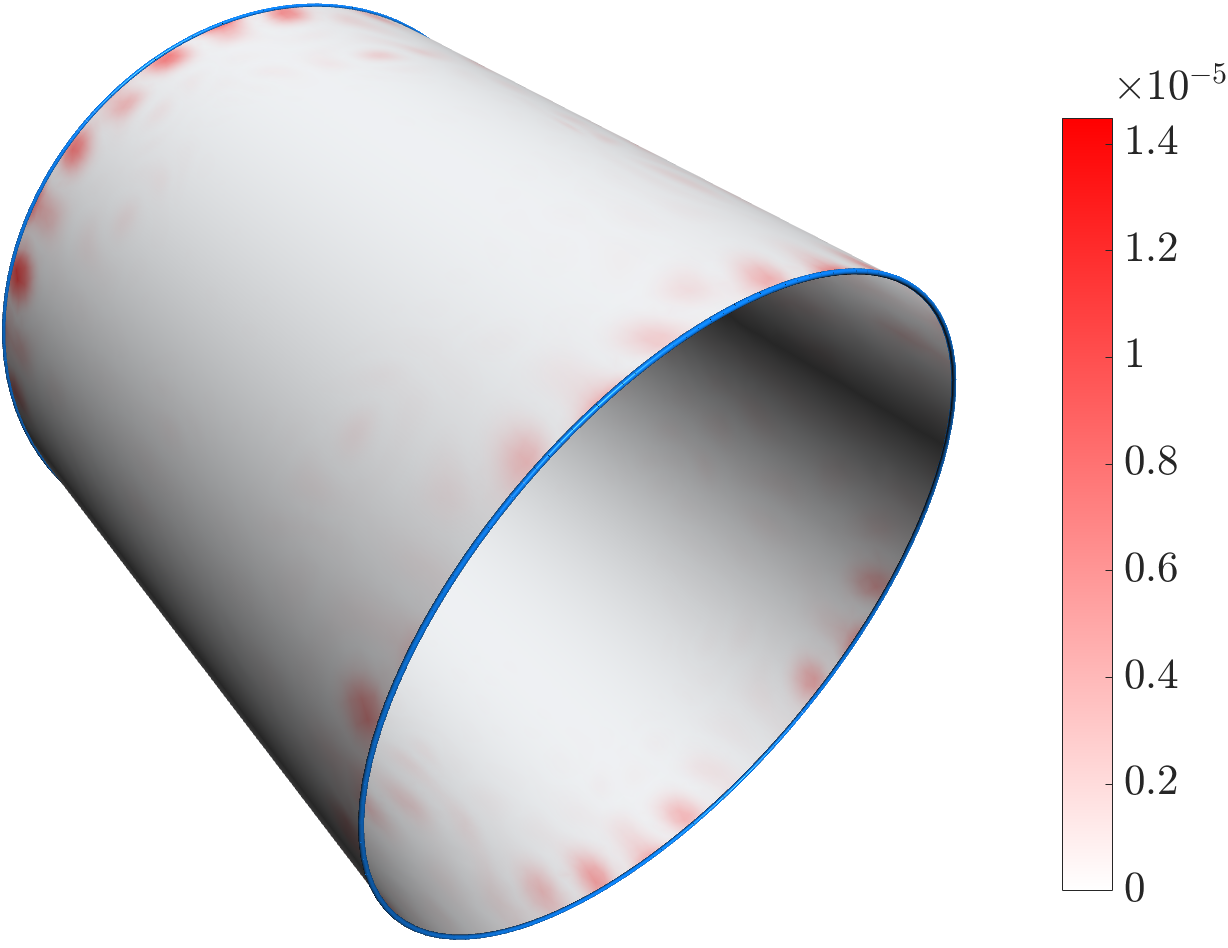}  \caption{}  \end{subfigure} ~
  \begin{subfigure}[b]{0.28\textwidth} \centering \includegraphics[width=\textwidth]{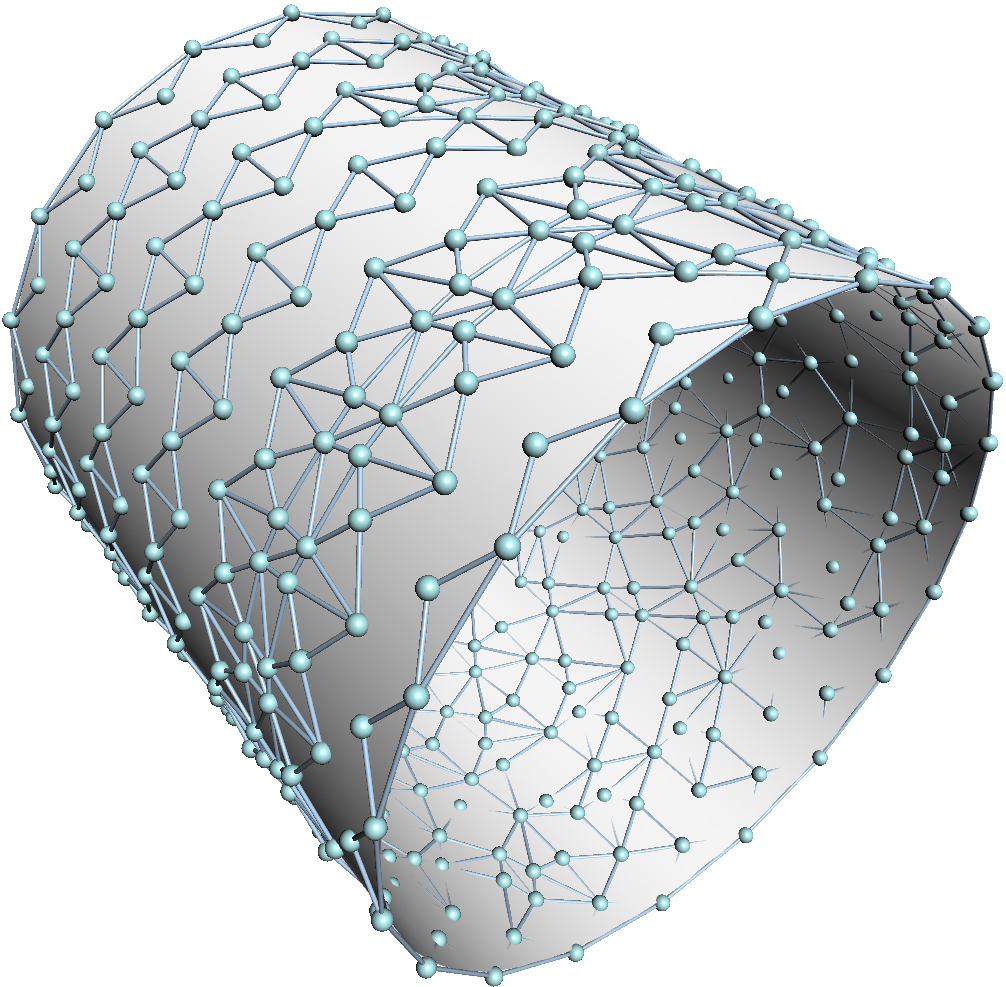}  \caption{}  \end{subfigure} ~
  \begin{subfigure}[b]{0.34\textwidth} \centering \includegraphics[width=\textwidth]{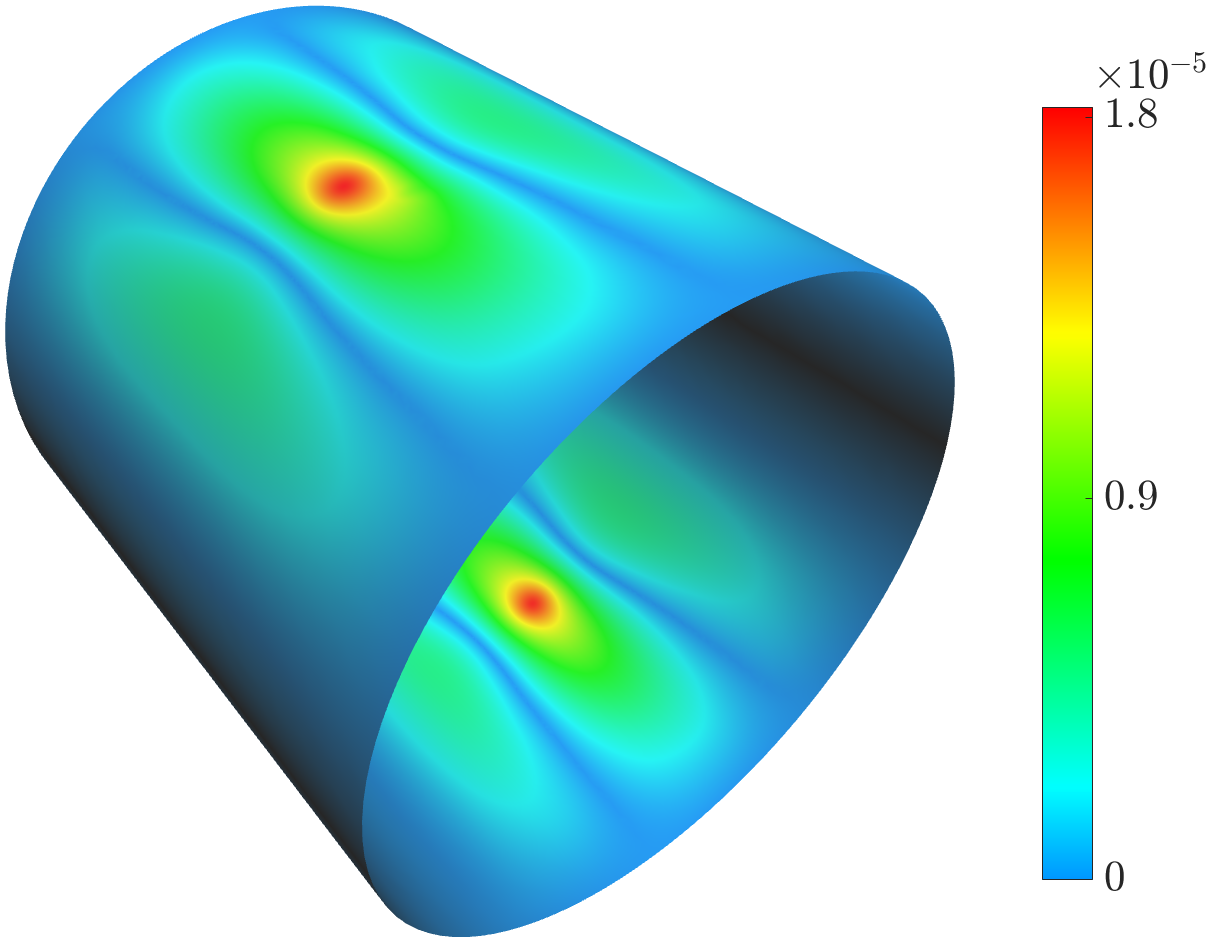}  \caption{}  \end{subfigure} ~
  \begin{subfigure}[b]{0.34\textwidth} \centering \includegraphics[width=\textwidth]{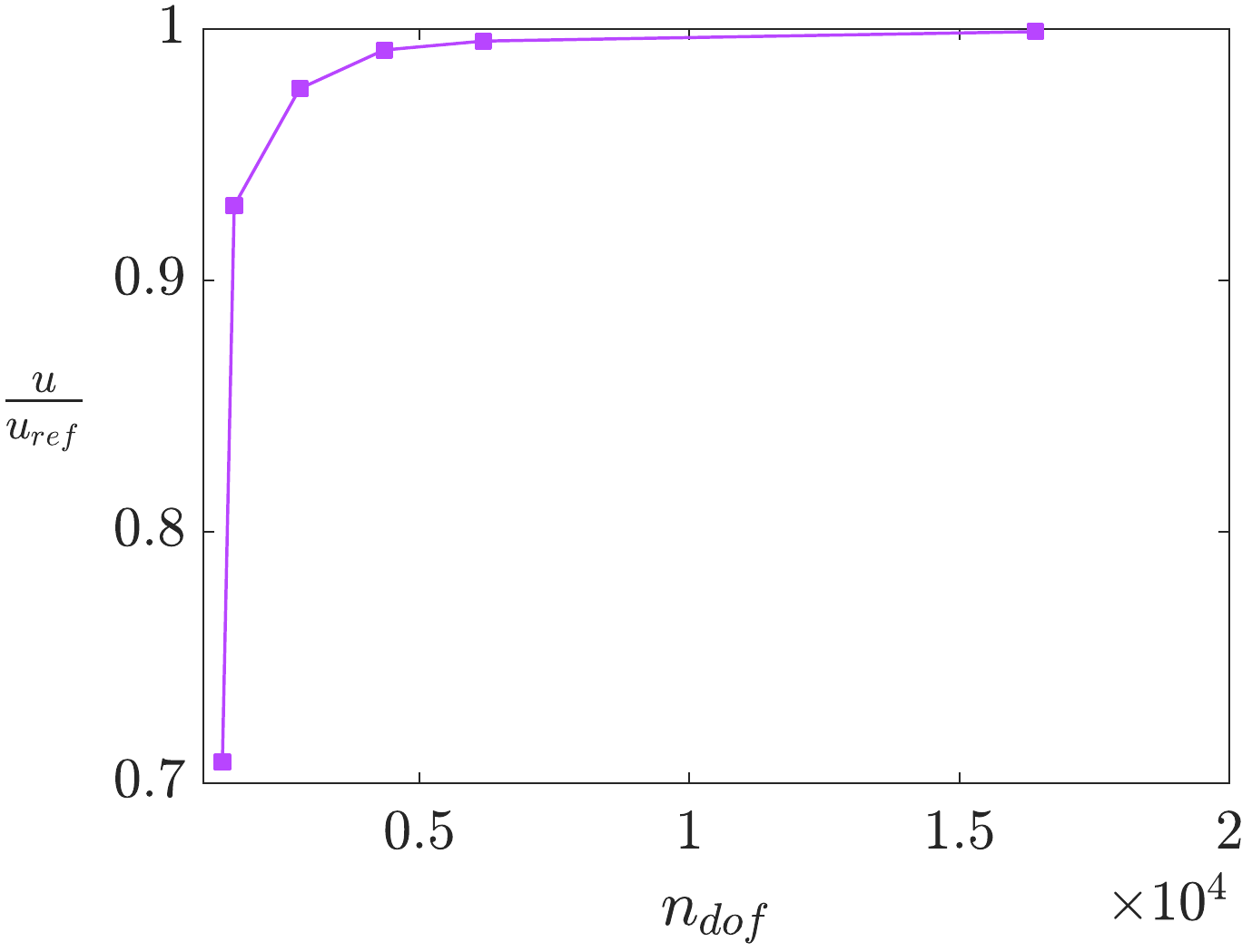}  \caption{}  \end{subfigure}
 \caption{The pinched cylindrical shell. (a) The geometry for the pinched cylindrical shell; (b-d) the knot mesh, fitting error and control net for the fitting result; (e) the final analyzed displacement magnitude; and (f) the convergence of normalized displacement.} \label{fig:pinched cylindrical}
\end{figure}

\subsubsection{Pinched cylindrical shell}
In the second problem of the shell obstacle course, the pinched cylindrical shell, supported by rigid diaphragms at the ends, is subjected to a pair of opposing radial point loads $F = 1.0~\rm{N}$ in the middle; see Fig.~\ref{fig:pinched cylindrical}(a).  Its geometric parameters are the radius $R = 300.0~\rm{mm}$, the length $L=600.0~\rm{mm}$, and the thickness $t=3.0~\rm{mm}$. The material properties are the Poisson's ratio $\nu = 0.3$ and Young's modulus $E=3.0\times 10^6~\rm{N/{mm^2}}$. The reference solution $u_{ref} = 1.8248 \times 10^{-5}$ is given as the radial displacement under the concentrated point loads~\cite{Belytschko:1985:CMAME}; see the points $A$ in Fig.~\ref{fig:pinched cylindrical}(a).

We can naturally unfold and scale the cylindrical model to a rectangular domain $\Omega = [0,1] \times [0, 1/\pi]$. To represent the cylindrical model, we use a periodic TCB-spline surface defined over $\Omega$ by repeating the knots and corresponding LTP across the periodic interface, i.e., the left and right boundaries. An example of a knot mesh is shown in Fig.~\ref{fig:pinched cylindrical}(b), where repeated initial triangles in the LTP are filled with gray. Based on the periodic knots, the reconstructed TCB-spline surface has repeated control points and marginal fitting errors; see Fig.~\ref{fig:pinched cylindrical}(c\&d). During the adaptive refinement in analysis phase, the knot mesh is refinement by setting $\eta_1=0.4$ and $\eta_2=0.6$. After several iterations of knot refinement, we obtain a converged result, as shown in Fig.~\ref{fig:pinched cylindrical}(e), where the radial displacement is visualized using color-coded representation. The convergence curve of the normalized radial displacement versus the number of degrees of freedom in the refinement iteration is shown in Fig.~\ref{fig:pinched cylindrical}(f), where the converged solution is $1.8245 \times 10^{-5}$ at point $A$, which is slightly smaller than the reference value $u_{ref}$.

\subsection{Nonlinear problem}

\subsubsection{Pipe junction \label{Pipe_junction_problem}}

\begin{figure}
  \centering
  \graphicspath{{figures/}}
 \hspace{-0.5cm}
  \begin{subfigure}[b]{0.27\textwidth} \centering \includegraphics[width=\textwidth]{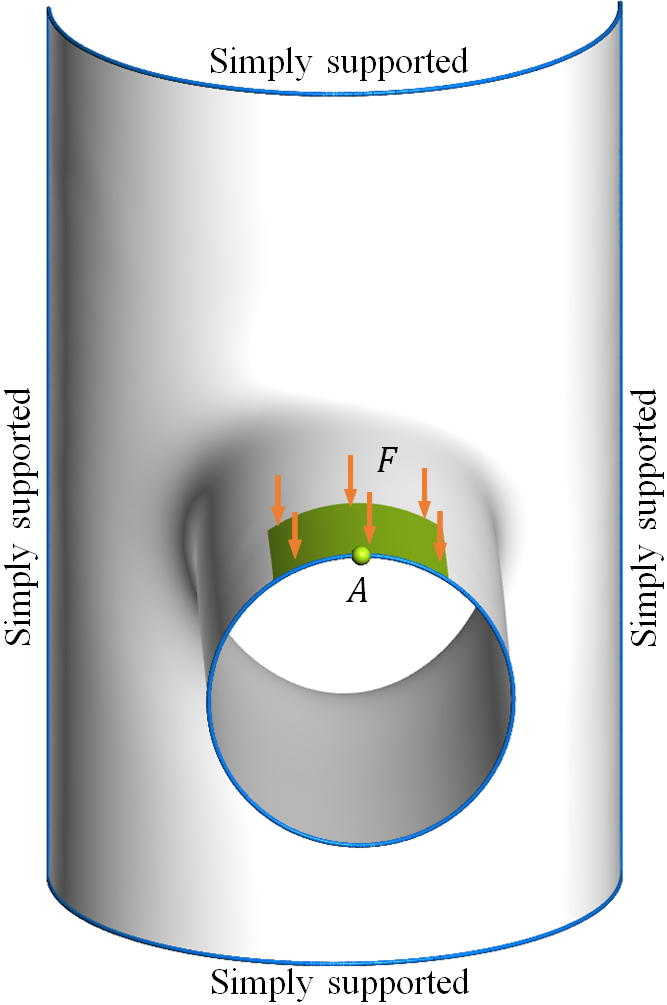} \caption{}  \end{subfigure} \hspace{2.9cm}
  \begin{subfigure}[b]{0.5\textwidth} \centering \includegraphics[width=\textwidth]{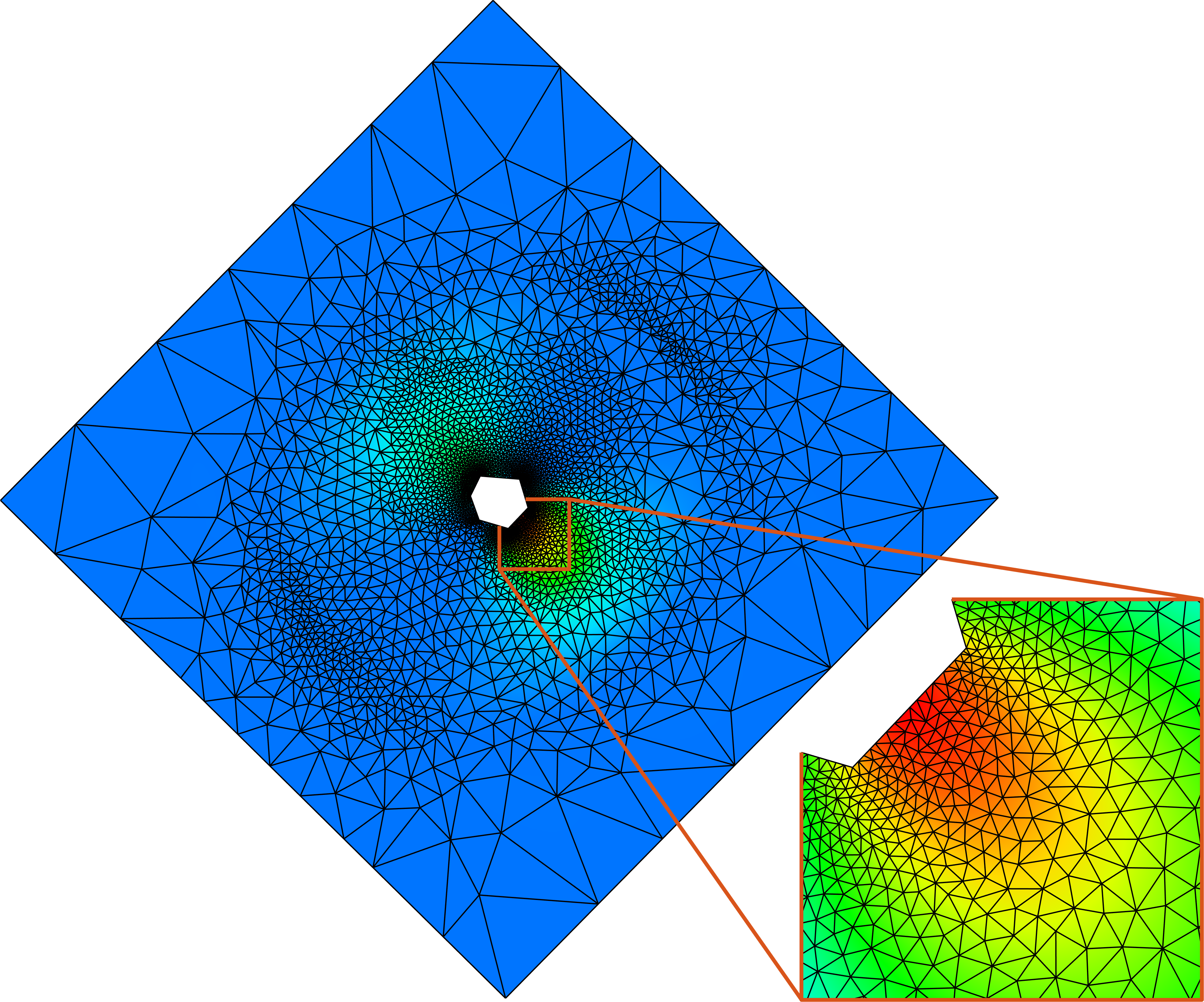} \caption{}  \end{subfigure} \hspace{3cm}

  \begin{subfigure}[b]{0.4\textwidth} \centering \includegraphics[width=\textwidth]{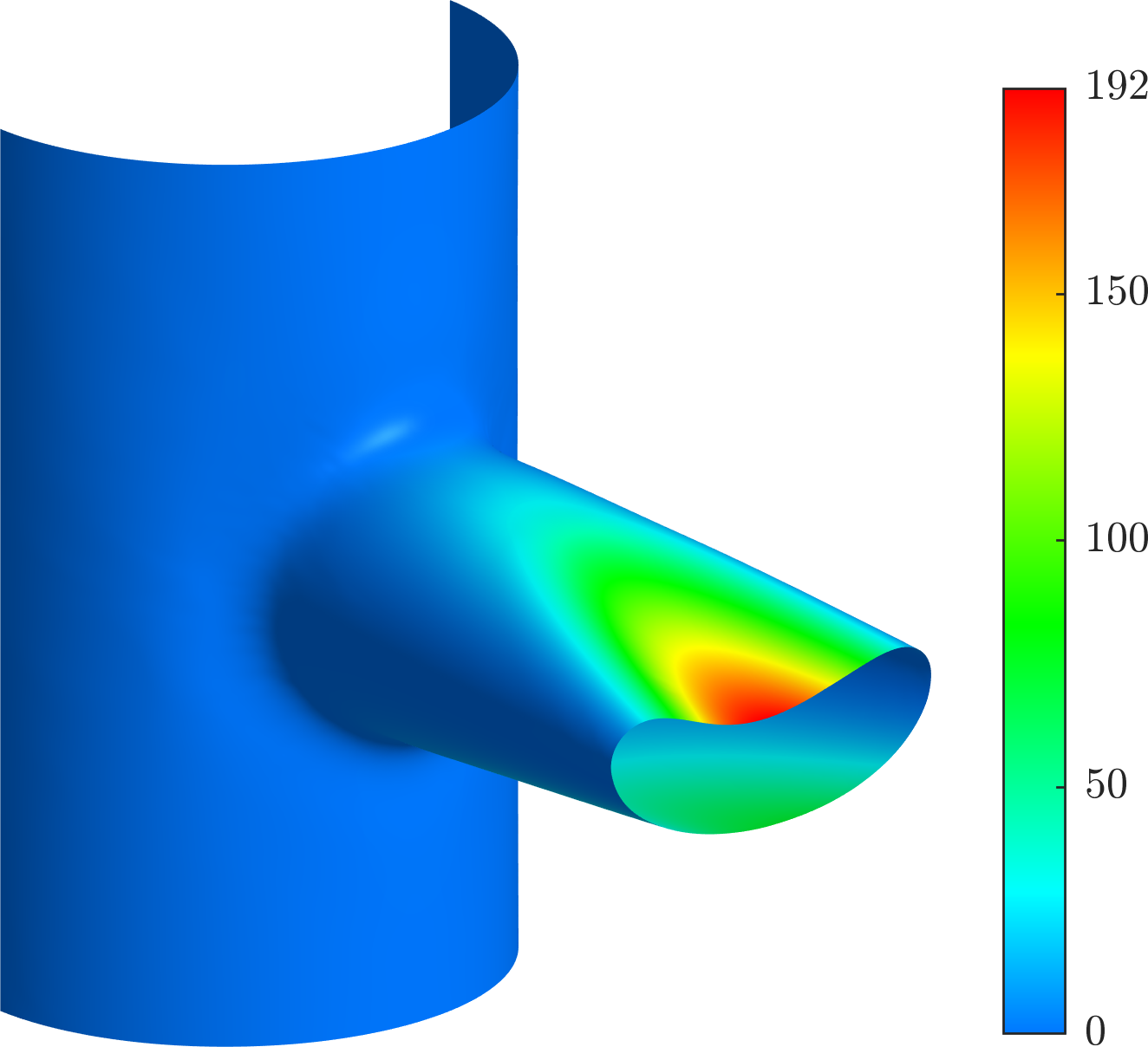} \caption{}  \end{subfigure}  \hspace{1.8cm}
  \begin{subfigure}[b]{0.45\textwidth} \centering \includegraphics[width=\textwidth]{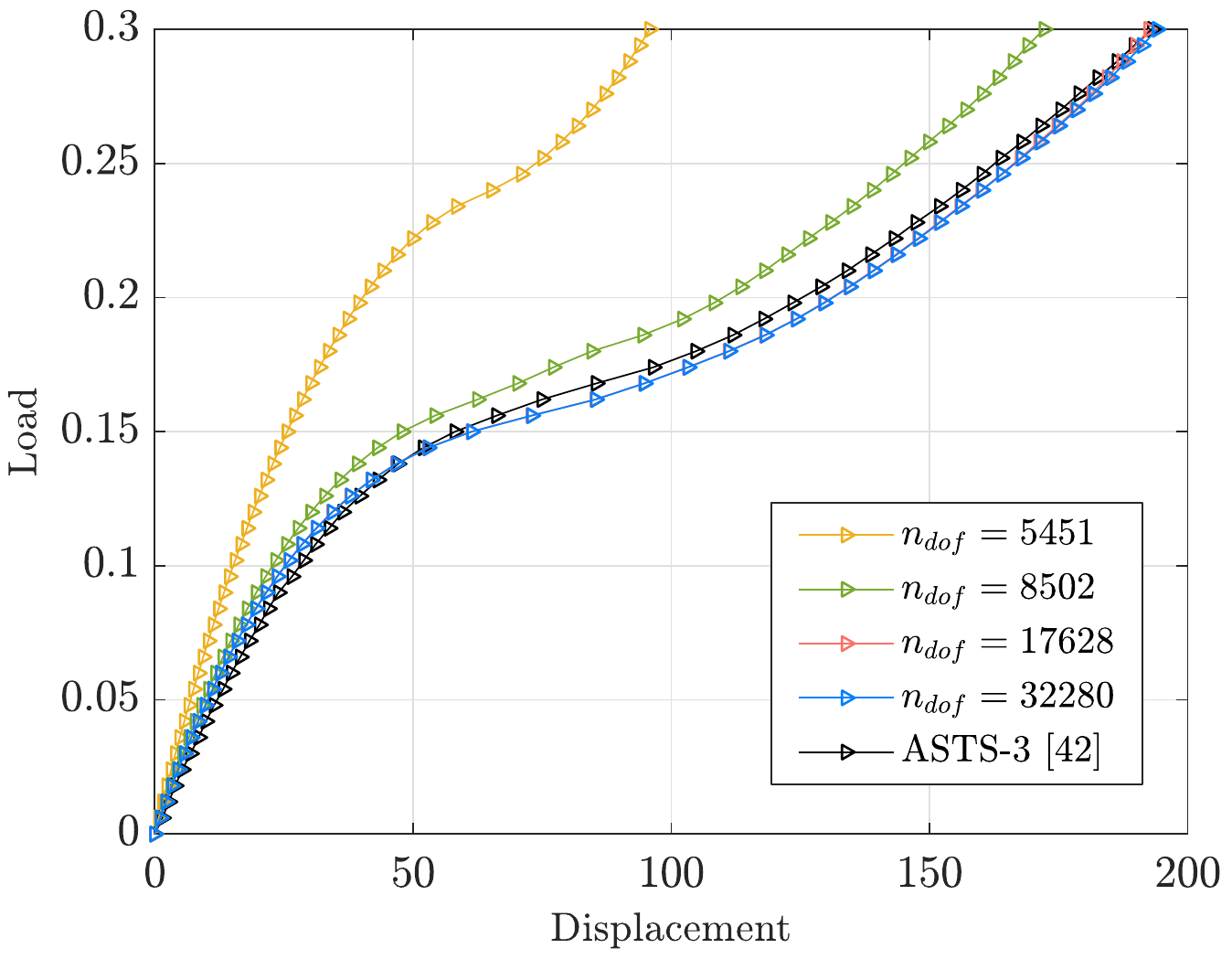}  \caption{}  \end{subfigure}
\caption{The pipe junction shell. (a) The geometry for pipe junction; the final analyzed knot mesh and the corresponding deformed geometry with color-coded displacement magnitude in (b) and (c), respectively; and (d) the load-deflection curves.\label{fig:pipe_shell_nonlinear}}
\end{figure}

\begin{figure}
  \centering
  \graphicspath{{figures/}}

  \begin{subfigure}[b]{0.43\textwidth} \centering \includegraphics[width=\textwidth]{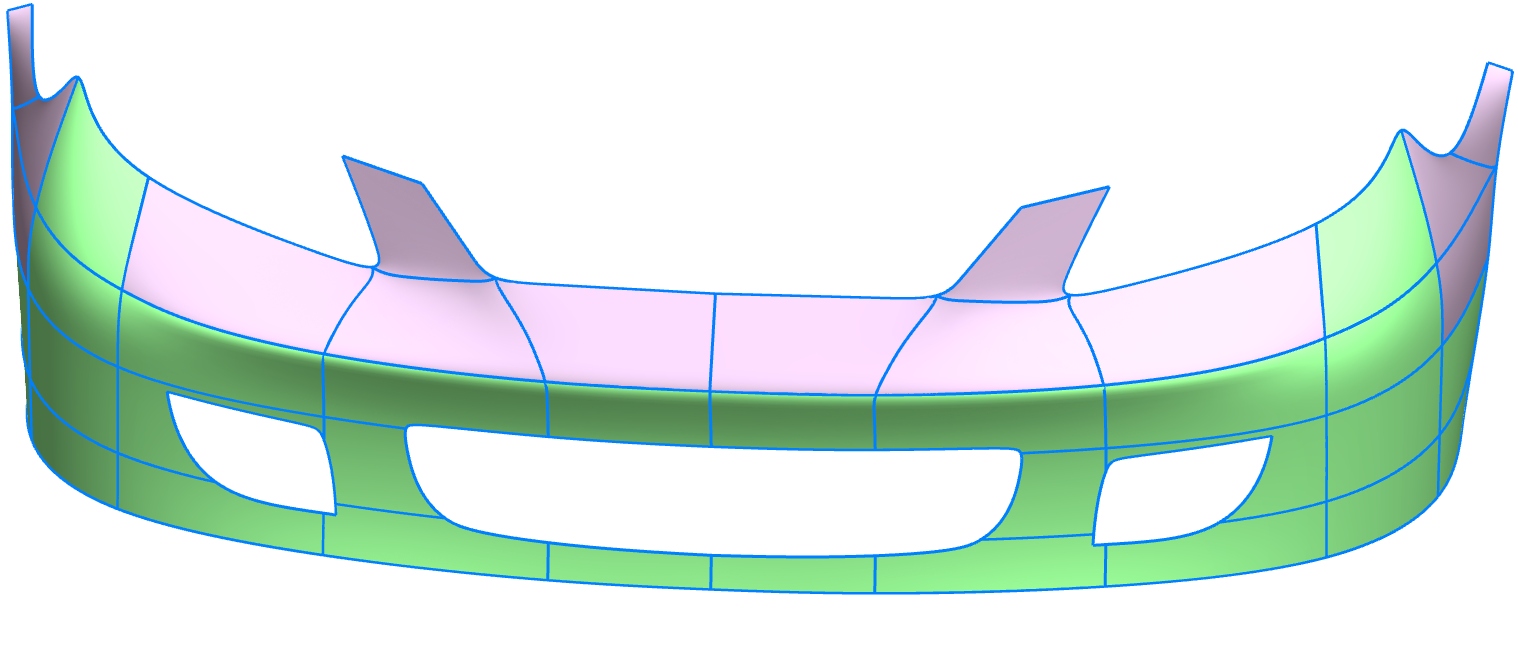} \caption{}  \end{subfigure} \qquad
   \begin{subfigure}[b]{0.45\textwidth} \centering \includegraphics[width=\textwidth]{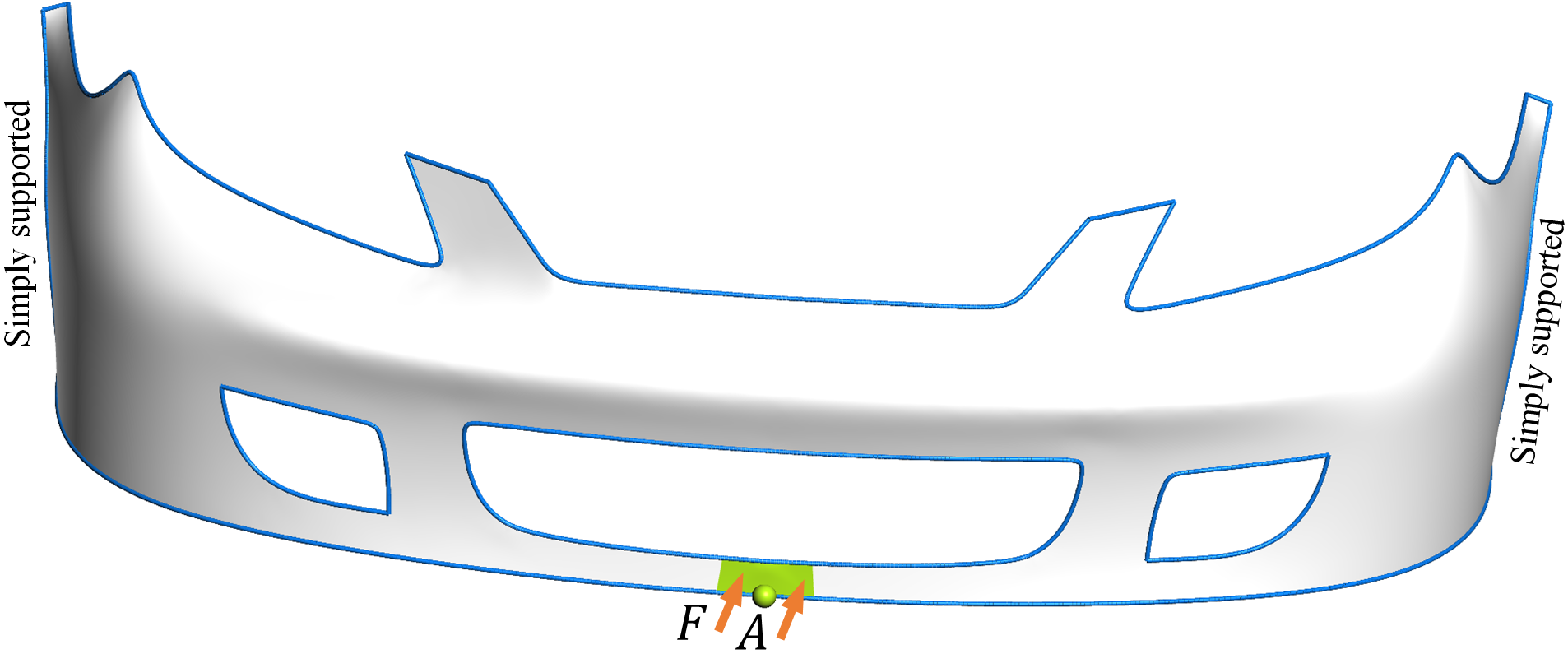} \caption{Problem geometry}  \end{subfigure}

  \begin{subfigure}[b]{0.45\textwidth} \centering \includegraphics[width=\textwidth]{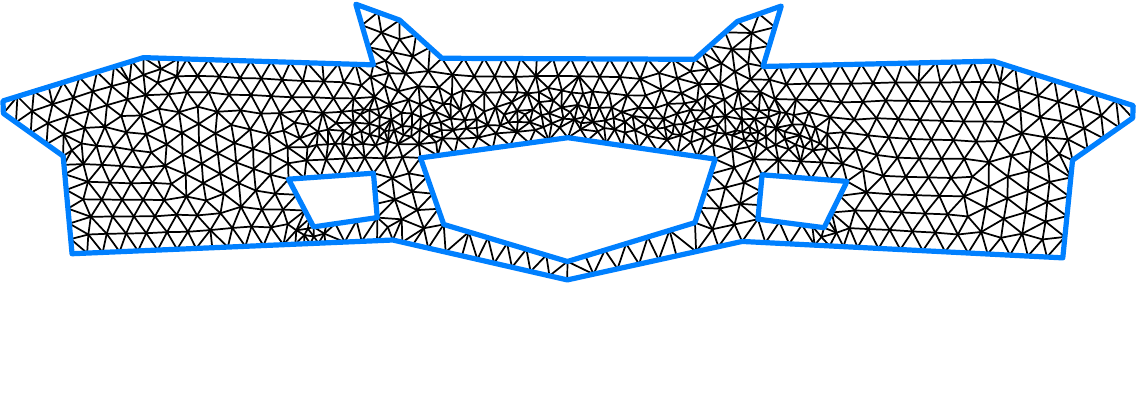} \caption{}  \end{subfigure} \quad
  \begin{subfigure}[b]{0.45\textwidth} \centering \includegraphics[width=\textwidth]{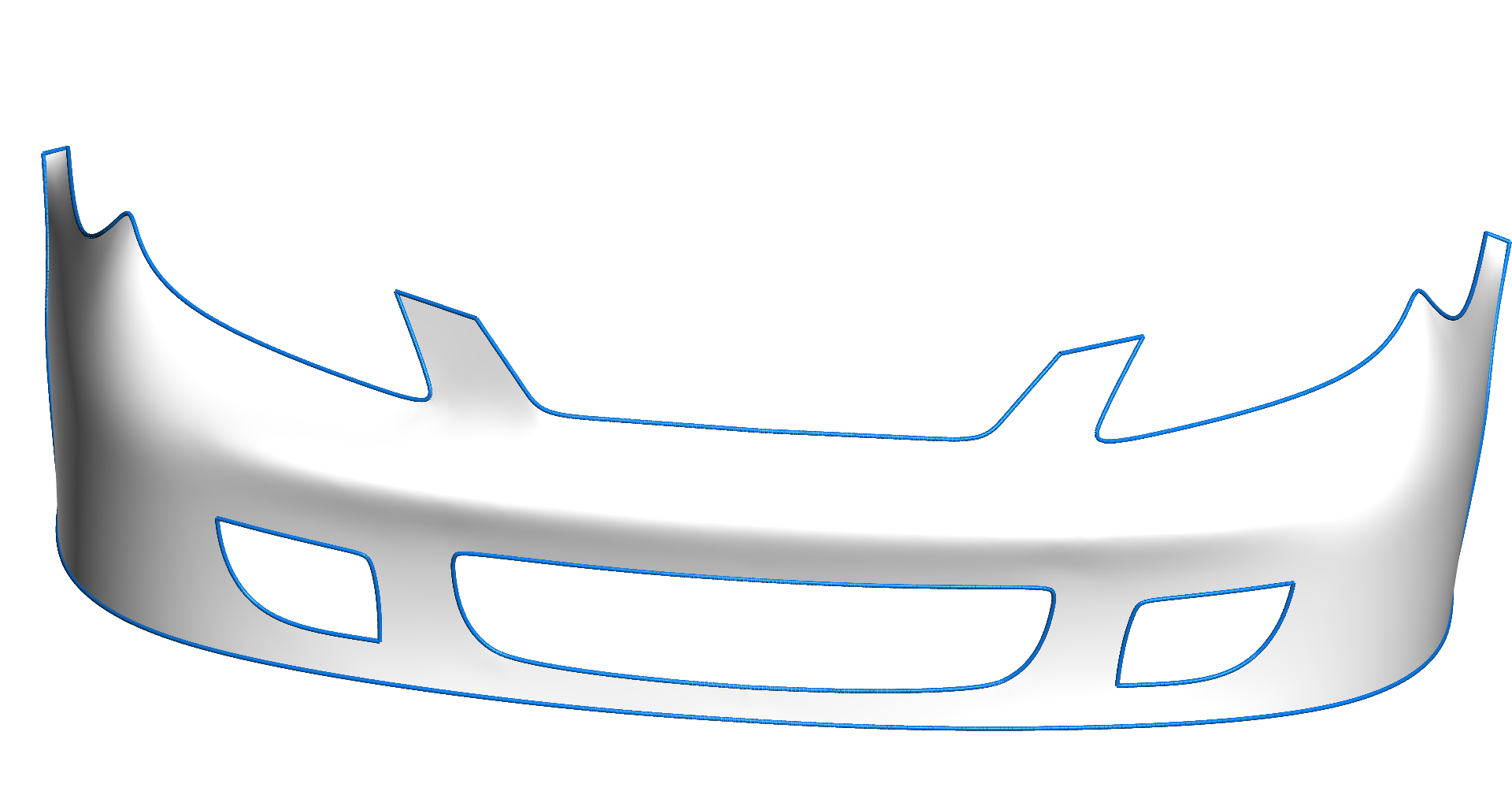} \caption{}  \end{subfigure} \quad

  \begin{subfigure}[b]{0.45\textwidth} \centering \includegraphics[width=\textwidth]{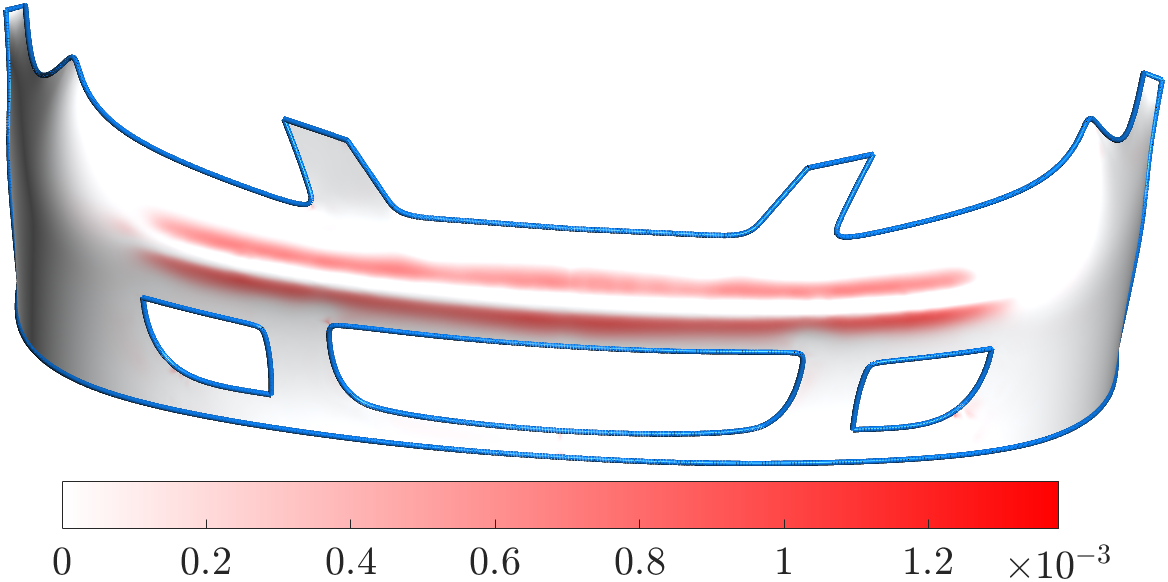} \caption{}  \end{subfigure} \quad
   \begin{subfigure}[b]{0.45\textwidth} \centering \includegraphics[width=\textwidth]{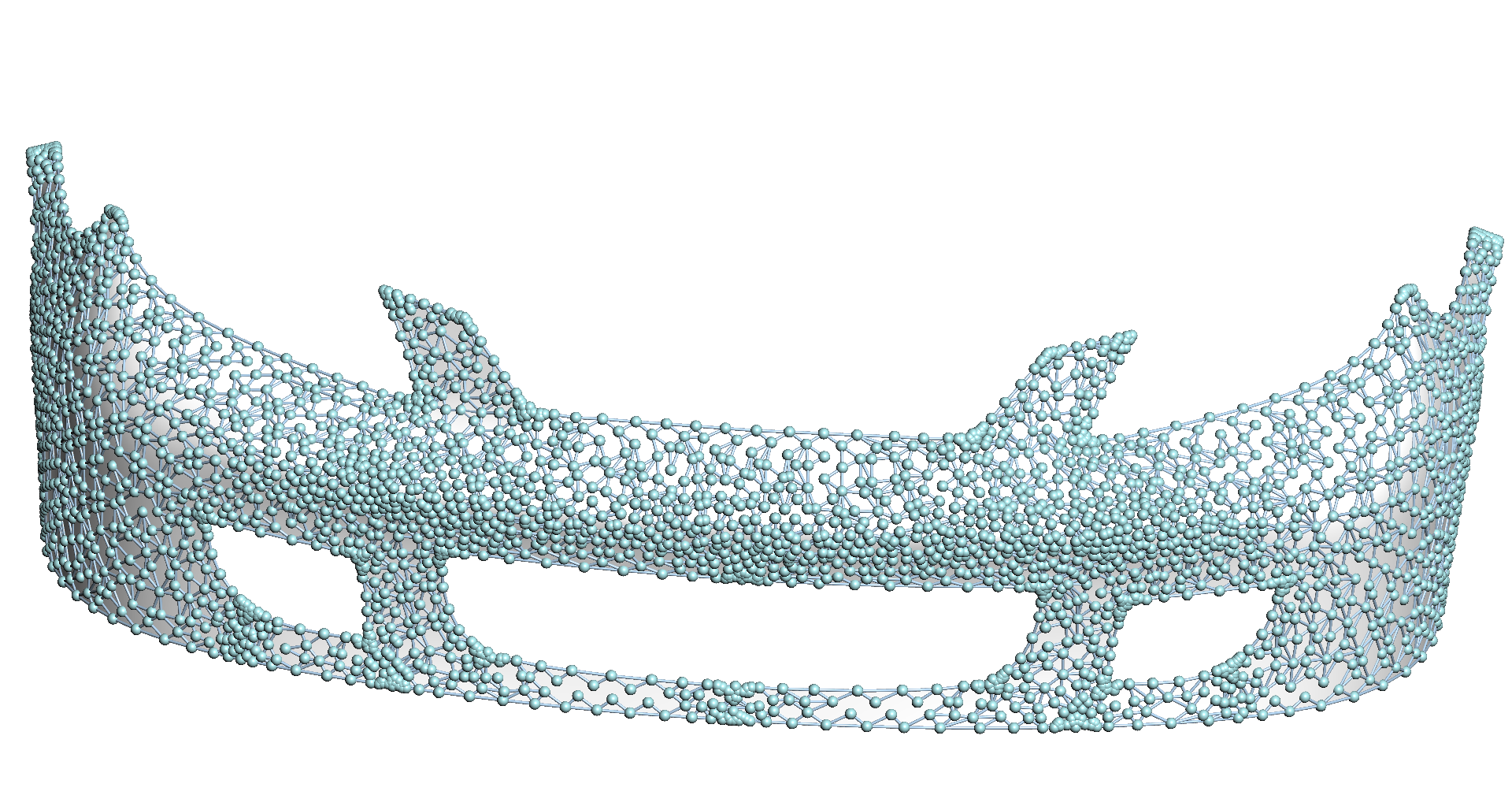} \caption{}  \end{subfigure} \quad

  \begin{subfigure}[b]{0.45\textwidth} \centering \includegraphics[width=\textwidth]{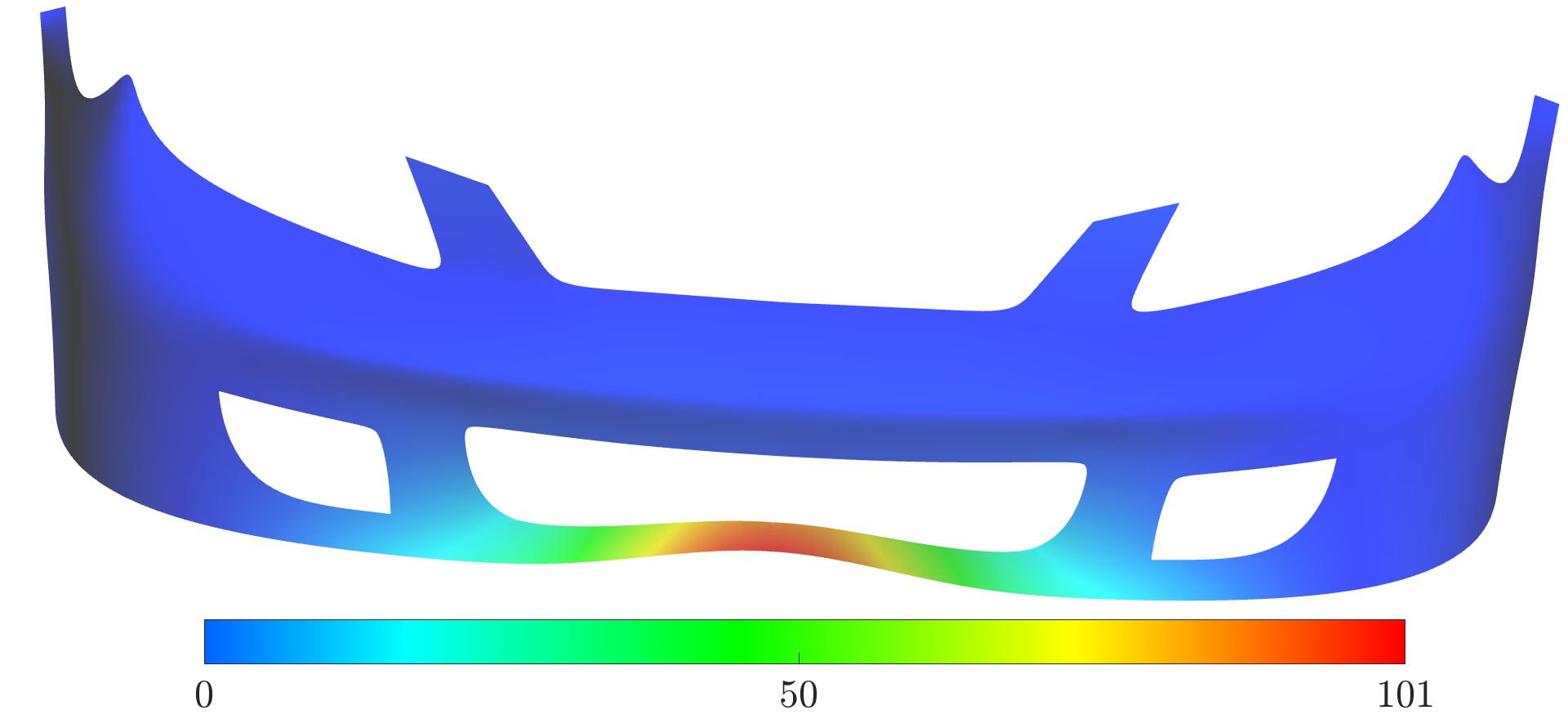} \caption{}  \end{subfigure} \hspace{1cm}
  \begin{subfigure}[b]{0.35\textwidth} \centering \includegraphics[width=\textwidth]{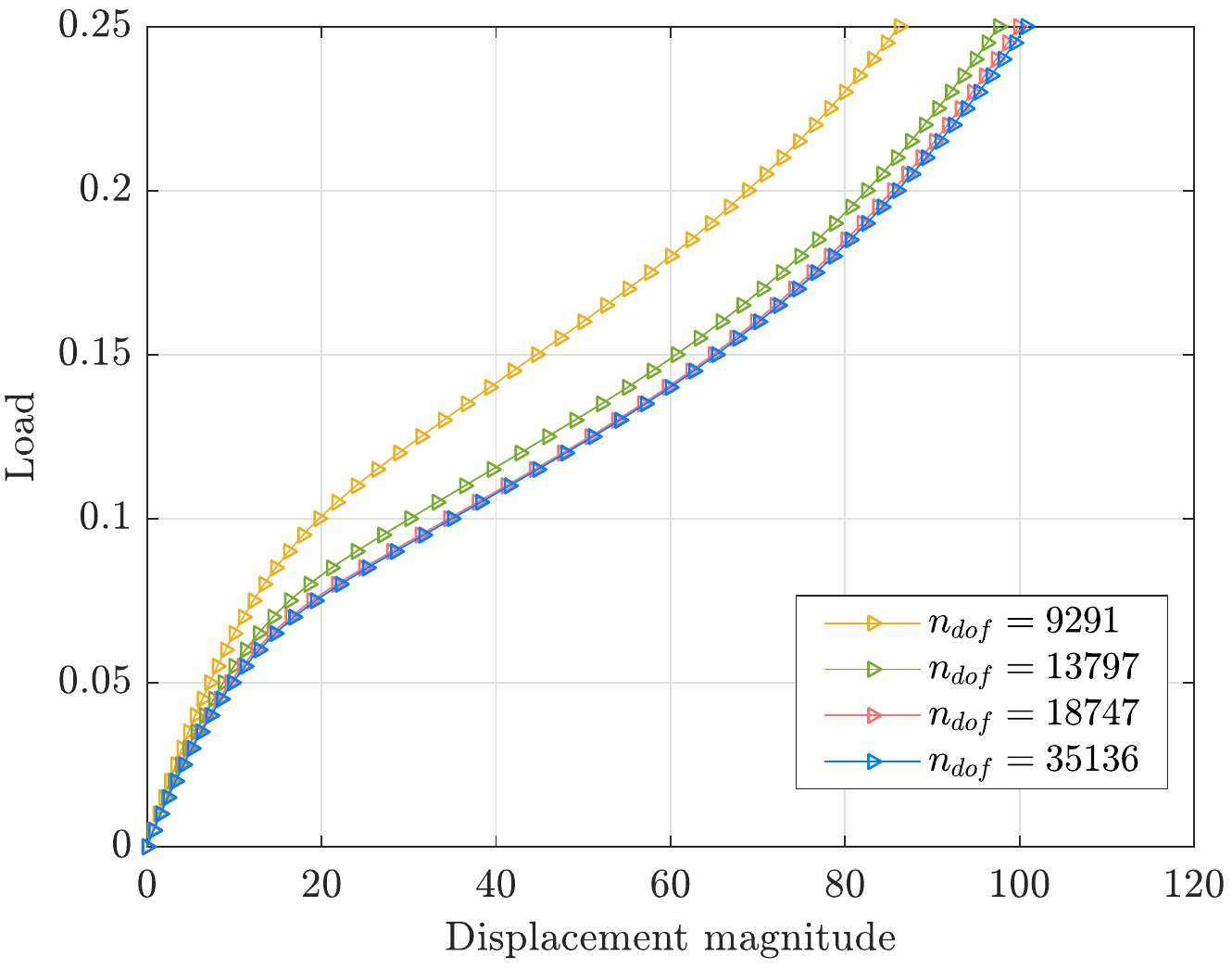} \caption{}  \end{subfigure} \quad

\caption{The front bumper shell. (a) The given CAD model with 42 patches, including 14 bivariate quintic NURBS patches (marked by pink) and 28 bivariate cubic NURBS patches (marked by green), and 5314 control points; (b) geometry, boundary conditions, and distributed load applied for the front bumper; (c-f) the knot mesh, remodel surface, fitting error and control net for the final fitting result; (g) the final analyzed deformed geometry with color-coded displacement magnitude; (h) the load-deflection curves. \label{fig:fbc_fitting}}
%

%

%
\end{figure}

In this example, we consider the pipe junction model consisting of four trimming patches, as shown in Fig.~\ref{fig:fit_algorithm}(a). A similar shape is successfully represented by a single piece of AST-spline surfaces in~\cite{Casquero:2020:CMAME}, introducing two extraordinary points with a valence of $6$.  We here represent this model using a single piece of TCB-spline surface that is defined over a quadrilateral polygonal region with a hexagonal hole in the middle, where no extraordinary point is introduced in the interior of the parametric domain; see Fig.~\ref{fig:fit_algorithm}(d\&l).

We consider this model subjected to downward loads $F=0.3~\rm{N}$, as depicted in Fig.~\ref{fig:pipe_shell_nonlinear}(a) and the external boundary of the model is supported, where the load region is marked in green. Its parameters are given as the thickness $t=0.04~\rm{mm}$, the Poisson's ratio $\nu = 0.25$, and the Young's modulus $E=2.1\times 10^5~\rm{N/{mm}^2}$. The adaptive knot meshes during the selected iterations in the fitting and analysis phases of the model are depicted in Fig~\ref{fig:knot_refinement}(a-d) and Fig~\ref{fig:knot_refinement}(e-h), respectively. The magnitudes of displacement of the model are visually represented with color-coding in Fig.~\ref{fig:pipe_shell_nonlinear}(b) and Fig.~\ref{fig:pipe_shell_nonlinear}(c), respectively. In this example, we set $\eta_1=0.85$ and $\eta_2=0.85$ and apply $50$ equal load steps in the process of nonlinear analysis. The load-deflection curves of the load versus the displacement magnitude at the point $A$ in  Fig~\ref{fig:pipe_shell_nonlinear}(a) are plotted in Fig.~\ref{fig:pipe_shell_nonlinear}(d), implying the convergence of the result. To achieve convergence, our method utilizes $32,280$ degrees of freedom, while the ASTS-3~\cite{Casquero:2020:CMAME} employs $52,836$ degrees of freedom. It should be noted that the slight difference in geometric model and load region compared to~\cite{Casquero:2020:CMAME} may result in slight variations in the convergence plot of our results; see Fig.~\ref{fig:pipe_shell_nonlinear}(d).

\subsubsection{Front bumper}

To demonstrate the versatility and applicability of our method in real-world scenarios, we showcase a car front bumper structure in Fig.~\ref{fig:fbc_fitting}(a). The original model consists of $42$ patches, including $14$ bivariate quintic NURBS patches and $28$ bivariate cubic NURBS patches, and $5,314$ control points. We consider this model subjected to distributed load $F=0.25~\rm{N}$ and simply supported
at the left and right sides, as depicted in Fig.~\ref{fig:fbc_fitting}(b). The parameters are defined as the thickness $t=1.5~\rm{mm}$, the Poisson's ratio $\nu = 0.25$, and the Young's modulus $E=1.6\times 10^6~\rm{N/{mm^2}}$.
 
A polygonal domain with three holes is constructed as the parametric domain for TCB-splines, see Fig.~\ref{fig:fbc_fitting}(c). After employing adaptive refinement, we obtain a TCB-spline surface that fits over the knot mesh as shown in Fig.~\ref{fig:fbc_fitting}(c). Fig.~\ref{fig:fbc_fitting}(d-f) illustrate the fitting surface, the normalized fitting error, and the control net, respectively. The adaptive refinement is performed during the analysis phase with parameters $\eta_1=0.9$ and $\eta_2=0.9$. A distributed load is applied in $50$ equal load steps. The resulting deformed geometry, with color-coded displacement magnitudes, is presented in Fig.~\ref{fig:fbc_fitting}(g). Furthermore, we measure the displacement magnitude at point $A$ in Fig.~\ref{fig:fbc_fitting}(b) for each refinement knot mesh. It is observed that the load-deflection curve shows convergence as the knot mesh is refined.

\section{Conclusions \label{conclusions}}

This paper develops a new framework for representing and analyzing KL shells with complex geometries using TCB-splines in the context of isogeometric analysis. The input CAD model, which may consist of multiple patches of trimmed NURBS, is remodeled as a single piece of TCB-spline surface using the linear least-squares fitting method. The flexibility of local refinement of TCB-splines makes the analysis efficient. The feasibility and accuracy of our approach are validated by solving some benchmark problems, including both linear and nonlinear deformation. The applicability of the proposed approach to shell analysis is further exemplified by performing geometrically nonlinear KL shell simulations of a pipe junction and a front bumper.

One limitation of our current scheme is that we only focus on smooth mid-surfaces that are topologically homomorphic to an open disk with a finite number of holes. Shells in practical applications may have a  mid-surface with more general topology and locally containing geometric features with different continuity orders. Hence, it is desirable if our framework can handle shells with more general shapes. One necessary step in improving our scheme is to generalize our TCB-spline-based fitting method to models of complicated topology and features. It is theoretically possible but practically challenging and hence needs further investigation. Besides, we remodel the mid-surface without considering the model's symmetry for simplicity. It is also interesting to tailor knot placement and control point updating methods to enable the remodeled shapes to have strict symmetric geometry.

\section*{Acknowledgments}
{The research of Juan Cao was supported by the National Natural Science Foundation of China
(No. 62272402), the Xiamen Youth Innovation Funds (No. 3502Z20206029) and the Fundamental Research Funds for the Central Universities (No. 20720220037). The research of Zhonggui Chen was supported by the National Natural Science Foundation of China (No. 61972327) and Natural Science Foundation of Fujian Province (No. 2022J01001). Xiaodong Wei was supported in part by the National Natural Science Foundation of China (No. 12202269). We thank Dr. Angran Li for providing us the data of the pipe junction for the comparative study.}


  \bibliographystyle{elsarticle-num}
  \bibliography{mybibfile}





\end{document}